\documentclass{tufte-handout}
\usepackage[utf8]{inputenc}
\usepackage{graphicx}
\usepackage{amsmath,amsthm}
\usepackage{amsfonts}
\usepackage{mathtools}

\usepackage{tikz-cd}
\usepackage{tikz}
\usepackage{tkz-euclide}
\usepackage{pgfplots}
\usepackage{tikz-3dplot}
\usetikzlibrary{shapes}
\usetikzlibrary{calc}
\usetikzlibrary{fit}
\usetikzlibrary{positioning}
\usetikzlibrary{decorations.markings,arrows,arrows.meta,bending}
\usetikzlibrary{math} 
\tikzset{->-/.style={decoration={markings,
  mark=at position #1 with {\arrow[line width=2pt]{>}}},postaction={decorate}}}
\DeclareMathOperator{\im}{im}

\tikzset{
  pics/torus/.style n args={3}{
    code = {
      \providecolor{pgffillcolor}{rgb}{1,1,1}
      \begin{scope}[
          yscale=cos(#3),
          outer torus/.style = {draw,line width/.expanded={\the\dimexpr2\pgflinewidth+#2*2},line join=round},
          inner torus/.style = {draw=pgffillcolor,line width={#2*2}}
        ]
        \draw[outer torus] circle(#1);\draw[inner torus] circle(#1);
        \draw[outer torus] (180:#1) arc (180:360:#1);\draw[inner torus,line cap=round] (180:#1) arc (180:360:#1);
      \end{scope}
    }
  }
}

\newtheorem*{theorem}{Theorem}
\usepackage{framed}

\colorlet{shadecolor}{gray!15}

\newenvironment{thm}
  {\begin{shaded}\begin{theorem}}
  {\end{theorem}\end{shaded}}

\title{Topology in Sound Synthesis and Digital Signal Processing - DAFx2022 Lecture Notes}
\author{Georg Essl --- University of Wisconsin - Milwaukee}
\date{November 2022}

\begin{document}

\maketitle

\section{Introduction}

These are lecture notes accompanying a two part tutorial presented at DAFx2022 in Vienna on the topic of topology in digital signal processing and sound synthesis.

The purpose of the tutorial and hence the notes is to acquaint the attendees and reader with basic notions of topology that has proven useful in Digital Signal Processing and Sound Synthesis.

Topology and specifically algebraic topology is not a topic that is standard to a typical engineering mathematics education. Hence even basic concepts of this topic area are unknown to many practitioners in DAFx. The tutorial seeks to provide a rapid introduction in the topic while also focusing on why a DAFx researcher might benefit from learning about topology and using it in their work.

\section{Part I: Understanding Topology}

\begin{marginfigure}
    \centering
    \includegraphics[width=.95\marginparwidth]{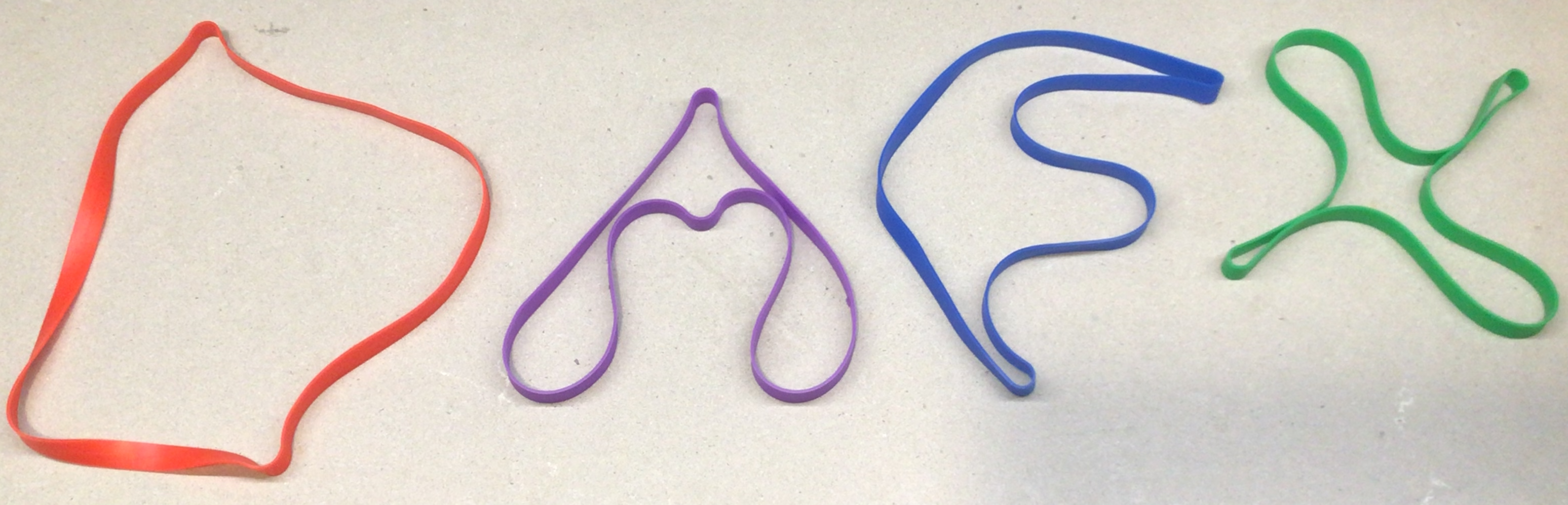}
    \caption{Rubber is a physical instance of material that is deformable, stretchable, while keeping its shape, such as that of a loop.}
    \label{fig:rubber}
\end{marginfigure}
%
What is topology? A useful intuition is to think of topology as "rubber-band" geometry (Figure \ref{fig:rubber}). Imagine idealized rubber that can be stretched arbitrarily. There are certain properties the rubber maintains under deformation, and there are some that it does not. In particular, the stretching changes distances, while a rubber band in a loop configuration does not change the loopiness by stretching. Many useful topological constructions can sensibly be visualized using this intuition, despite its lack of formality. At times, other physical metaphors can also be useful as we will see later.

\marginnote{Ways of thinking about topology.}Slightly more precisely, one can think of topology as being concerned with the following: (i) Modeling connectivity, (ii) global invariants in a geometric setting (iii) independence from metric choices (or flexible choice of metric).

Practitioners in Digital Audio are very familiar regarding (i). We constantly model connectivity when we design filter structures, networks, use visual programming languages, and the like. The connectivity part of these constructions is known as graph theory, which is a combinatorial low dimensional theory in topology. However, both (ii) and (iii) may not be familiar concepts though it is quite likely that a practitioner in DAFx has seen them but perhaps not interpreted them as a notion related to topology.

\subsection{Changing and Keeping Topology}

\marginnote{Cutting and Gluing can change Topology.}For rubber we recognize that simply deforming and stretching a rubber band loop does not change that it is a loop. However, if we apply glue, or cut with scissors this changes things. A cut loop ends up being a string and can no longer hold jar lids in place. A glued rubber band may end up having more than one loop. Hence, adding connectivity and removing connectivity can change topology.

More strictly, however, this is a helpful intuition, but is not always correct. For example one can glue a rubber band in such a way that it ends up remaining a single loop. One can then conversely cut the glued band and again arrive back at the original loop. Hence if we think of topology as numbers of loops, we see that gluing and cutting may or may not change topology, and we have to be careful. However, in practice and with some experience, it remains a useful intuition because without gluing and cutting you are safe. Your rubber band does not change topology.

\marginnote{Topological Equivalence.}More formally, one wants to define when we have a topological equivalence. There are in fact numerous notions of topological equivalence that have different properties. A further approach is to categorize topologies computationally by some notion and define equivalences by those that have the same categorization. The second way of doing this is actually more practical and we will learn about it in Part II under the name {\em Homology}.

\begin{marginfigure}
\centering
\includegraphics[width=.95\marginparwidth]{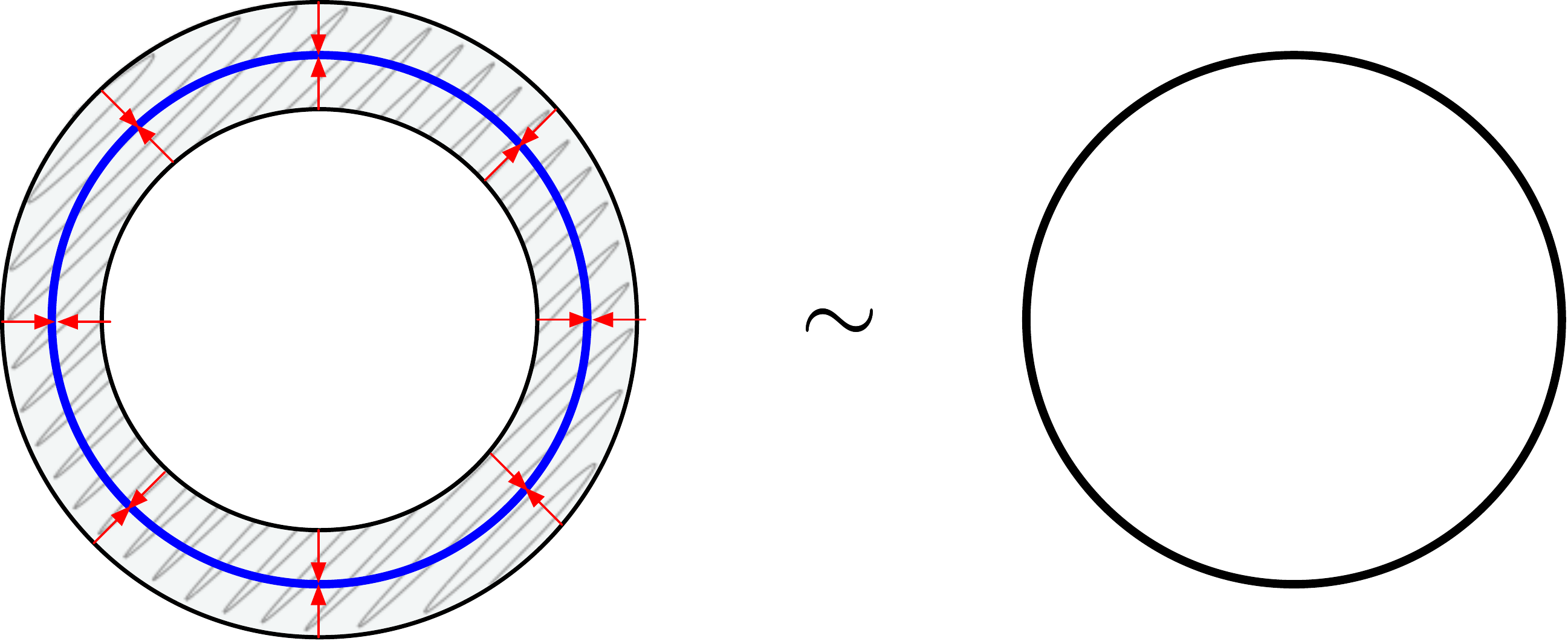}
\caption{A deformation retract of an annulus to the circle.}\label{fig:deformationretract}
\end{marginfigure}

\marginnote{Homotopy as sliding a rubber string over a topological space.}For an example of topological equivalence we will get a sense of the notion of {\em Homotopy}. Intuitively a homotopy is described by taking a piece of rubber string and moving it along a topological space. If you hit boundaries and are forced to wrap it around loops or other voids are formed, these constitute different homotopies. However moving along a surface otherwise is homotopic. This leads to the idea of a {\em deformation retract} which is shrinking all the space down that can be homotoped without obstruction. In other words "deformation retracts" capture  changes in the space that do no change the {\em homotopy type}. For example if you have a disk you can shrink that to a point. Figure \ref{fig:deformationretract} shows the deformation retract from an annulus to a circle.

We will not look at homotopy computationally. Instead it is the generally assumed underlying equivalence that tells us if two topological configurations are the same or different. We will at times augment these equivalences with extra information, but the base equivalence will be homotopy throughout our discussions. Mathematicians like to say that we are working "up to homotopy".\marginnote{We work "up to homotopy."}

\subsection{Making Loops: Quotient Topologies}

\marginnote{Making loops by taking quotients.}Repetition are important in almost anything we do in Digital Audio and a loop is a characterization of repetition. Hence loops are important, and it is interesting to know how to make loops.

Taking a rubber band we can glue its ends together to form a loop. We can formalize this by "taking the quotient" or alternatively forming a "quotient topology".

\marginnote{$[0,1]$ as a model of a flat line.}By taking two entities and declaring them identical, one makes things previously separate into one. To illustrate this, we start with a piece of string and further assume that we can think of it as the line segment with the interval $[0,1]\in\mathbb{R}$. This metric choice is for our mental convenience, as well as useful, as it relates to many practical settings, but it is not a restriction. If we chose an identification of the beginning of the line segment at $0$ with its end at $1$ we form an equivalence relation $0=1$. \marginnote{Equivalence classes are entities that are defined to be identical.}More generally, we can relate whole classes under equivalences, aptly called {\em equivalence classes}.

By having identified $0$ and $1$, they are no longer separate entities, hence our domain changed $[0,1)$ (as the $1$ is now just the $0$). 

\begin{marginfigure}
    \centering
    \includegraphics[width=.95\marginparwidth]{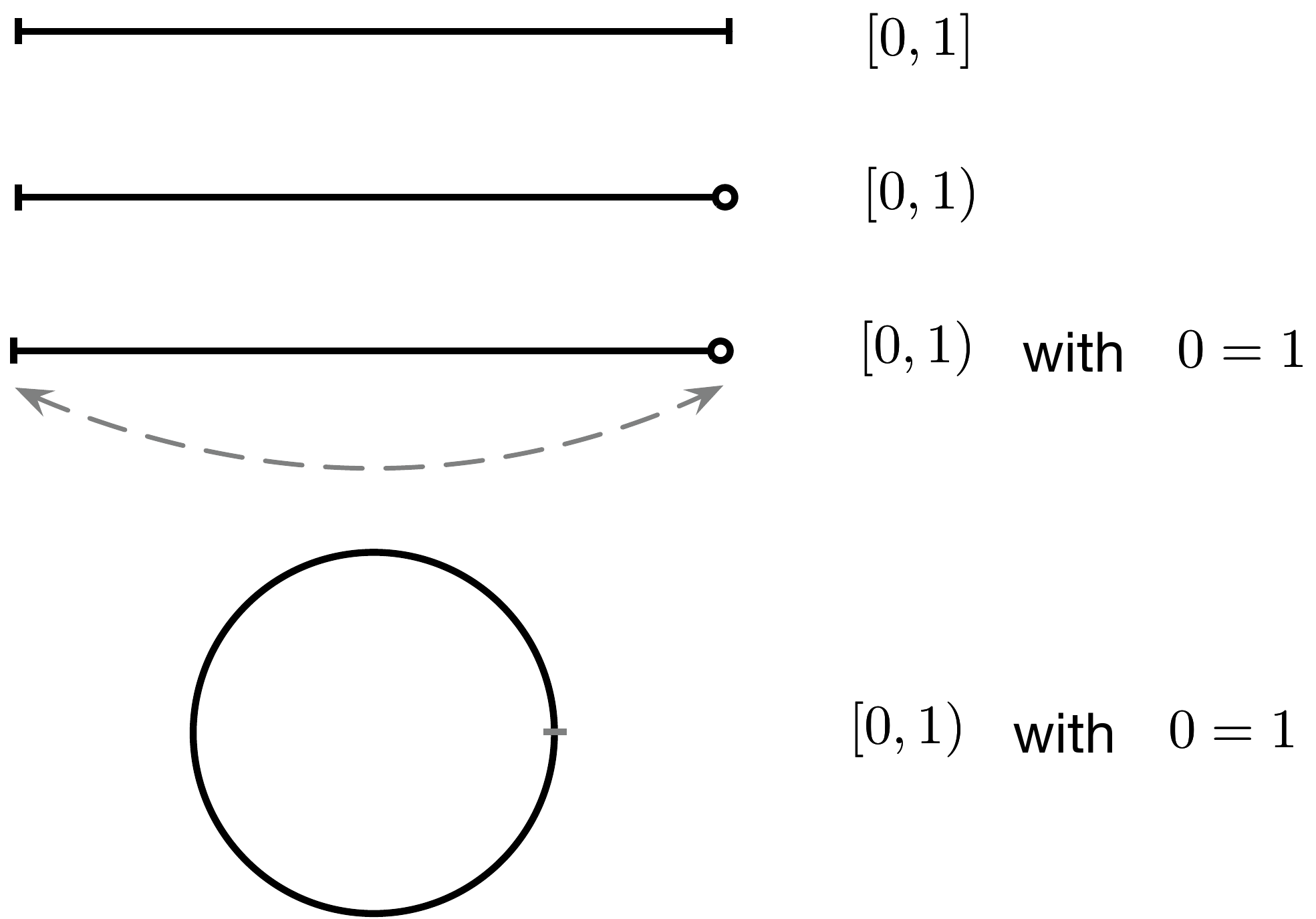}
    \caption{Forming a flat circle by identifying two ends of a line segment.}
    \label{fig:quotient}
\end{marginfigure}
There are a few ways one can think of this construction as seen in Figure \ref{fig:quotient}. If we do not insist on the flatness of the depiction of the domain, we see that it can be thought of as a parametrization of a circle. The identified flat line has an apparent jump while the circle is nicely continuous. It is also not obvious how one could apply a kind of rubber band analogy to describe the relationship.

One way out of this is by a relationship that we will call (un)rolling. Another way to form the same intuition is by (un)winding.

\begin{figure}
    \centering
    \includegraphics[width=.95\columnwidth]{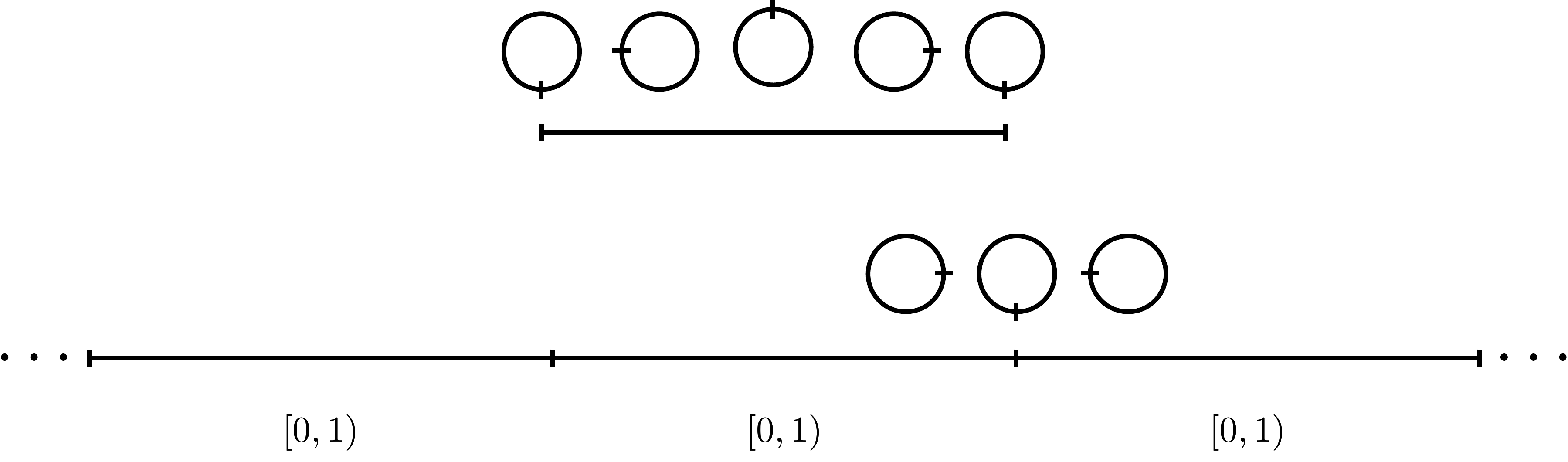}
    \caption{Rolling a circle on a line. The circle can roll over repeated segments of the line indicating that the circumference continues without obstruction.}
    \label{fig:cover}
\end{figure}

At the top of figure \ref{fig:cover} we see the circle rolling along the line and a one-to-one correspondence between the line contact and the circumference position on the circle. Hence a repeated "unrolled" line actually captures the continuous nature of the circle in a flat fashion. The winding intuition comes from holding the circle fixed and winding an infinite rubber string around the circle and marking the positions where the winding repeats. If we then unwind again and lay down the string we arrive at the identical situation as depicted in figure \ref{fig:cover}. In the winding case it is more obvious that the wound string {\em covers} the circle. If we just wind a finite amount of times we would call the result just a {cover}. \marginnote{Universal Cover of the circle is the infinite line with segmented repetition.}If we create a winding that encapsulates all possible covers (we wind indefinitely as there is no obstruction to winding) we arrive at what is called the {\em universal cover}. 

\begin{marginfigure}
    \centering
    \includegraphics[width=.95\marginparwidth]{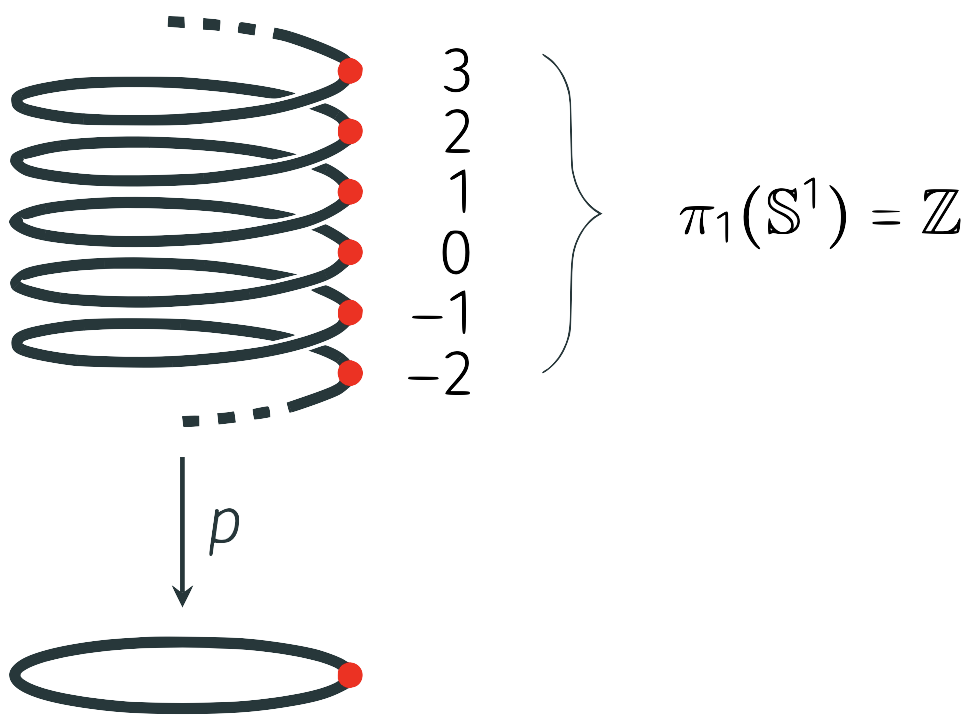}
    \caption{Fundamental group of the circle.}
    \label{fig:fundamentalgroup}
\end{marginfigure}

While we have been discussing flat lines and circles at this point it is important that this argument is --- pun intended --- substantially more flexible. Imagine a square with a circumference of $1$ and we wind a rubber string around it as before and we mark repetitions. We arrive at the identical image as before. In fact we can do this form any path-like loop. Imagine that there are concave parts and we force the rubber into its nooks but then unwind to check the cover: we again arrive at the same cover. In short, these pictures are topological and are robust under deformation.
\begin{figure}
    \centering
    \includegraphics[width=.95\columnwidth]{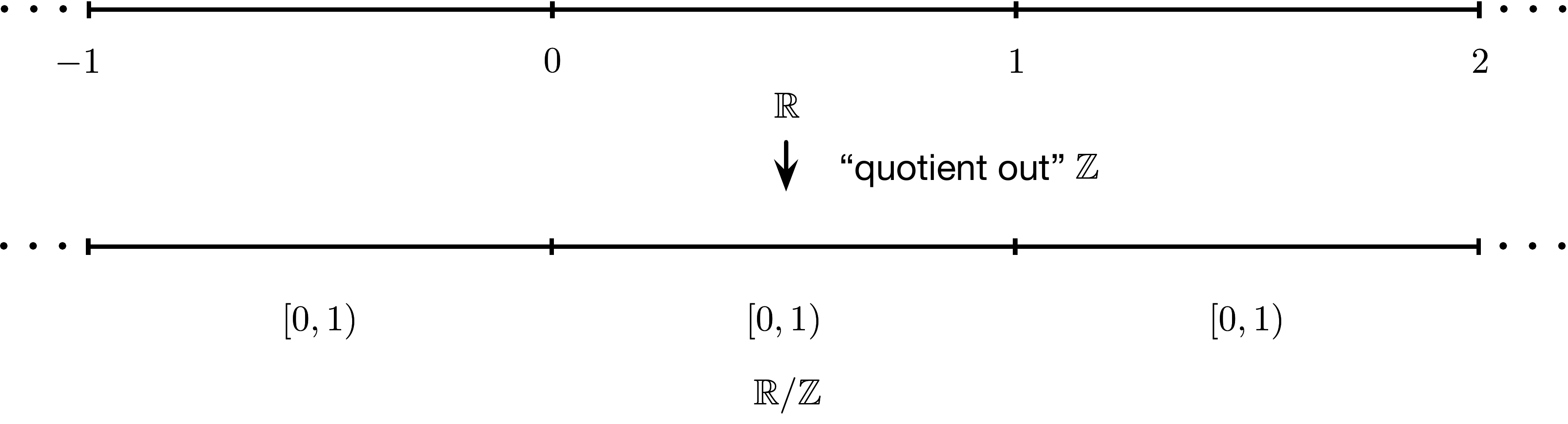}
    \caption{"Modding out" $\mathbb{Z}$ from $\mathbb{R}$ ($\mathbb{R}/\mathbb{Z}$) creates an infinite segmentation of identical intervals $[0,1)$.}
    \label{fig:modout}
\end{figure}
We observe from Figure \ref{fig:modout} that the line segment repeats, but also that we can keep track of how many windings (or rolls) we have performed. So we can count the windings. This is a way to characterize a circle. Mathematicians call this count the {\em fundamental group} and a typical depiction of it can be seen in figure \ref{fig:fundamentalgroup}. 
Notice that this is actually the same picture as the relationship of the circle and it rolling on an infinite line, except that we picture the line as a winding helix above the circle. This allows us to depict the counting as the points on the helix above a base point on the circle. The helix can be projected onto the circle (map $P$) which also motivates the use of the symbol $\pi_1$ for the fundamental group. The integers $\mathbb{Z}$ under addition capture counting (and form a group, hence the name).

Again, this depiction does not depend on the precise shape of the circle but really applies to topological loops. We actually have good intuitions for this. We understand that for race tracks that are in a loop configuration, even if it is not a perfect circle, we can observe the number of laps a racer has performed.

\marginnote{$[0,1)$ is $\mathbb{R}/\mathbb{Z}$ is $\mathbb{S}^1$.}By forming the quotient $\mathbb{R}/\mathbb{Z}$ (say "$\mathbb{R}$ mod $\mathbb{Z}$", or we "mod out $\mathbb{Z}$") which means we identify all possible windings, we arrive at the same circle and again generate an interval domain of $[0,1)$. For this reason, we consider the following interchangable: $[0,1)$, $\mathbb{R}/\mathbb{Z}$, and $\mathbb{S}^1$. This is computationally relevant, as it indicates that if we parametrize a loop by $[0,1)$ computations on the loop follow arithmetic operations "mod $1$". \marginnote{The flat circle.}We will coin a circle that looks like the interval $[0,1)\in \mathbb{R}$ the {\em flat circle} given that its local metric looks like the flat line but suggesting that we are dealing with a circular topology.

\subsection{Dynamics on Loops}

\marginnote{Topological space as the domain on which dynamical evolution operates.}One area of application of the above loop topologies is in dynamics. We want to evolve time over a space that repeats or is periodic. The canonical example is sinusoidal oscillation which can be thought of as the orthogonal projection of a dynamics on the circle.

\begin{equation*}
    \sin(2\pi\cdot n\cdot t) =\mathfrak{Re}\left( e^{i2\pi\cdot n\cdot t}\right)
\end{equation*}
\marginnote[-0.9cm]{Sine as orthogonal projection of the circle in the complex plane.}

\begin{marginfigure}
    \centering
    \includegraphics[width=.95\marginparwidth]{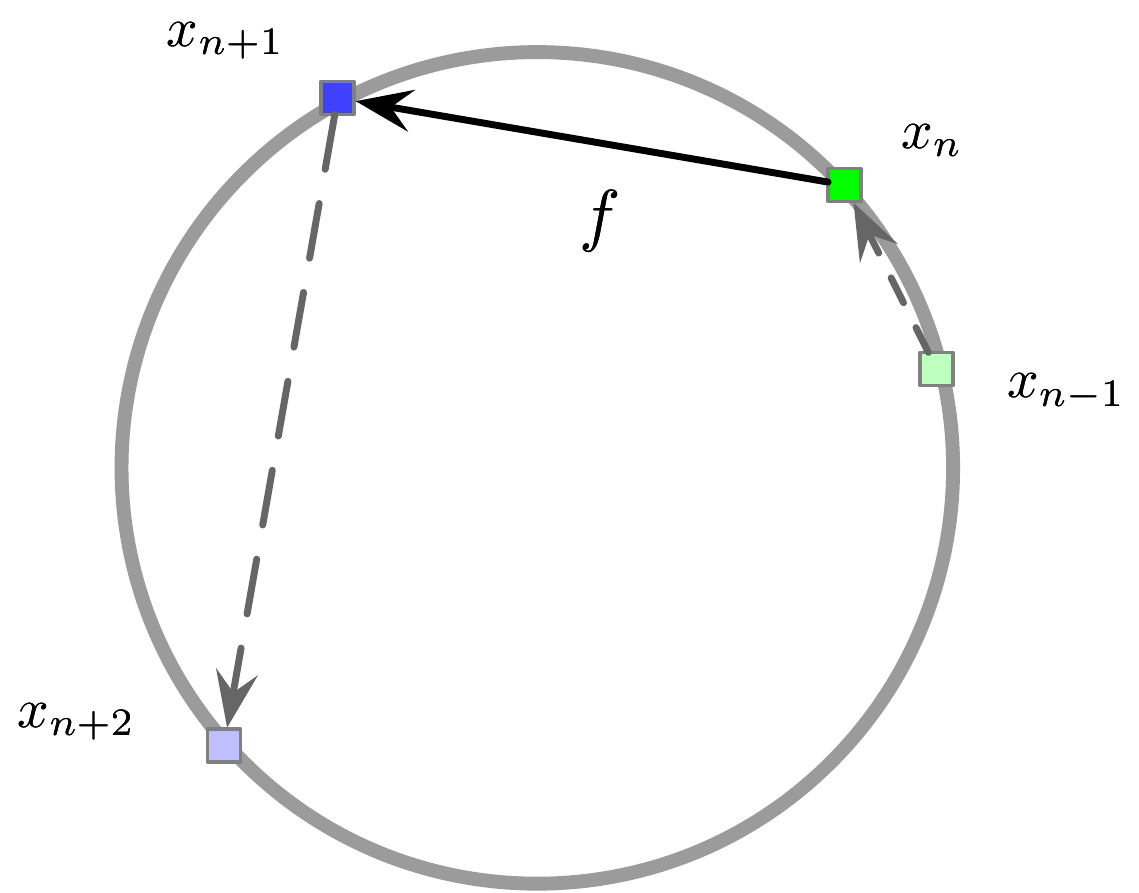}
    \caption{Iterative phase functions $f$ describe discrete steps from a position on the circle to another.}
    \label{fig:circlemap}
\end{marginfigure}

In the study of dynamical systems it is typical to consider evolution on some domain. This is implied in the above notation but not explicit. We can make it explicit in the following fashion. Our domain is a topological space equivalent to $\mathbb{S}^1$ parametrized by $[0,1)$. This parametrization we can variably think of as a position on the circumference, or as the angle. A dynamical process is a map from one domain to another. In our case we want to describe the dynamics on a loop (for now a circle). Because we want to stay close to discrete computational methods we only consider discrete time dynamics. Further we will only consider time stepping that only depend on the previous time location. Hence we can describe a future position on the domain $x_n$ as a function $f$ of the previous position $x_{n-1}$, which we write as $x_n=f(x_{n-1})$. From our previous discussion we know that "mod $1$" arithmetic will enforce the circle topology. We will call the discrete time functions {\em iterative phase functions} (see Figure \ref{fig:circlemap}).\marginnote[-0.9cm]{Iterative phase functions describe maps from the circle to itself in terms of discrete time steps.} A discrete time series algorithm analogous to the sine oscillator above could hence be written as follows:

\begin{align}
    \underbrace{y_n}_{\text{Time Series}}=\underbrace{p}_{\text{Projection}}(\underbrace{x_n=f(x_{n-1}) \overbrace{\text{mod } 1}^{\text{Circle Topology}}}_{\text{Iterative Phase Function}}) \qquad \label{eq:discreteoscillator}
\end{align}

Given the discussion of the interchangable description of flat circles we can define these functions in multiple equivalent ways:
\begin{align*}
f: \mathcal{S}^1\rightarrow\mathcal{S}^1  \qquad
f: \mathbb{R}/\mathbb{Z}\rightarrow\mathbb{R}/\mathbb{Z} \qquad
f: [0,1)\rightarrow[0,1)
\end{align*}

Study of these maps are known as {\em circle maps}.\marginnote[-0.9cm]{Circle maps are maps from the circle to itself.} Observe that in equation (\ref{eq:discreteoscillator}), we have abstracted what would be the orthogonal projection of a sine oscillator into an unspecified projection function. This is because written this way we see that projection and computation of dynamics on the circle topology are actually independent of each other, hence we are freed up to pick other projections. 

We can recover a discrete version of the sinusoidal oscillator the following way:

\begin{align*}
    y_n&=\sin(2\pi\cdot\phi_n) & \phi_n: x_n&=x_{n-1}+\Omega \mod 1\label{eq:phaseamplitude}
\end{align*}

where $\Omega$ is a constant. Hence we step around the circle with constant step sizes.

A large class of oscillatory techniques in sound synthesis can be formulated in this way \cite{essl2021iterative}. Our goal here is to focus on the topological aspects and hence we will not have time to discuss them. An important topological generalization worth observing is that given that we have iterative phase function on the interval $[0,1)$, these functions are in principle applicable to any homotopy of that interval! An application of this insight was be presented at DAFx2022 during the paper session.

Another topological notion worth mentioning is that arithmetic "mod $1$" is {\em topologically stable} in the following sense: Given that a circle is compact and that the arithmetic forces all computations onto the compactness of the circle, the computation can never grow out of bounds. This is not true for iterative maps that are not defined on a circle domain such as the logistic map, which can become unstable when parameters do not fall into a controlled range.\marginnote[-1.5cm]{Topological stability denotes the stability of a numerical algorithm due to the underlying topological domain. Maps from the circle to itself are topologically stable.}

\subsection{Folded Circle}

The orthogonal projection of the circle as computed by the sinusoid has a property that we can study topologically.
Due to the orthogonal projection, we get two extrema of the circle. This gives us some extra structure that is not captured by homotopy. 
\begin{marginfigure}
    \centering
    \includegraphics[width=.95\marginparwidth]{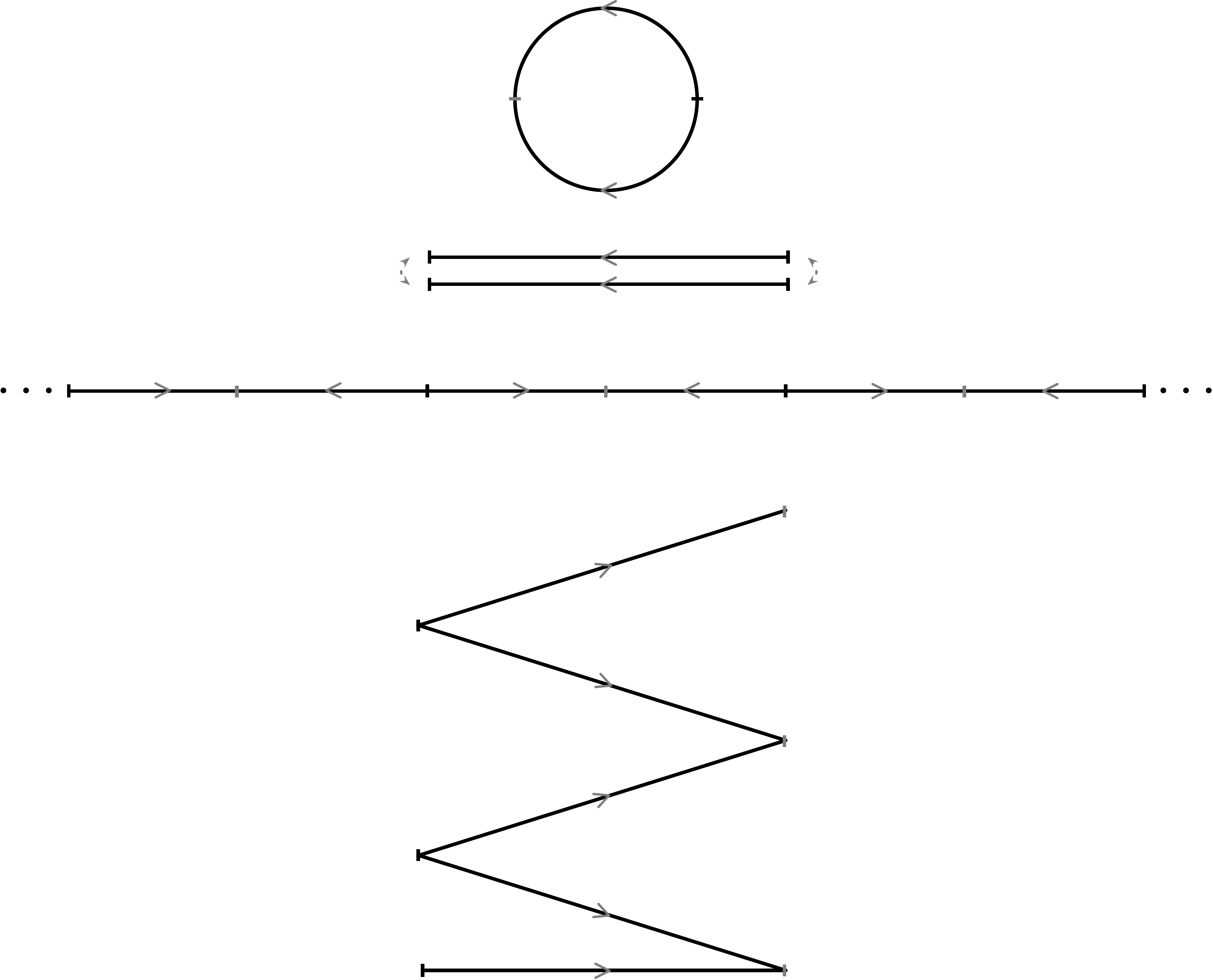}
    \caption{Different depictions of the Folded Circle.}
    \label{fig:foldedcircle}
\end{marginfigure}
We will capture this notion by marking the two extrema, but also by adding markings to the two paths connecting the extrema as depicted in Figure \ref{fig:foldedcircle}. As we roll this circle on the real line we observe that the two sides alternate. If we want to construct an analogy to the flat circle from before, we can simply flatten out the circle and we arrive at a two flat lines with sharp folds at each end (giving this construction its name). Here an analogy to folding and creasing paper is useful. We can further construct a version of a helix-like cover that keeps the fold points marked. This picture is again just a different way of depicting the real line with folding points marked.
If we skip every other fold point, we arrive at the standard flat circle. In other words the folded circle is a flat circle with "extra structure".

\subsection{Aliasing and Sampling}
\begin{marginfigure}
    \centering
    \includegraphics[width=.95\marginparwidth]{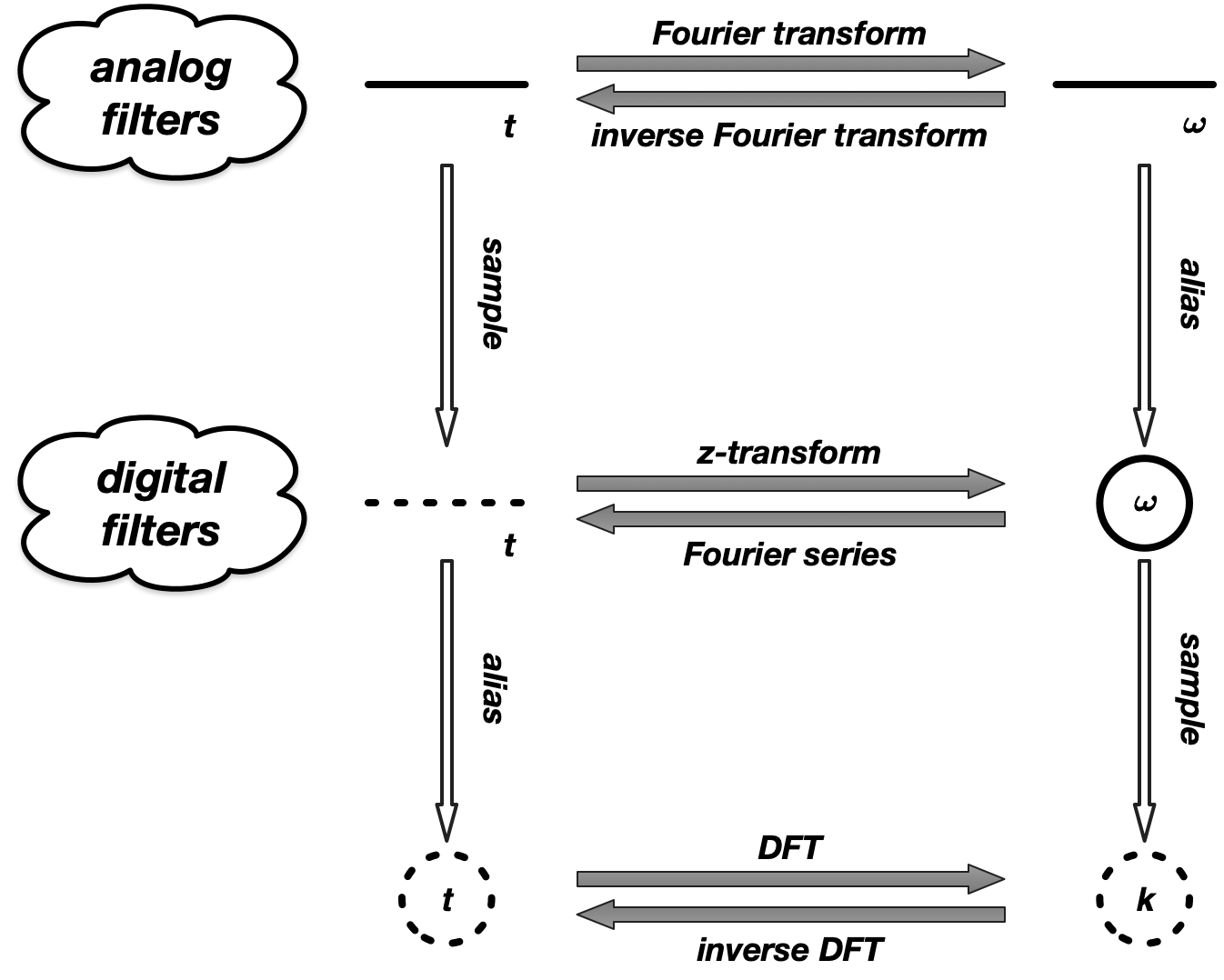}
    \caption{Six domains of Signal Processing.}
    \label{fig:sixdomains}
\end{marginfigure}

Steiglitz's {\em Six Domains of Signal Processing}\cite{steiglitz1997digital}  shown in Figure \ref{fig:sixdomains} provide a very nice pictorial relationship of aliasing, sampling, and duality under Fourier-style transforms for various forms of signal processing. It depicts the relationship between infinite continuous time signals (also known as {\em analog} signals), the sampled discrete signal thereof, and short-term windowed signals (with a periodicity assumption). Via duality of the respective transforms we arrive at the respective frequency domain versions and one observes that sampling is dual to aliasing.

It is rather typical to think signal first, hence we think sampling first. However, for our purpose, aliasing is the more interesting phenomenon as we have a topological way to describe the phenomenon now. We observe that aliasing is the consequence of "modding out $\mathbb{Z}$" hence leading to the fundamental group $\mathbb{Z}$ counting the multiplicity of reaching any point on the quotient topology. The map in Steiglitz's diagram labeled "aliasing" moves a non-quotiented entity onto a quotiented one and this means that one repeatedly wraps around the former around the latter (one can again use the winding or rolling metaphor here).

As is well known to practitioners in digital audio, sampling and aliasing are related. They are dual under a Fourier-style transform. This relationship can be seen nicely in the six domains. It turns out that this duality is a special case of a more general duality: the {\em Pontryagin duality}, which holds for locally compact topological groups. 

%
%
%
\begin{marginfigure}
    \centering
    \includegraphics[width=.95\marginparwidth]{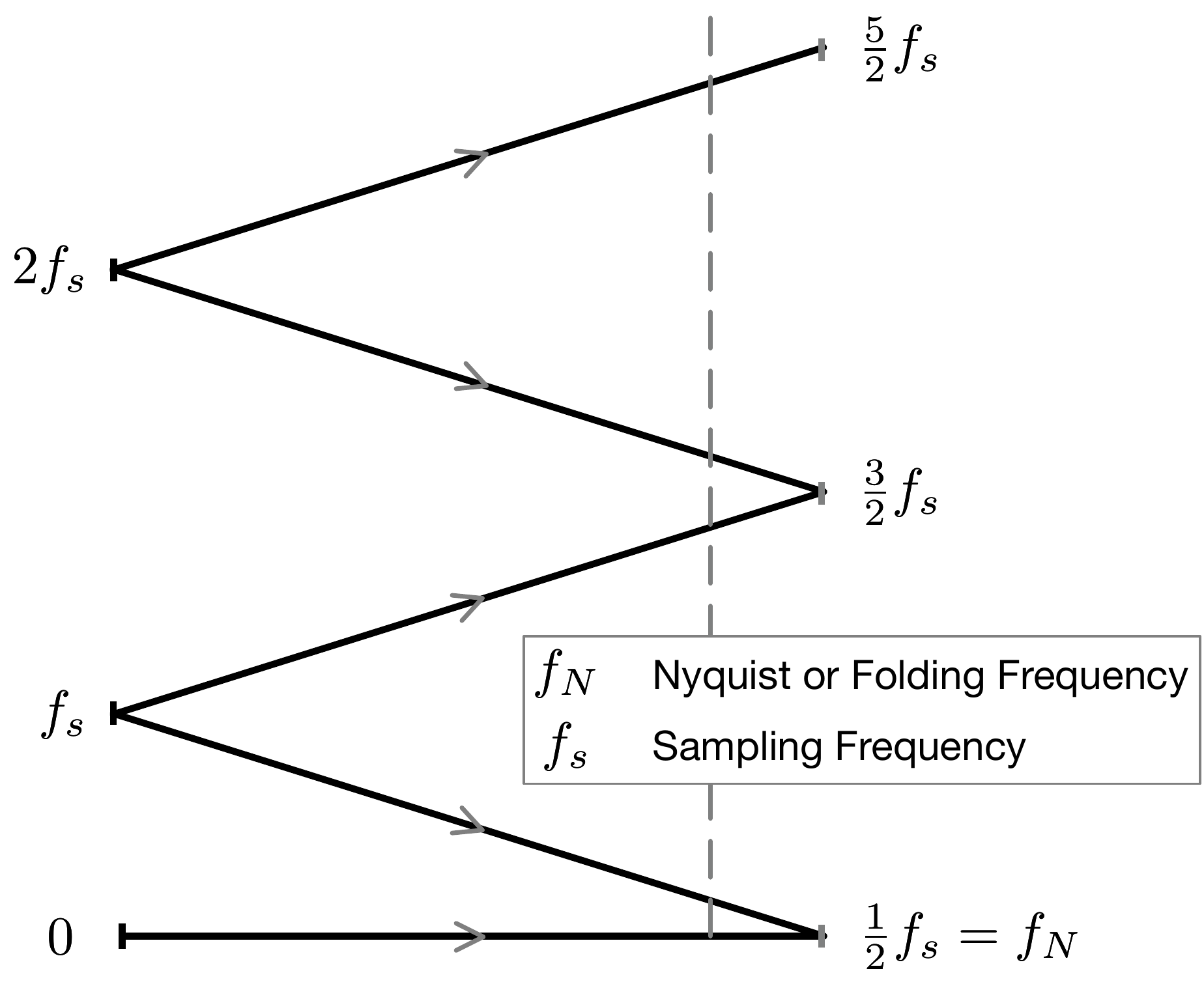}
    \caption{The folding diagram is a metric version of the cover of the folded circle.}
    \label{fig:foldingdiagram}
\end{marginfigure}
%
Given that the aliasing picture in Figure \ref{fig:sixdomains} involves the complex circle and its real projection, we are dealing with the topology of a folded circle.
It is well known in signal processing that frequencies just above the Nyquist (sometimes also call folding) frequency "folds over" as it passes that frequency. By the same token it also requires that negative frequencies "fold back" at the bottom of the spectrum in a similar fashion. Folding diagrams (Figure \ref{fig:foldingdiagram} shows an example) describe this property metrically, but topologically they look like the folded cover of Figure \ref{fig:foldedcircle} (bottom).  Hence we can understand the observed folding behavior to stem from the topology of the folded circle.

\subsection{Fourier Analysis of the Wave Equation as Aliasing and Sampling}

\begin{marginfigure}[-16\baselineskip]
    \centering
    \includegraphics[width=.95\marginparwidth]{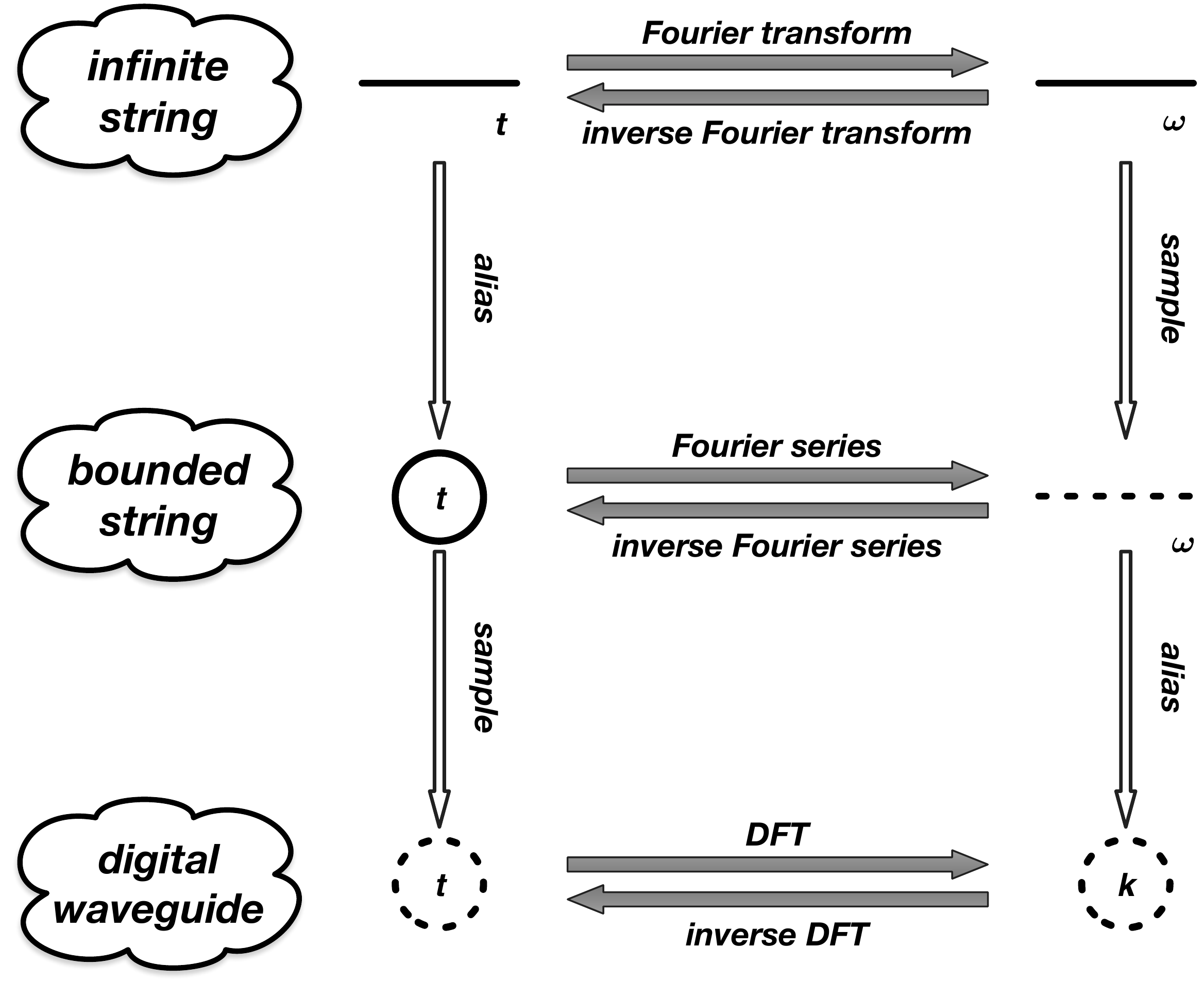}
    \caption{Six domains of the Wave Equation.}
    \label{fig:sixdomainswave}
\end{marginfigure}

\begin{marginfigure}
    \centering
    \includegraphics[width=.95\marginparwidth]{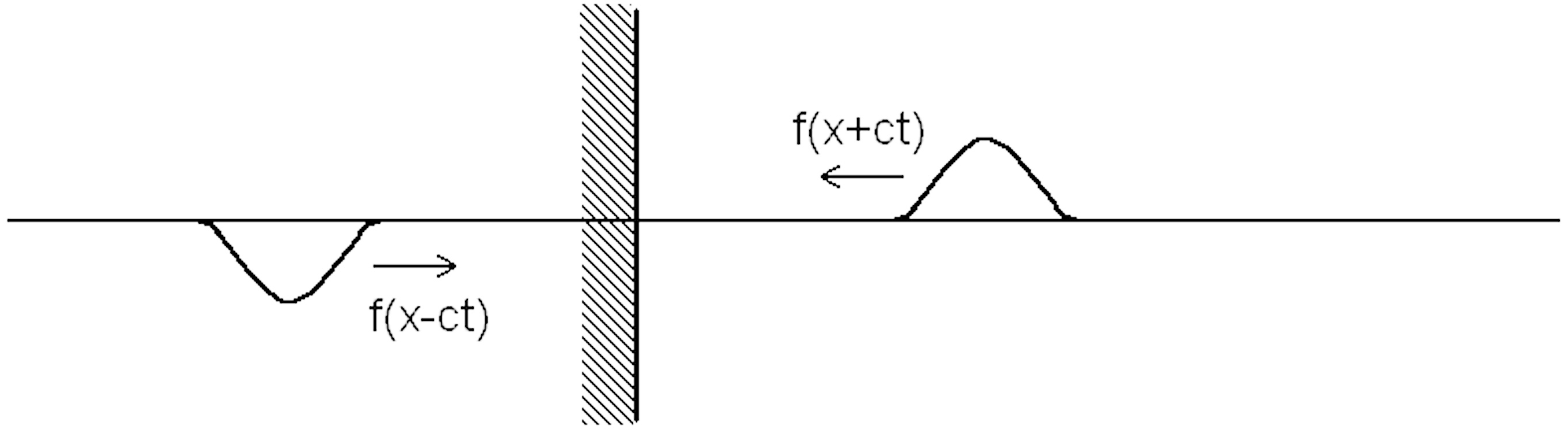}
    \caption{Reflection of a traveling wave at a boundary. A virtual image equivalently travels towards the reflection from behind the reflection point.}
    \label{fig:wavereflection}
\end{marginfigure}

\begin{marginfigure}
    \centering
    \includegraphics[width=.95\marginparwidth]{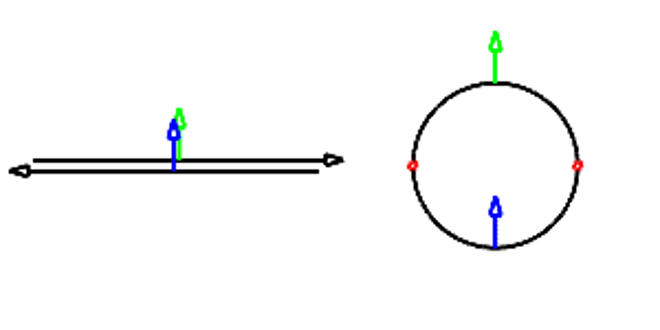}
    \caption{An impulse load on a waveguide in flat configuration as well as in a lift onto a circle.}
    \label{fig:waveguideloop}
\end{marginfigure}

Rather than sample first one can also chose to alias first. Much of what we just discussed pertains unchanged though it deserves a different physical interpretation. This process then again leads to six domains, which this time we call {\em Six Domains of the Wave Equation} as depicted in Figure \ref{fig:sixdomainswave}.
Introducing boundary conditions leads to a folded circle, for which the dual is samples. However, now the time and frequency domain is switched, and the resulting picture captures the well-known fact that the finite string has a discrete (though infinite) spectrum. By sampling the finite string, we arrive at a waveguide-style discretized model of the wave equation which has an aliased discrete spectrum.
The unfolding of the bounded string is a well-known property in specific settings. Taking the traveling wave solution one can use the boundary condition to derive the reflection function, which turns out to have an equivalent image behind the boundary that travels towards the reflection point from behind as a wave to be reflected travels towards (see Figure \ref{fig:wavereflection}). This pattern repeats for each length of the string. We should now recognize this as nothing but the "unrolled" string, i.e. its universal cover. 
There is a third way to depict this situation (Figure \ref{fig:waveguideloop}). This move is similar with an unfold, though we merely "lift" the flat propagation directions into a circle. This change already has a noteworthy effect. If arrows move left and right on each branch, in the circle depictions their apparent motion is always forward, hence the unfolding has turned abrupt bounces into smooth motion.
\subsection{Using Covers to Study Dynamics of Reflections}

\begin{marginfigure}
    \centering
    \includegraphics[width=.95\marginparwidth]{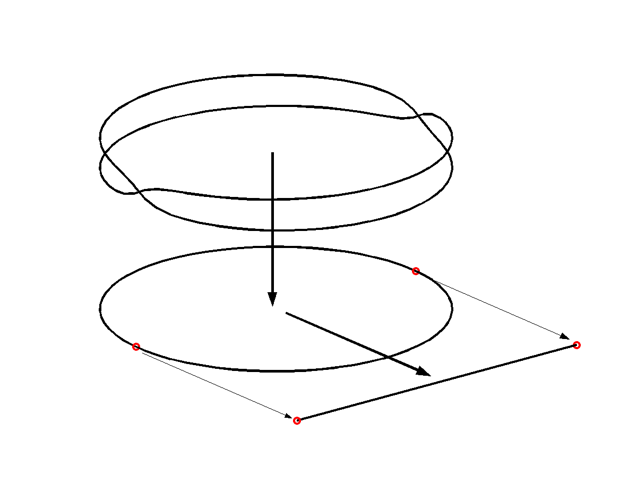}
    \caption{The up/down state of traveling impulses is captured as a lifted cover over the circle, yielding a double cover.}
    \label{fig:doublecoverlift}
\end{marginfigure}

\begin{figure}
    \centering
    \includegraphics[width=.95\columnwidth]{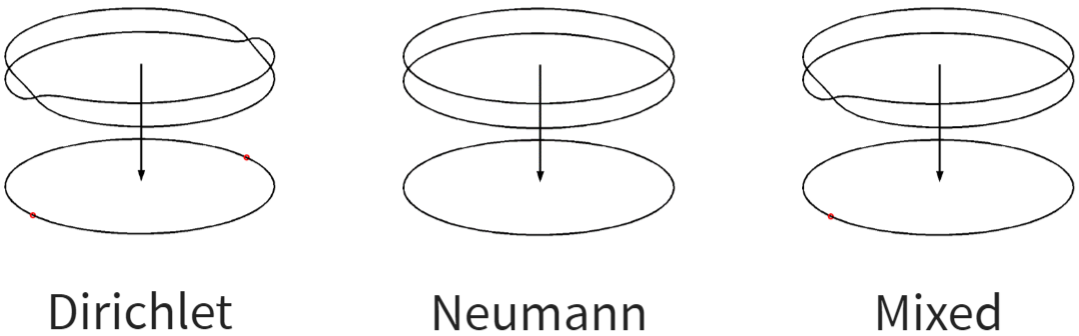}
    \caption{Double Covers for Dirichlet, Neumann, and Mixed boundary conditions. The last is equivalent to the cover of a M\"obius strip.}
    \label{fig:doublecovers}
\end{figure}

Figure \ref{fig:waveguideloop} illustrates the circle topology of the bounded string, but it is rather specific in its impulsive propagation. A way to think about topology is ways to capture behavior that no longer relies on specific details. Here we give a construction that generalizes traveling impulses and gives us a topological interpretation of the impact of boundary conditions. Rather than model the impulse propagation, we instead merely capture the state of an impulse over the range of the circle between boundaries. We observe that the up/down direction of the impulse does not actually change between boundaries and that this is true even under variation. If the impulses were damped or travelled irregularly they would still retain their up/down orientation between boundaries. So rather than model the impulse, we will simply create a lift of a path for each impulse color (green and blue) that captures its up/down state between the range. This construction yields the configuration depicted in Figure \ref{fig:doublecoverlift}.
This is a topological picture that captures a range of changes in impulses, but it also allows us to clarify that of course we require that the travel-topology, i.e. the paths that the impulses can take do not change under an assumed variation. For example if there is a scatterer introduced, that would change the way impulses travel and the above construction does not contain this case. This depiction of wave reflection is, despite its simplicity, still insightful, as can be seen in Figure \ref{fig:doublecovers}.
Moving around the circle once corresponds to moving along the bounded wave domain twice, and we observe that both Dirichlet and Neumann conditions share that property. However, the mixed boundary condition leads to a double cover that requires rotation about the circle twice to return to the starting point in the same state. This is the well-known behavior of the M\"obius strip. This also leads to comparative behavior. Given that it takes twice as long to complete a "period", the mixed boundary should be half the fundamental pitch. This is of course well known for instruments such as clarinets or one-sided capped organ pipes compared to their open counterparts such as flutes and open-ended organ pipes. The elegance of this argument is that it is quite general. As long as we have no scattering we can expect this behavior, hence we have found a description that we expect to hold even if we do not have a precise solution to compute a specific case.

\subsection{The Flat Torus}

The flat torus is a fairly straightforward generalization of the flat circle we have already discussed. Those of us old enough to know the computer game Pac-Man know that the character and its ghost enemies are allowed to tunnel through the sides and top ends of the map to re-enter at the opposite side.
\begin{marginfigure}[-32\baselineskip]
    \centering
    \includegraphics[width=.95\marginparwidth]{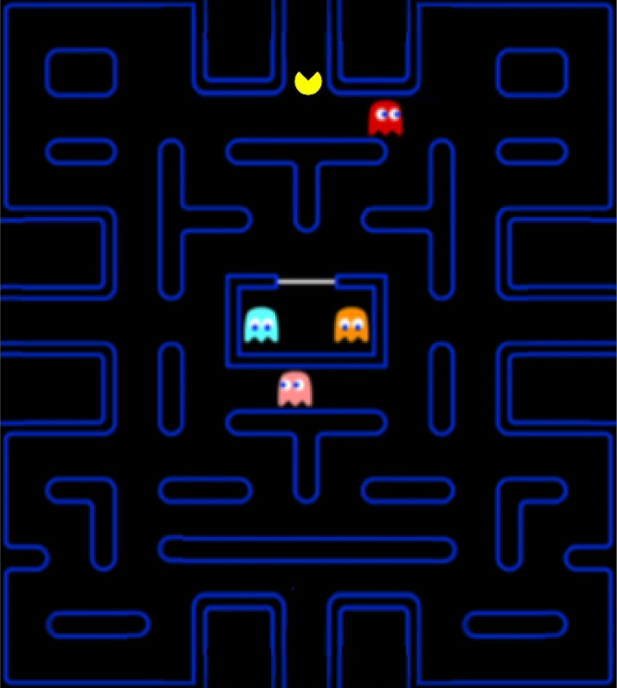}
    \caption{Pac-Man lives in a planar world with identifications between opposite tunnels (Image by Ajeet Gary).}
    \label{fig:pacmantorus}
\end{marginfigure}
In the spirit of the quotient topology construction already discussed we can think of the top edge identified with the bottom edge and the left edge identified with the right. In order to depict this analogous to our circle construction we again draw the identifications together to first form a cylinder, and then by identifying the cylinder ends, we form a torus. Pac-Man's world looks flat to him, hence we call this torus the {\em flat torus}. Depicting Pac-Man's world on the more familiar circular torus shape again highlights how this depiction removed the discontinuity of the identification of his flat rectangular world. We can unroll or unwind the flat torus just as we did the circle. 
\begin{marginfigure}[-12\baselineskip]
    \centering
    \includegraphics[width=.95\marginparwidth]{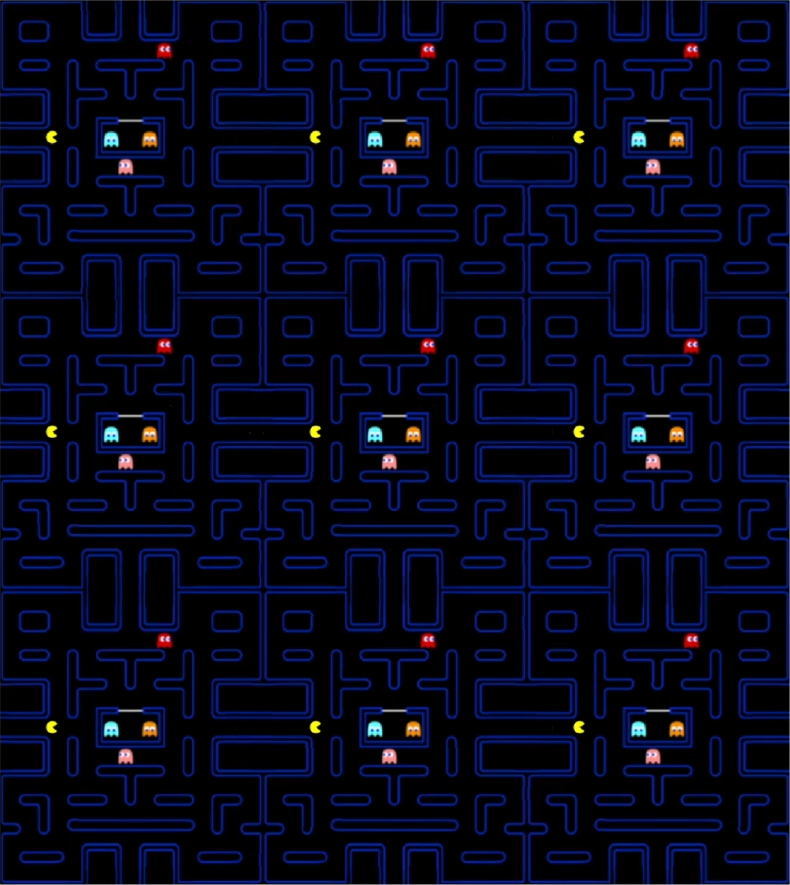}
    \caption{Cover of the torus. (Image by Ajeet Gary).}
    \label{fig:toruscover}
\end{marginfigure}
This yields a tiling of the plane with identical copies of Pac-Man's world. If we continue the tiling to infinity, we arrive at the universal cover of the torus (Figure \ref{fig:toruscover}). The unique space Pac-Man lives in is the {\em fundamental domain} $[0,1)\times[0,1)$.

 The cover tiling is a repetition of this domain with integer differences "quotiented out" in both directions $(\mathbb{R}/\mathbb{Z})\times(\mathbb{R}/\mathbb{Z})$ and we can construct this plane by viewing the intervals as circles and rolling it along either direction $\mathbb{S}^1\times\mathbb{S}^1$. Hence we get the torus as a product of the circle in each of the three views we discussed.
\marginnote{$[0,1)\times[0,1)$ is $(\mathbb{R}/\mathbb{Z})\times(\mathbb{R}/\mathbb{Z})$ is $\mathbb{S}^1\times\mathbb{S}^1$. The torus is the Cartesian product of the three pictures derived from the quotient topology construction of the circle.}

\subsection{The Image-Source method as an example of the Folded Torus}

Pac-Man serves as a familiar illustration of the flat torus structure but has no real reference to signal processing. An example much closer to signal processing is given by the well-known Image-Source method introduced by Allen and Berkeley\cite{allen1979image}. The following Figure \ref{fig:allenberkeley} is a reproduction of their original paper with some structure in it highlighted.
\begin{marginfigure}
    \centering
    \includegraphics[width=.95\marginparwidth]{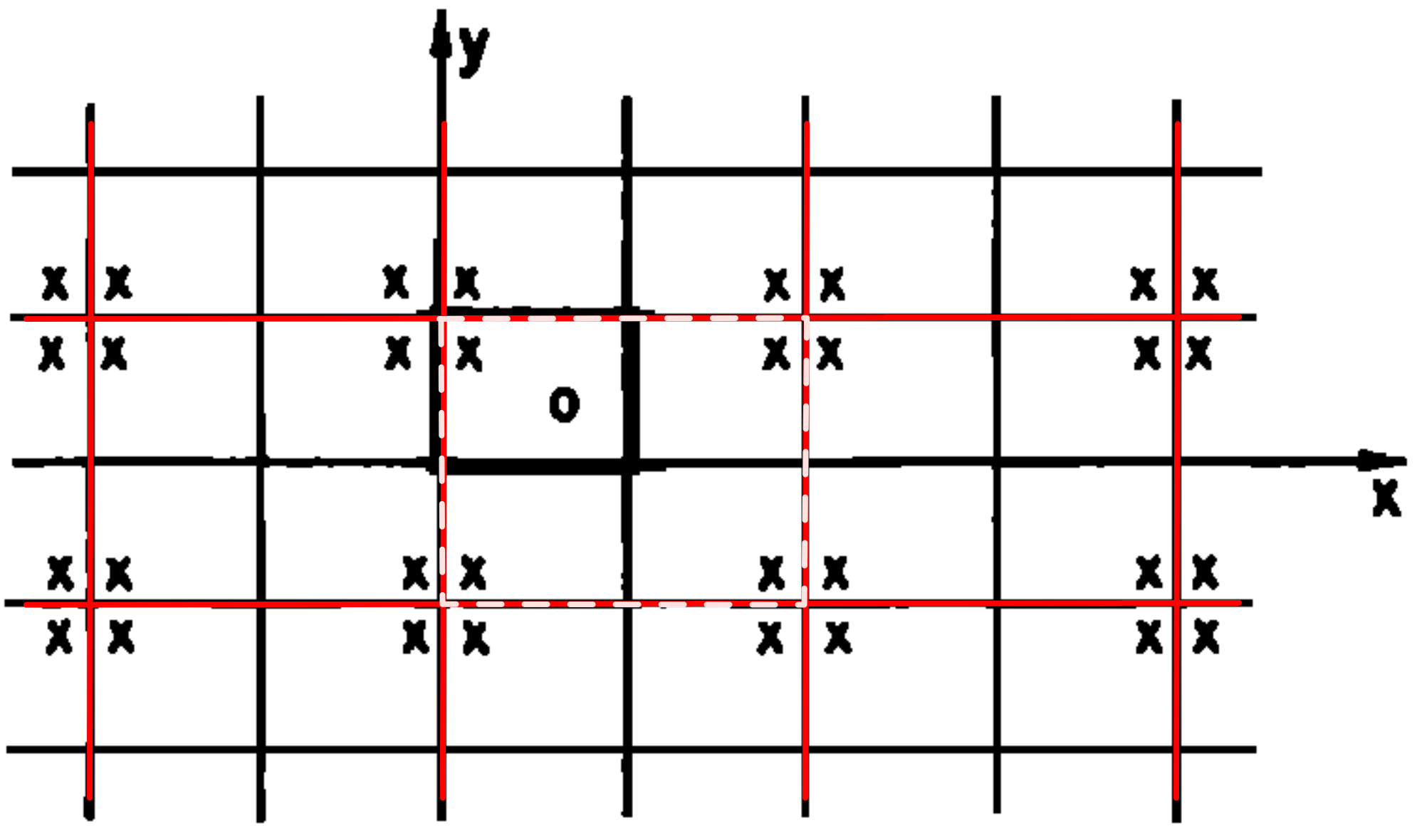}
    \caption{The torus topology underlying the Image-Source method of Allen and Berkeley. The dashed red line marks one of four possible fundamental domains, and the red tiling shows the tiling repetition of the universal cover.}
    \label{fig:allenberkeley}
\end{marginfigure}
Surrounded by the red dashed line, we see that the pattern formed by the original "x" markings is periodically repeated in the plane, forming a planar rectangular tiling in the plane marked by the solid red lines. In our new language, we observe that the marked dashed rectangle is the fundamental domain of a flat torus, and that the plane tiling is the universal cover of this torus. While one can pick the original room with the circular listener in it as each of the four sub-segments of the fundamental domain, the fundamental domain has the same substructure of extra reflections inside this domain. Allen and Berkeley observed that the image behind a reflective wall makes the traveling path between listener and reflected source straight, hence giving a simple computation of the impulses from distance in this patterned plane.

One can construct the torus that has this property by taking a rectangular piece of paper and folding the piece of paper twice. The first fold yields a flat cylinder, and the second fold forms the {\em Folded Torus}.\marginnote{The folded torus.}  As we see, the folded torus is a flat torus with an additional fold in each direction, hence we have the product of two folded circles.

\begin{figure}
    \centering
    \includegraphics[width=.95\columnwidth]{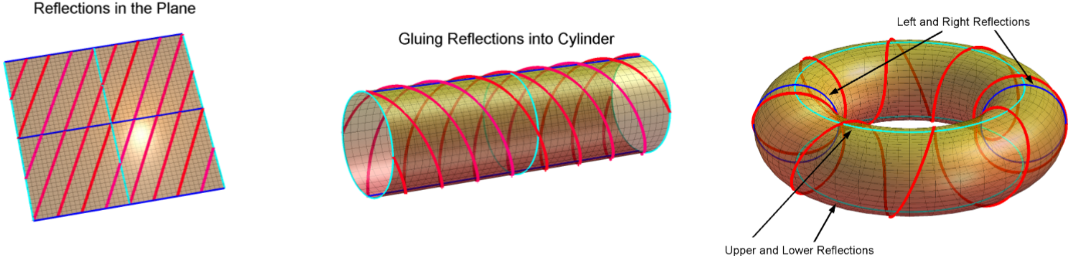}
    \caption{The torus construction maintaining fold information (Blue and Cyan). Red depicts a path under reflection.}
    \label{fig:goldentorus}
\end{figure}

Both the toroidal construction by identifications and the effect of straightening paths on the torus is shown in Figure \ref{fig:goldentorus}.
Observe that reflections become linearized by unfolding the plane, but furthermore, they become windings first around the cylinder then around the torus. The winding number counts the number of times it winds around the circle in each direction. If these two numbers are integer and relative prime, they describe a closed circular path on the torus. If the quotient of these two numbers is not in the integers, but an irrational number, the path will eventually fill the whole torus. This state of affairs is known by a few names including {\em Kronecker fibration}. Integer windings describe various configurations of what are known as {\em Torus knots}.

\begin{marginfigure}
    \centering
    \includegraphics[width=.95\marginparwidth]{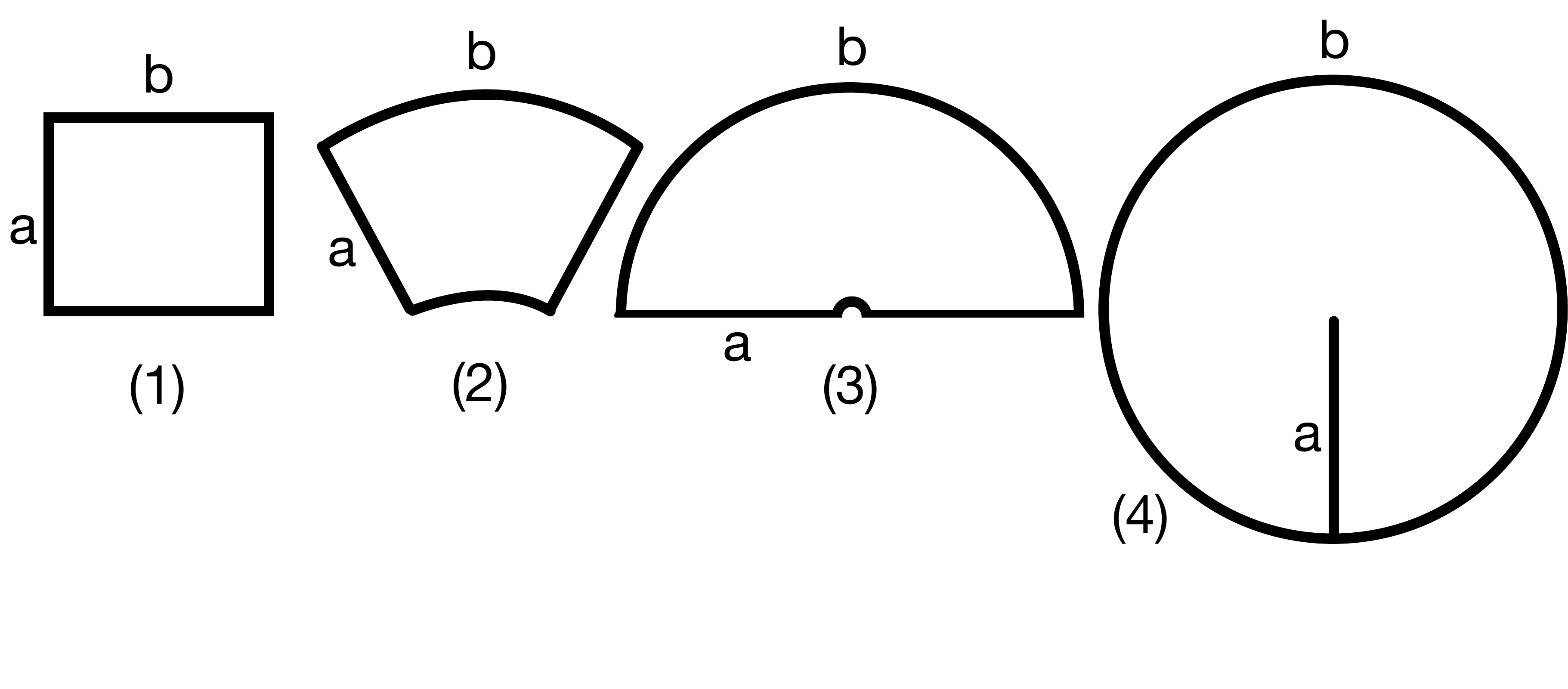}
    \caption{Deforming a square into a disk maintaining separate coordinates.}
    \label{fig:square2circe}
\end{marginfigure}

This kind of construction is not exclusive to square domains. Circular and Elliptical domain shapes also reflect like a torus, a fact used by Keller and Rubinow \cite{keller1960asymptotic} to model resonances on these shapes. The key ingredient is that they can be dscribed by two independent coordinates and a coordinate preserving deformation from the square to the circle is shown in Figure \ref{fig:square2circe} (redo figure).
\section{Combinatorial Construction of Topological Spaces}

\begin{marginfigure}
    \centering
    \includegraphics[width=.95\marginparwidth]{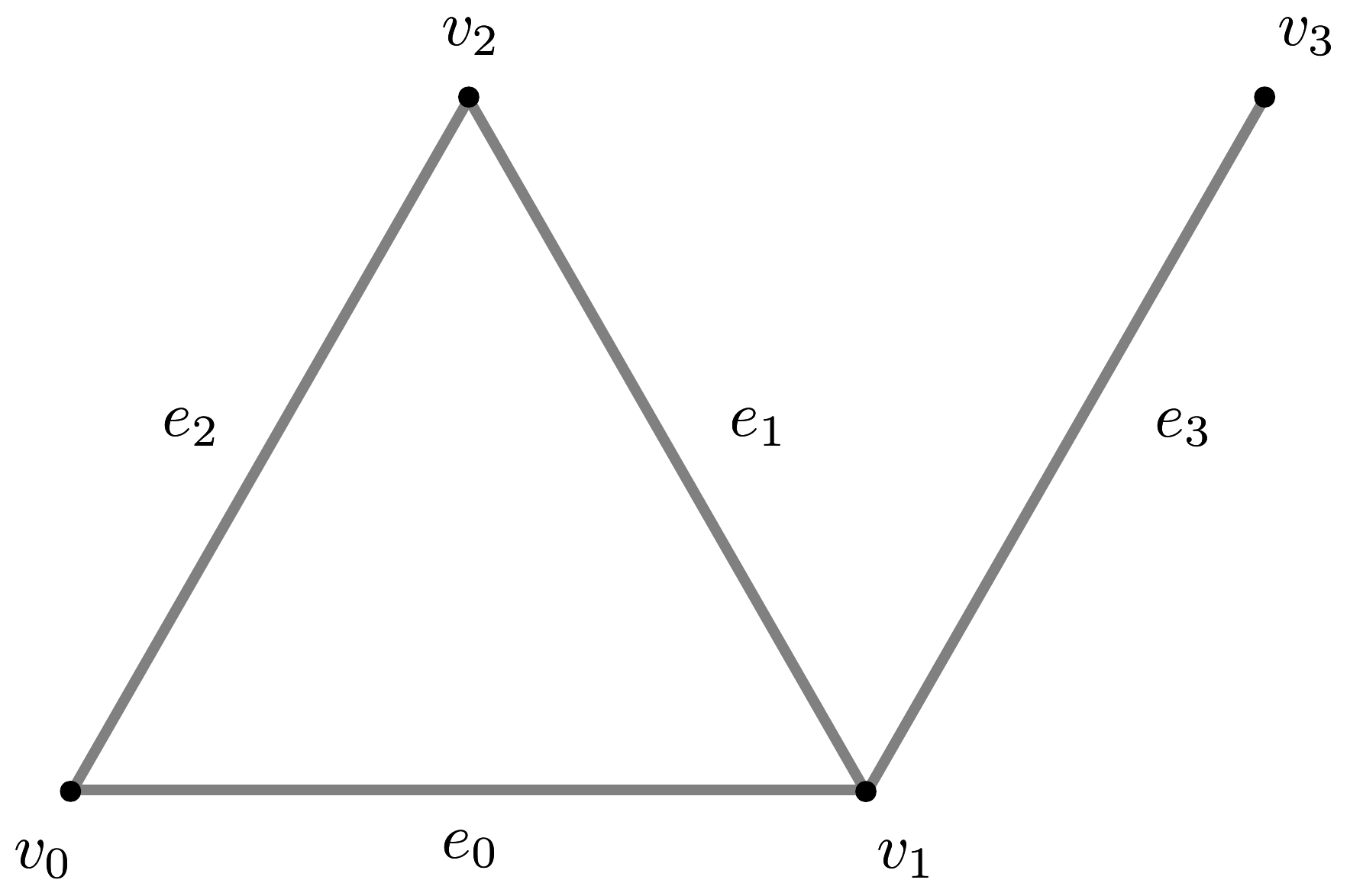}
    \caption{A simple Graph.}
    \label{fig:graph}
\end{marginfigure}

In many applications topological spaces naturally arise as one creates interconnectivity or combinations of entities. The case most familiar case for practitioners in signal processing are graphs, which then are made more specific as networks or dataflows, depending on the application. Graphs themselves just model connectivity. It is customary to define graphs by their collection of vertices and edges written as $G=(V,E)$ with $V=\{v_0,v_1,\ldots,v_n\}$ and $E=\{e_0,e_1,\ldots,e_m\}$. An interesting observation about this definition is that it treats edges and vertices as completely separate, though one usually thinks of edges as connecting vertices. The relationship of edges and vertices is introduced via some connectivity construction. Matrices can be used to capture this information. Two commonly used examples of matrices capturing connectivity information are the adjacency and the incidence matrices. Consider the following simple graph depicted in Figure \ref{fig:graph}.
The {\em adjacency matrix} captures a vertex to vertex relationship and has $1$ as entry if an edge is present between a pair of vertices. Hence we can full out the adjacency matrix as follows:%
							\[A_G=\bordermatrix{
  ~ & v_0 & v_1 & v_2 & v_3 \cr
  v_0 & 0 & 1 &  1 & 0 \cr
  v_1 & 1 & 0 & 1 & 1 \cr
  v_2 & 1 & 1 & 0 & 0 \cr
  v_3 & 0 & 1 & 0 & 0 \cr
}\]
The {\em incidence matrix} relates vertices to edges, and has a $1$ entry if an edge is "incident" to an edge (or vice versa!) and for the example graph we get the following matrix:%
							\[B_G=\bordermatrix{
  ~ & e_0 & e_1 & e_2 & e_3 \cr
  v_0 & 1 & 0 & 1 & 0 \cr
  v_1 & 1 & 1 & 0 & 1 \cr
  v_2 & 0 & 1 & 1 & 0 \cr
  v_3 & 0 & 0 & 0 & 1 \cr
}\]
Matrix representation immediately invites a kind of algebraization of the graphs as we can now study properties of the matrix. This is in fact a way to enter into algebraic graph theory. In fact, the incidence matrix plays a much more important role in this algebraization. We will develop some of the reasons for this later.

\subsection{Simplicial Complex as Generalization of Graphs}

\begin{figure}[h]
    \centering
    \includegraphics[width=.95\textwidth]{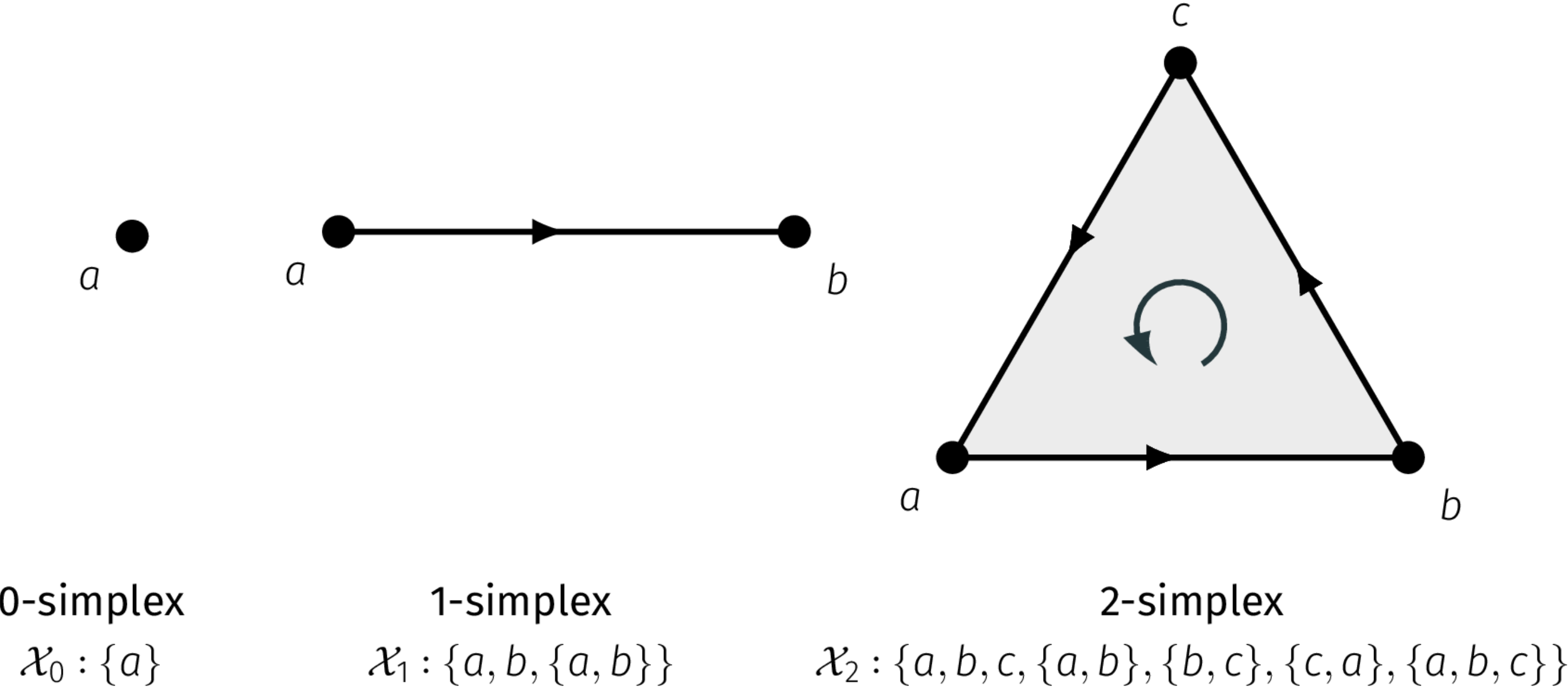}
    \caption{Low-dimensional simplices.}
    \label{fig:simplices}
\end{figure}

\begin{marginfigure}
    \centering
    \includegraphics[width=.95\marginparwidth]{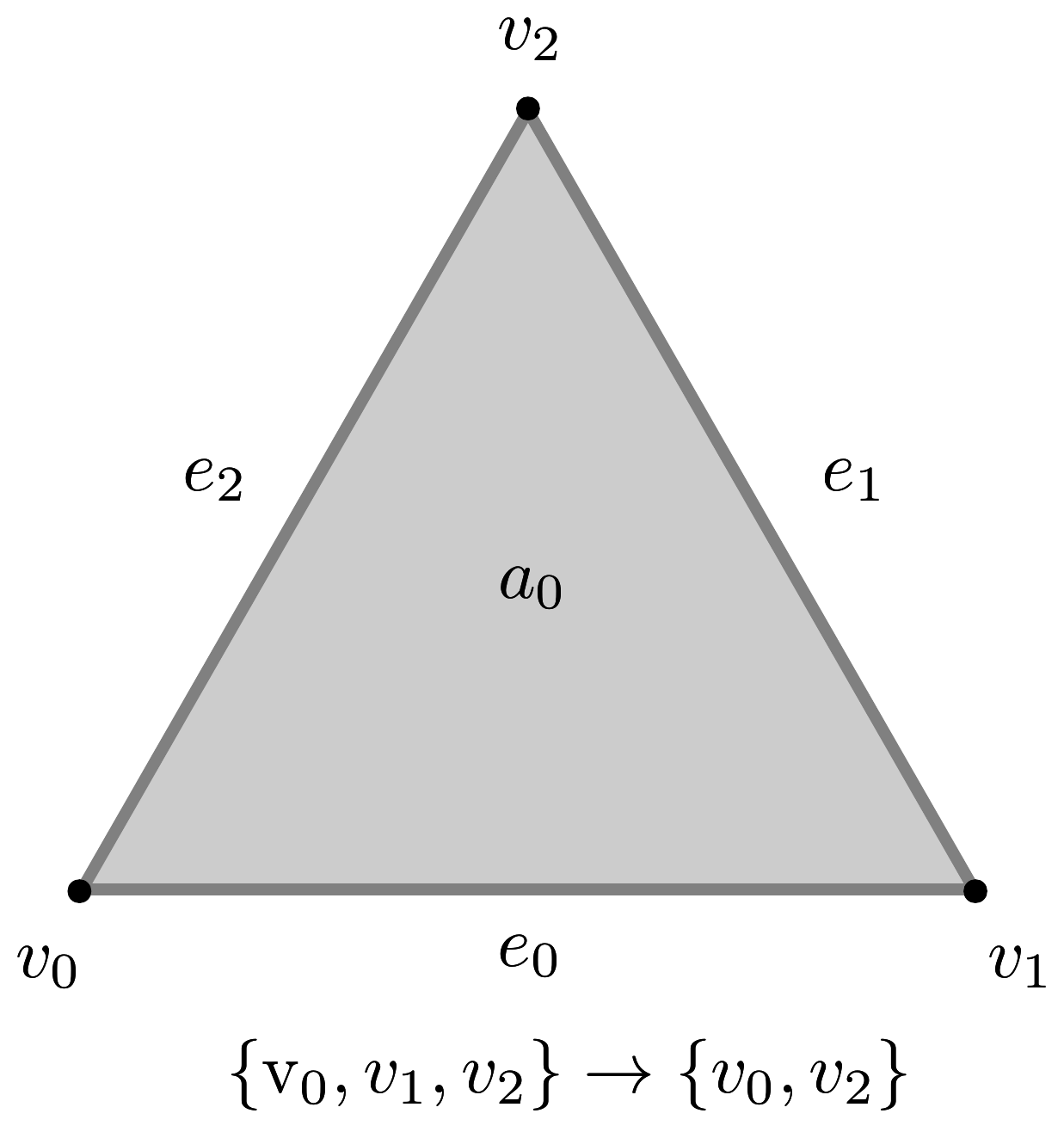}
    \includegraphics[width=.95\marginparwidth]{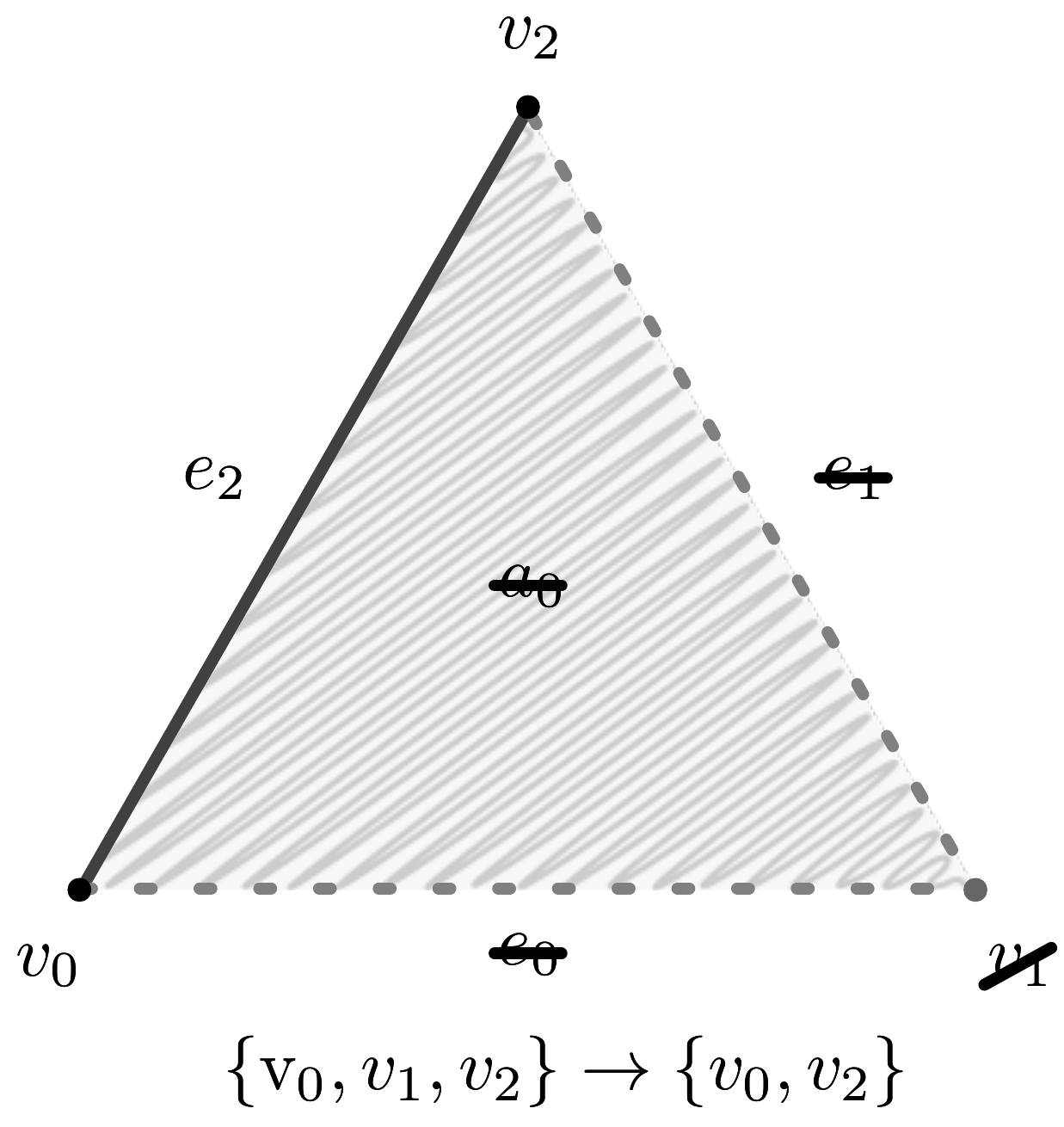}
    \caption{Deleting an entry in the set is the same as returning the face opposite the removed $0$-simplex.}
    \label{fig:facemap}
\end{marginfigure}

Simplices are higher-dimensional generalizations of vertices and lines. The next entity we need is something capturing the notion of area. The simplest possible combinatorical structure is that of an area captured by three points and surrounded by three edges (see Figure \ref{fig:simplices}).
This process can be continued. A volume can be constructed by surrounding it by four triangles and so forth into higher dimensions. This can be described by sets just as we did for graphs, though we will organize our sets slightly differently. This justifies giving things a new name. A vertex will be called a $0$-simplex and it's set is simply a label for each vertex $\mathcal{X}_0=\{a\}$. For all higher-order simplices we require that they include the sub-simplices from which it is built. In this setup, an edge also contains the vertices it connects and is now called a $1$-simplex and its set description hence is $\mathcal{X}_1=\{a,b,\{a,b\}\}$. This is just a somewhat different version of collecting vertex and edge information compared to our graph sets. However, we are now ready to write down the 2-simplex $\mathcal{X}_2=\{a,b,c,\{a,b\},\{b,c\},\{c,a\},\{a,b,c\}\}$. Observe that all the lower-order simplices in this set are just subsets of the largest one. This leads to data compression, because we can always construct the lower order simplices by set deletions. Set deletion also gives us a way to get sub-simplices as depicted in Figure \ref{fig:facemap}. We can get a map to a face of an $n$-simplex by deleting one entry from its set, and it will return the $n-1$-simplex opposite to the one we deleted. (Co)face maps is all we need to navigate a single simplex.
This map is hence named face-map and it takes us from an $n$-simplex to one of its $n-1$-simplex faces. Sometimes, it is convenient to go in the other direction and the map is then called coface-map (the co-prefix refers to inverting the direction a map is taken).

\begin{marginfigure}
    \centering
    \includegraphics[width=.95\marginparwidth]{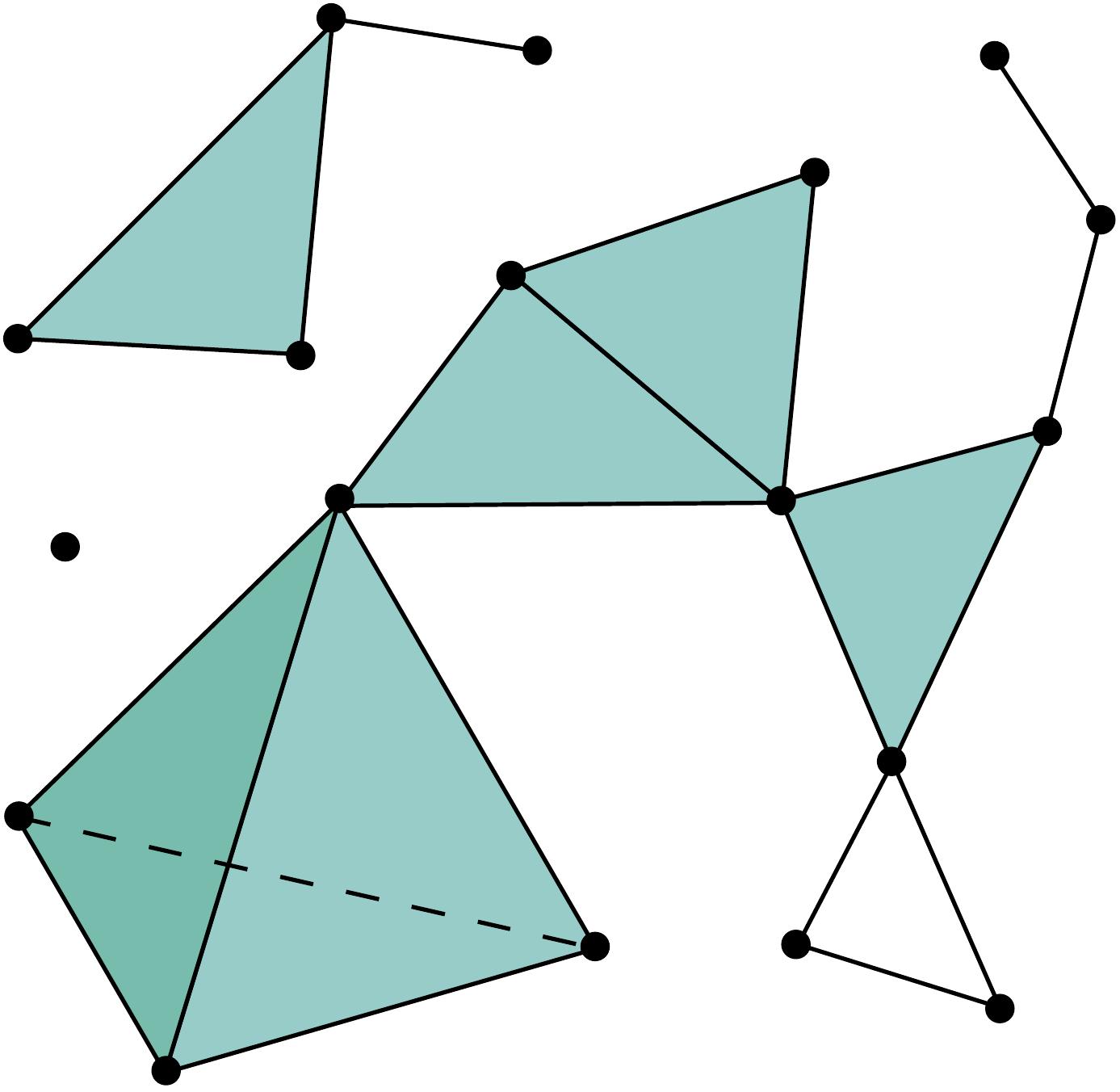}
    \caption{A simplicial complex.}
    \label{fig:simplicialcomplex}
\end{marginfigure}

An abstract simplicial complex consists of simplices that are connected together by the rule that they can only connect along shared (sub)simplices as seen in Figure \ref{fig:simplicialcomplex}. The word abstract here denotes that we do not think of a particular geometric configuration of the simplices in some space, so the notion of a simplex touching or penetrating another is not well defined. This should be familiar from graph theory as used for data-flow. Strictly we are only using that two points in a flow are connected, and some graphical depiction of the connection (whether it is for example a straight, or squiggly line or arrow) is an arbitrary choice to make this connection more visual or geometric.
\section{Computational Techniques}
Next we will discuss two specific computational techniques from the realm of computational topology. These are selected from a wider range of algorithms for their breath of applicability to signal processing and sound synthesis.

\subsection{(Persistent) Homology}

Persistent homology is perhaps the mash hit of computational topological methods. At its heart is homology, which itself is extremely useful. In fact understanding homology is the key, and it has its uses outside of persistence. Hence we will start off discussing homology first.

\subsection{Homology}

\begin{figure}[h]
    \centering
    \includegraphics[width=.95\textwidth]{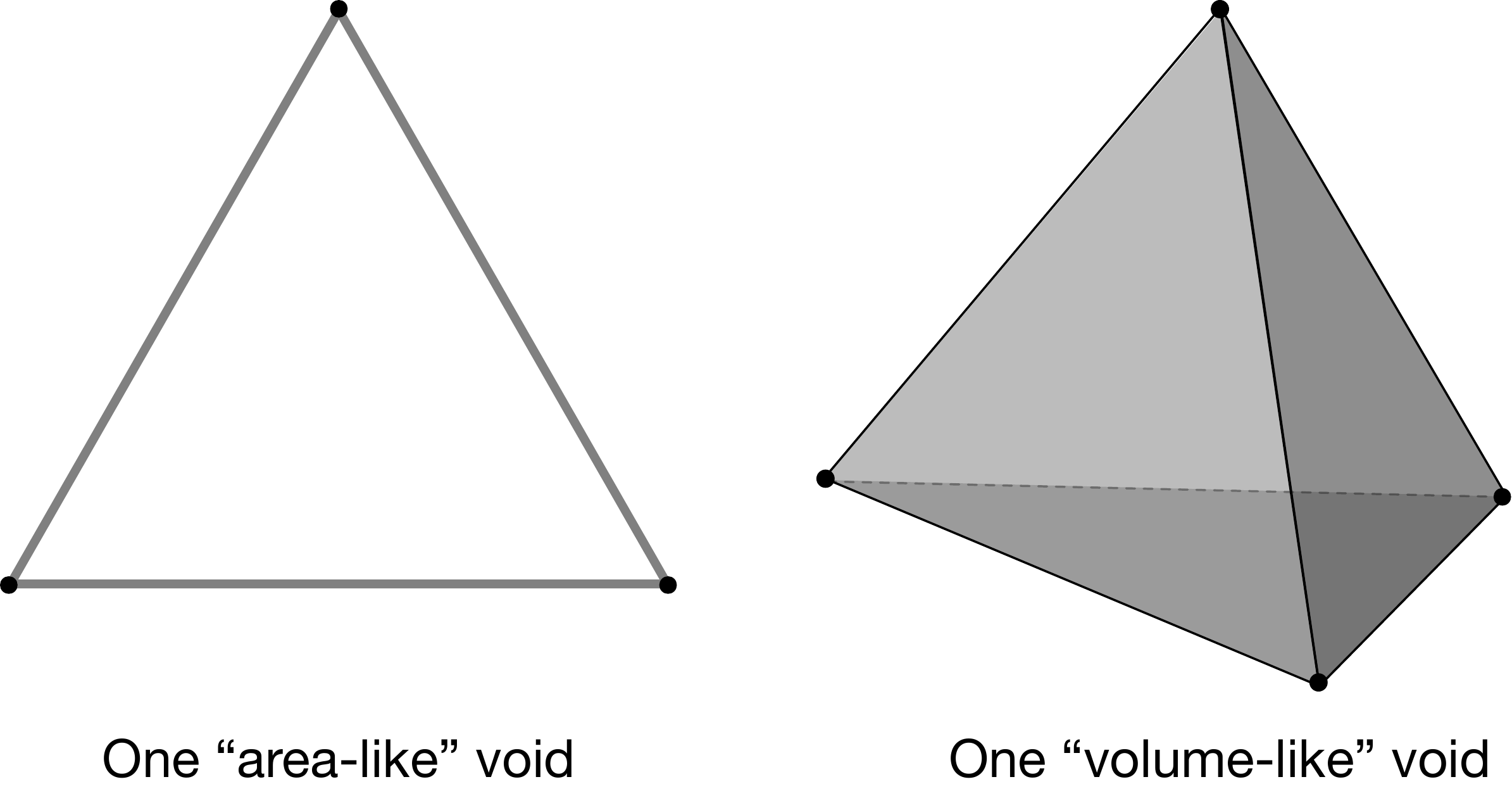}
    \caption{A triangle can either be filled or empty inside. If it is empty inside we call this a {\em void}. The inside of an empty tetrahedron is also a void but of different dimensions. The first is "area-like" the second "volume-like".}
    \label{fig:voids}
\end{figure}

\begin{figure}[h]
    \centering
    \includegraphics[width=.95\textwidth]{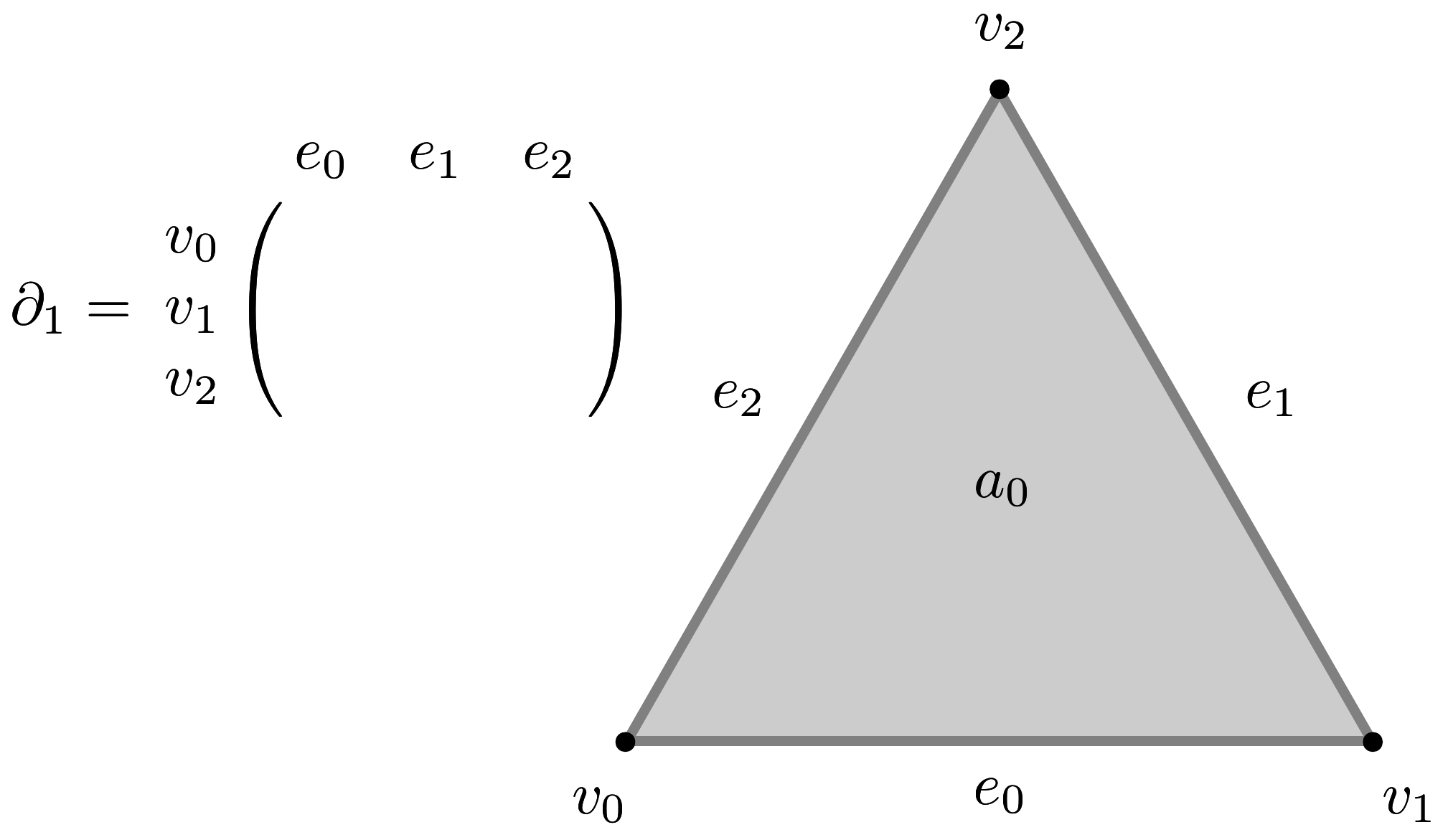}
    \caption{The boundary matrix of $1$-simplices in a simple simplicial complex relates edges to bounding vertices.}
    \label{fig:homologyempty}
\end{figure}

Homology counts $n$-dimensional voids as well as the number of connected components of a topological space. For us (and in computational settings) these topological spaces will be simplicial complexes. To understand what we mean by an $n$-dimensional void consider the example depicted in Figure \ref{fig:voids} showing two low-dimensional examples.
A triangle bounds an empty area inside if it is not filled. This is an example of an "area-like" void, thinking of the dimensionality of the simplex that would fill it. It takes three lines to fence in this void, and we observe that in fact one only gets a $n$-void if it is fully bounded by such an $n-1$ fence. We see that this also holds for the example of the tetrahedron shape (from the boundary of a $3$-simplex). The four $2$-simplices that bound it enclose an "volume-like" void inside (if it is not filled in).
It turns out that the algorithmic computations of these dimensional voids for simplicial complexes is fairly easy. Ultimately it turns out that we compute the rank information of two matrices, called boundary matrices to compute the number of voids. Stated like this, this may appear mysterious, but we have already noted that we only get a void when it is fully fenced in or bounded. But there is a second criterion: the fenced in area needs to indeed be empty and not be filled with an $n$-simplex. The aim here is to construct an algebraic setup in which we capture closed boundaries, which we will call {\em cycles} and denote by the letter $Z$ (likely from the German "Zyklus"). Let's see how we get a cycle from the construction of a {\em boundary matrix}. Here we construct the boundary matrix from $1-simplices$ to $0-simplices$. Recall that each $1$-simplex in isolation is bounded by two $0$-simplices. The boundary matrix captures this information for all $1$-simplices in the simplicial complex. For simplicity we will consider here only a single 2-simplex and its boundary matrix $\partial_1$ in Figure \ref{fig:homologyempty}.

\begin{figure}[h]
    \centering
    \includegraphics[width=.95\textwidth]{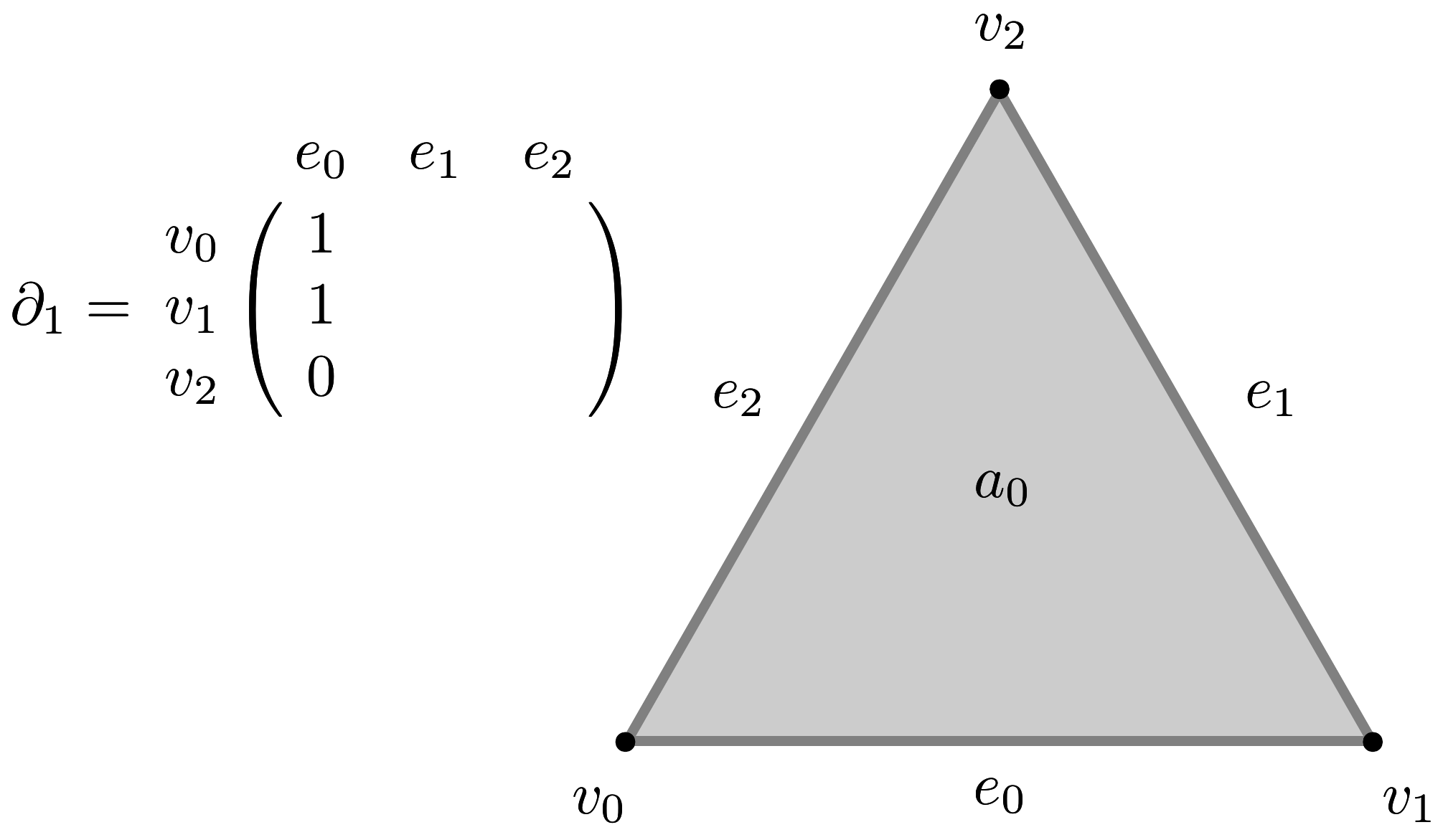}
    \caption{Two vertices $v_0$ and $v_1$ are bounding the edge $e_0$.}
    \label{fig:homologyonecolumn}
\end{figure}

\begin{figure}[h]
    \centering
    \includegraphics[width=.95\textwidth]{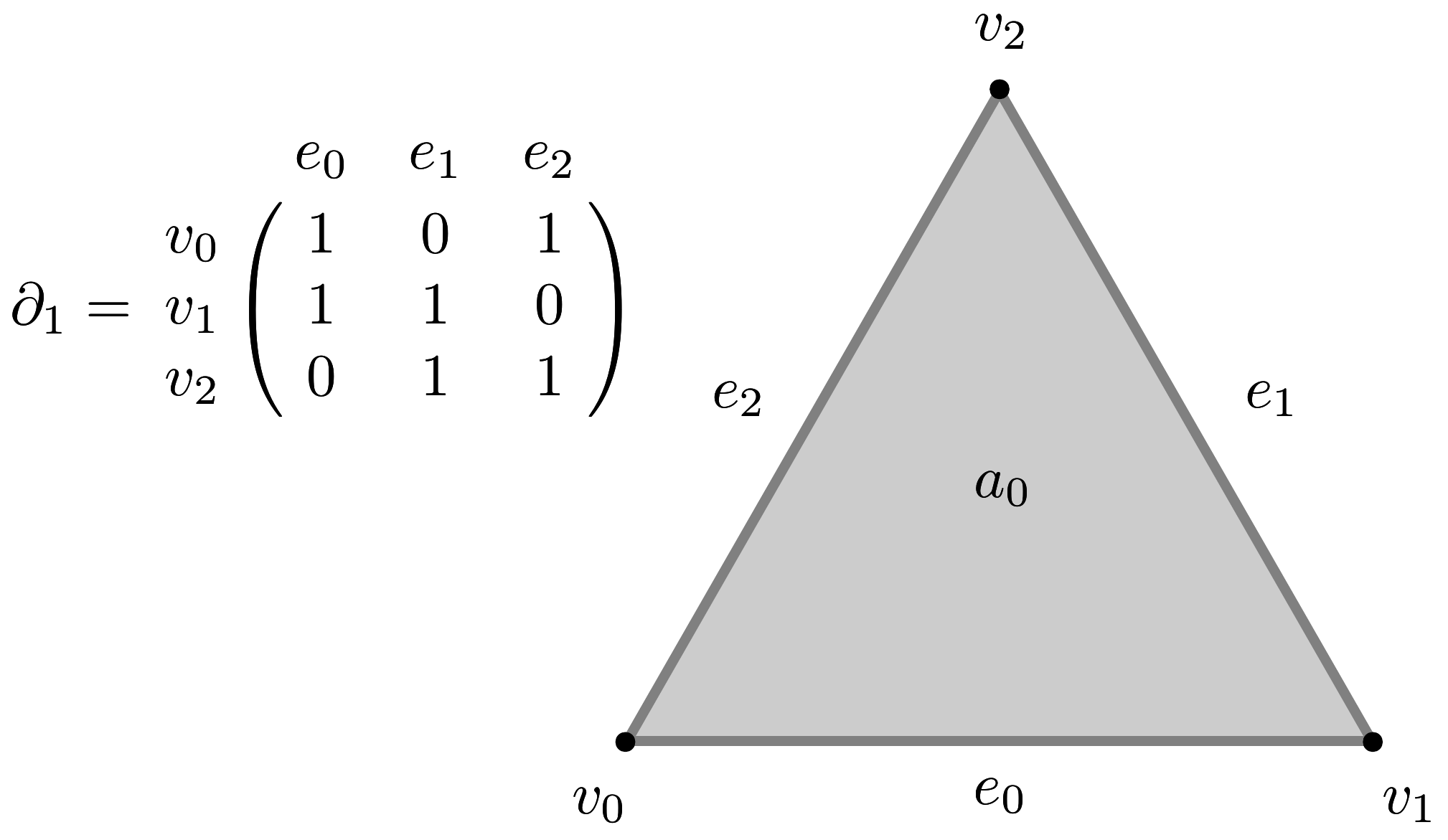}
    \caption{The completed boundary matrix from edges to vertices.}
    \label{fig:homologyfull}
\end{figure}

To populate our boundary matrix, we simply put a $1$ for each vertex $v_n$ that bounds a given edge $e_m$. We see that edge $e_0$ is bounded by vertices $v_0$ and $v_1$ and hence we get the following column entries (Figure \ref{fig:homologyonecolumn}).
We continue this process for all edges and arrive at a fully filled boundary matrix as depicted in Figure \ref{fig:homologyfull}.
While we simply entered $1$ entries here, we can make all my arguments in this simple binary setting without overflow. This means that all our additions and subtractions are carried out by xor operations. To mathematicians, this is known as the cyclic finite group $\mathbb{Z}/\mathbb{Z}_2$ and we can do linear algebra in this setting.
\begin{figure}[h]
    \centering
    \includegraphics[width=.95\textwidth]{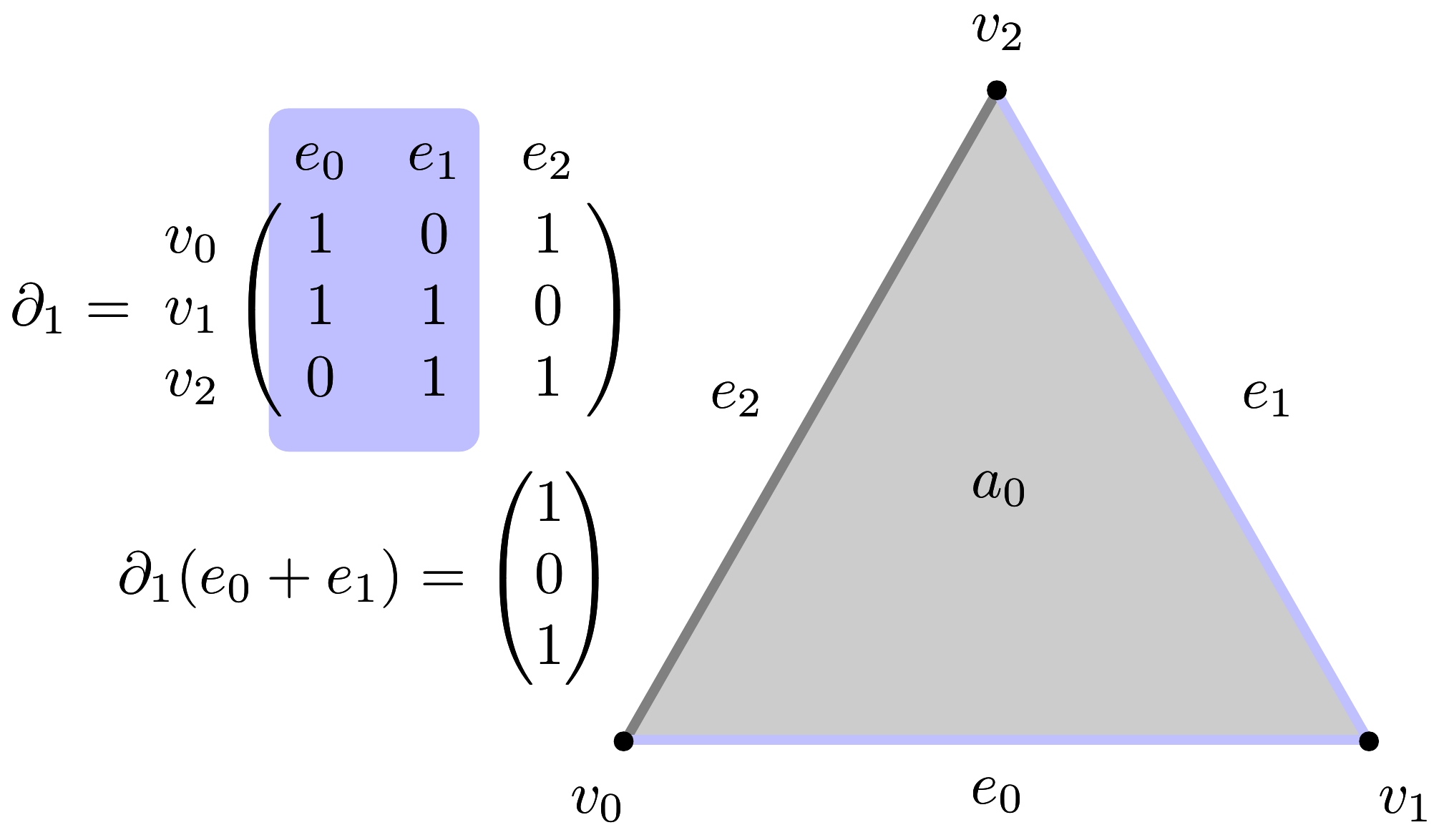}
    \caption{The linear combination of the first two columns.}
    \label{fig:homology}
\end{figure}
Given that we can do linear algebra, we can add and subtract columns (or rows). Let us add the first two columns (Figure \ref{fig:homology}). Observe that the resulting vector is actually identical to the remaining edge column $e_2$! That is these three columns are not linearly independent. Any pair of them is, but adding in the final column is not. We will not show a figure of this but you can easily check that the same holds for any chain of edges that eventually close on themselves in a cycle. In other words, if we sum over a closed chain of edges we get a zero vector and in our example we have $\partial_1(e_0+e_1+e_2)=(0,0,0)^T$. From linear algebra we know that linear dependence reduces rank. For every independent cycle (in the sense of linear algebra) we hence get a rank reduction. The size of the rank reduction is the size of the null-space of a linear map (here our matrix). This is the first key observation to computing homology.\marginnote{Independent cycles lead to rank reduction in the boundary matrix.}

\begin{figure}[h]
    \centering
    \includegraphics[width=.95\textwidth]{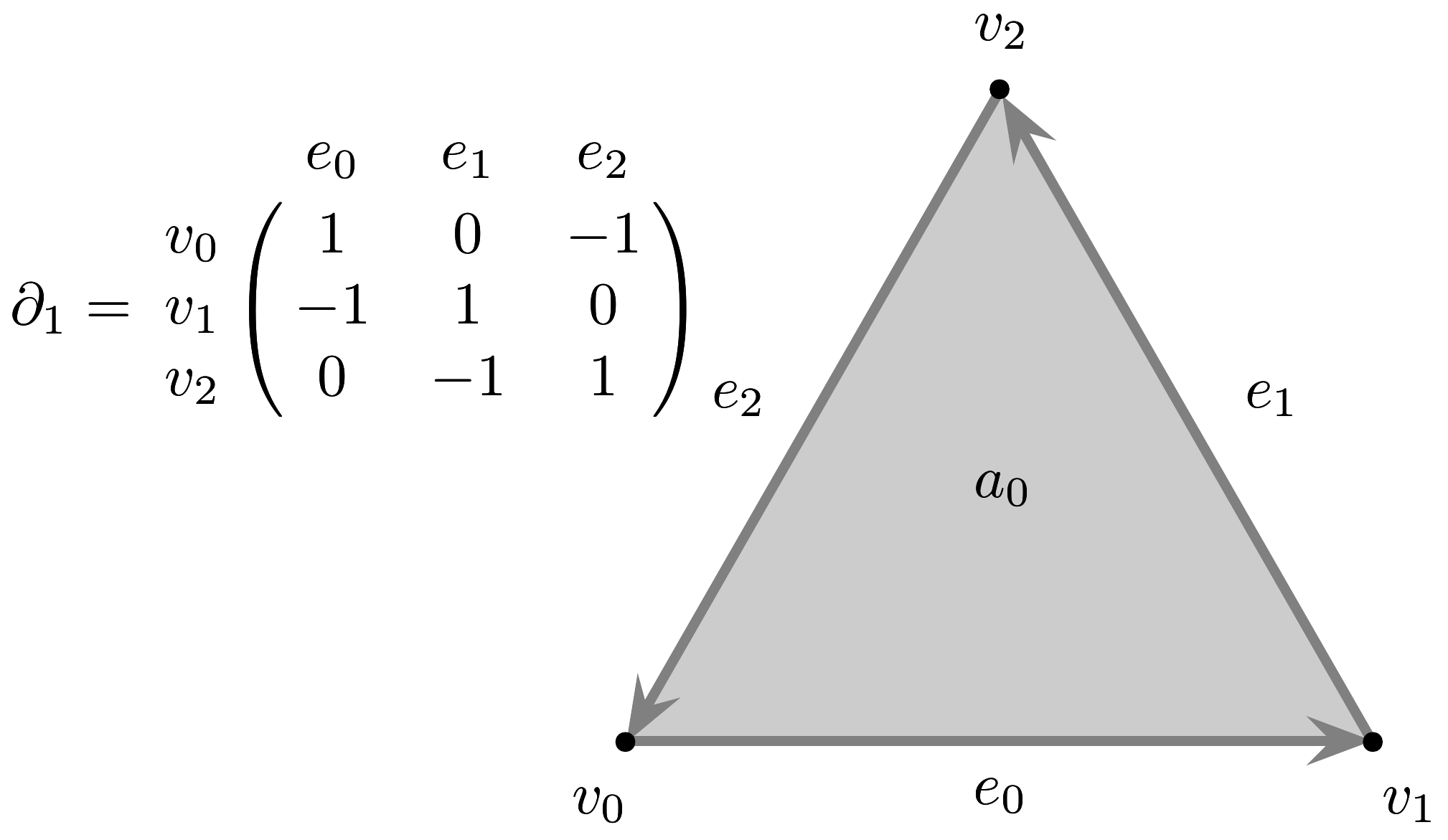}
    \caption{Boundary matrix of oriented $1$-simplices. The addition of all three columns again leads to a zero vector.}
    \label{fig:orientedhomology}
\end{figure}

Before we move on, let us consider a quick digression. We do this example in binary xor arithmetic. However the argument works for other coefficients as well. This allows us to capture more information. For example if we repeat the construction of the boundary matrix, but using a $-1$ for the tip of a directed simplex arrow, and $1$ for the base of such an arrow, we get the boundary matrix $\partial_1$ shown in Figure \ref{fig:orientedhomology}. It is easy to check that again the sum of the three column equals the zero vector. We will however stick to the simple xor example for the rest of our discussion.

\begin{figure}[h]
    \centering
    \includegraphics[width=.95\textwidth]{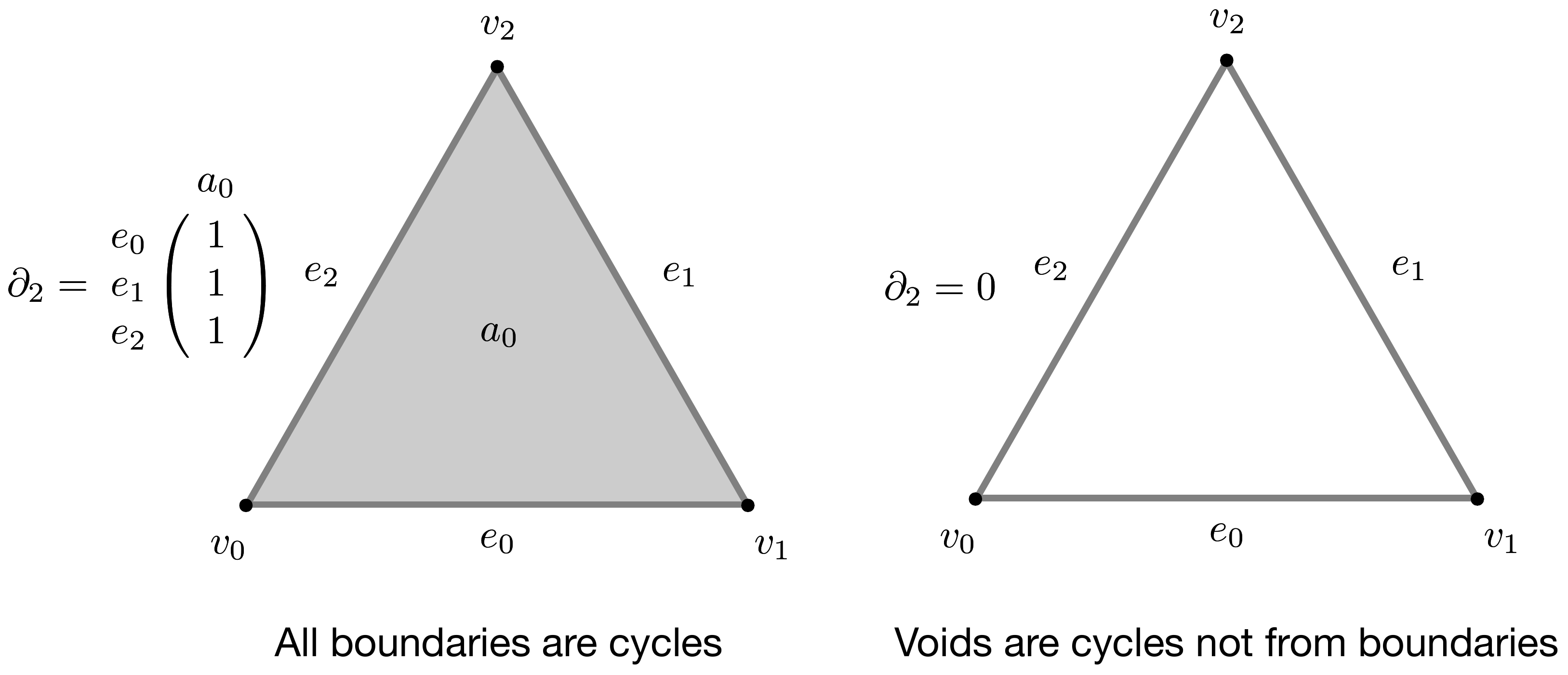}
    \caption{The boundary matrix $\partial_2$ is capturing if a cycle is a void or the boundary of a filling simplex?}
    \label{fig:homology2}
\end{figure}

We are however not done computing homology just by computing the number of cycles. After all the very example we have showed a filling area $a_0$, hence the the cycle is actually filled hence does not bound a void. This is precisely the condition we are missing. We need to be able to differentiate if a cycle is filled in or not. The condition is straightforward. Recall that each $n$-simplex by definition contains cycles that bound it. This information is encoded in the next higher-dimensional boundary matrix. So let us construct the $\partial_2$ boundary matrix for both possible cases. A cycle of three edges is either bounding a void or is the boundary of an area (or in the simplicial language a $2$-simplex). Both cases are shown in Figure \ref{fig:homology2}.

\begin{figure}[h]
    \centering
    \includegraphics[width=.95\textwidth]{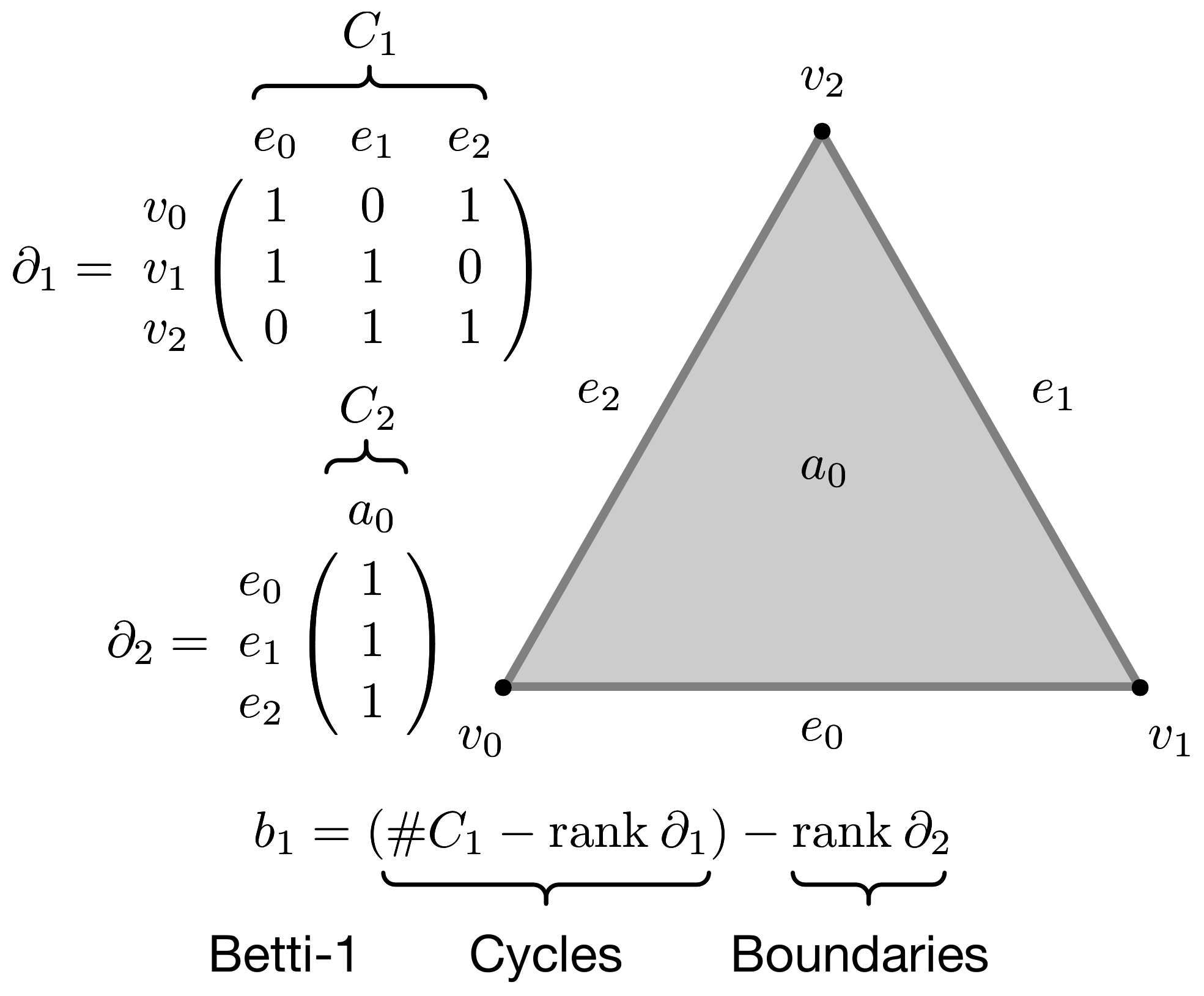}
    \caption{Computation of Betti-1 $b_1$, the number of area-like voids.}
    \label{fig:Betti1}
\end{figure}

\begin{figure}[h]
    \centering
    \includegraphics[width=.65\textwidth]{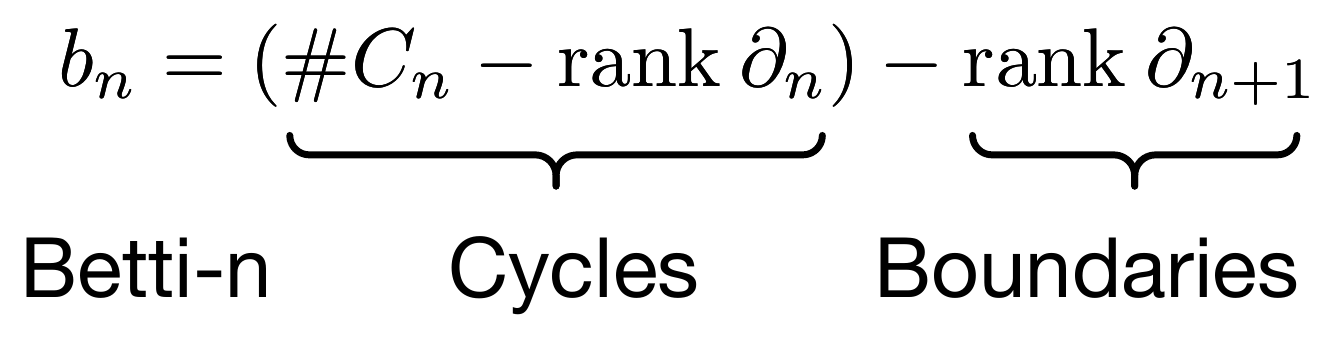}
    \caption{Computation of Betti-N $b_n$, the number of n-simplex-like voids.}
    \label{fig:Bettin}
\end{figure}

If the area is filled then the boundary map contains a map from the area to the three edges in its boundary. Given that we only have one area in this example we get only one column in the boundary matrix $\partial_2$. If, however, the cycle does not come from the boundary of an area there is no matching map in the boundary map. In other words cycles generated from being a boundary require a linearily independent entry in the boundary matrix, hence the rank must be containing all of them.
Now we have all the pieces to compute homology. The information if something is in the cycle is in the lower dimensional boundary matrix, specifically the size of its null space. The information if something is filled in or not is contained in the higher dimensional boundary matrix, specifically the rank of the boundary matrix.
All this is collected in this final Figure \ref{fig:Betti1} on the computation of the simplicial homology.
It turns out that these relationships hold for any dimensional simplicial complexes and their boundary matrices derived analogous to how we did it for $1$- and $2$-simplices. So we get the formula for Betti numbers as shown in Figure \ref{fig:Bettin}. 
This, in a nutshell is the computation of homology of a simplicial complex! It amounts to computing the rank (and by simple extension the size of the null-space) of two matrices. There are a plethora of ways to compute the rank of a matrix, and variations of Gaussian elimination should come to mind. A particular form of a matrix, the {\em Smith Normal Form} can be computed by a Gaussian elimination style reduction and it makes reading off rank easy. A point to keep in mind here is that the specifics of the columns of the matrix does not matter for these results, merely that they are linearily (in)dependent as we discussed before.
\subsection{Homology Groups}
What we have discussed so far is how one computes homology for a simplicial complex. However, virtually all modern textbooks on algebraic topology will describe homology through groups. To appreciate the connection of our discussion above in terms of constructed boundary matrices and rank-nullity computations here is a brief peek at the algebraic formulation of the same ideas.

The notion of an $n$-Chain $C_n$ captures our process of constructing boundary matrices and what they operate upon. Chains that are connected via a sequence of boundary operators are called a {\em Chain complex}.
\begin{figure}
\begin{equation*}
    0\xrightarrow{}C_n\xrightarrow{\partial_n}C_{n-1}\xrightarrow{\partial_{n-1}}\cdots\xrightarrow{} C_1\xrightarrow{\partial_{1}}C_0\xrightarrow{\partial_{0}}0
\end{equation*}

\centering
\begin{tikzpicture}[
node distance = 0 mm and 33mm,
     E/.style = {shape=ellipse, aspect=0.7,
                 minimum height=2mm+#1mm,
                 minimum width=#1mm,
                 draw, anchor=south,
                 node contents={}}
                    ]
    \node (n3a) [fill=black!5,E={9+21}];
    \node (n2a) [fill=black!10,E={9+14}];
    \node (n1a) [fill=black!15,E={9+7}];
    \node (n0a) [draw,fill=white,shape=circle,inner sep=0,minimum size =3pt] {};
    
    \node (m1a) [minimum width=3em,below=of n1a.north] {$\boldsymbol{B}_{n+1}$};
    \node (m2a) [minimum width=3em,below=of n2a.north] {$\boldsymbol{Z}_{n+1}$};
    \node (m3a) [minimum width=3em,below=of n3a.north] {$\boldsymbol{C}_{n+1}$};
    \node (m0a) [minimum width=3em,below=of n0a.south] {$0$};


    \node (n3b) [right=30mm,fill=black!5,E={9+21}];
    \node (n2b) [right=30mm,fill=black!10,E={9+14}];
    \node (n1b) [right=30mm,fill=black!15,E={9+7}];
    \node (n0b) [right=30mm,draw,fill=white,shape=circle,inner sep=0,minimum size =3pt] {};

    \node (m1b) [minimum width=3em,below=of n1b.north] {$\boldsymbol{B}_{n}$};
    \node (m2b) [minimum width=3em,below=of n2b.north] {$\boldsymbol{Z}_{n}$};
    \node (m3b) [minimum width=3em,below=of n3b.north] {$\boldsymbol{C}_{n}$};
    \node (m0b) [minimum width=3em,below=of n0b.south] {$0$};
    
    \draw[black] (n3a.north) .. controls +(right:10mm) and +(left:10mm) .. (n1b.north);
    \draw[black] (n2a.north) .. controls +(right:10mm) and +(left:10mm) .. (n1b.south);

    \draw[black] (n0a) -- node[midway, below] {$\xrightarrow[\partial_{n+1}]{\qquad}$} (n0b);

    \node (n3c) [right=60mm,fill=black!5,E={9+21}];
    \node (n2c) [right=60mm,fill=black!10,E={9+14}];
    \node (n1c) [right=60mm,fill=black!15,E={9+7}];
    \node (n0c) [right=60mm,draw,fill=white,shape=circle,inner sep=0,minimum size =3pt] {};

    \node (m1c) [minimum width=3em,below=of n1c.north] {$\boldsymbol{B}_{n-1}$};
    \node (m2c) [minimum width=3em,below=of n2c.north] {$\boldsymbol{Z}_{n-1}$};
    \node (m3c) [minimum width=3em,below=of n3c.north] {$\boldsymbol{C}_{n-1}$};
    \node (m0c) [minimum width=3em,below=of n0c.south] {$0$};

    \draw[black] (n3b.north) .. controls +(right:10mm) and +(left:10mm) .. (n1c.north);
    \draw[black] (n2b.north) .. controls +(right:10mm) and +(left:10mm) .. (n1c.south);
    \draw[black] (n0b) -- node[midway, below] {$\xrightarrow[\partial_{n}]{\qquad}$} (n0c);
%
\end{tikzpicture}

\begin{equation*}
    H_n=Z_n/B_n=\ker \partial_n/\im \partial_{n+1}
\end{equation*}
\caption{Homology in the language of abelian groups and $n$-chains. The sequence of chains is called a chain complex.}
\end{figure}
As we have seen, some subset of chains can form cycles, which here is captured by $Z_n$. Furthermore, some chains are present because they are in the boundary of some higher dimensional element in a $C_{n+1}$ chain, and therefore are in the {\em image} (or $\im$) of the boundary map from $\partial_{n+1}:C_{n+1}\xrightarrow{} C_{n}$. Furthermore recall that we noted that cycles $Z_n$ are characterized by falling into the null-space which is also known as the {kernel} (or $\ker$) of the boundary map $\partial_n$. Homology are the voids that are not from boundaries, hence we get the formula for the $n$th Homology group that is the cycles in the chain $Z_n$ with the boundaries $B_n$ i the chain "modded out" (that is removed), and we see that we can compute this information from the kernel and image of two boundary maps as we saw before.

Finally observe that the boundary of a boundary must always be zero, given that any boundary chains $B_n$ are fully included in the cycles $Z_n$ and all cycles are send to $0$ by the second boundary map. This fact is called the {\em fundamental lemma of Homology}.
\begin{figure}
\begin{equation*}
\partial_{n}\partial_{n+1}=0
\end{equation*}
\caption{The fundamental lemma of Homology states that a boundary does itself have no boundary.}
\end{figure}

\begin{marginfigure}
    \centering
    \includegraphics[width=.95\marginparwidth]{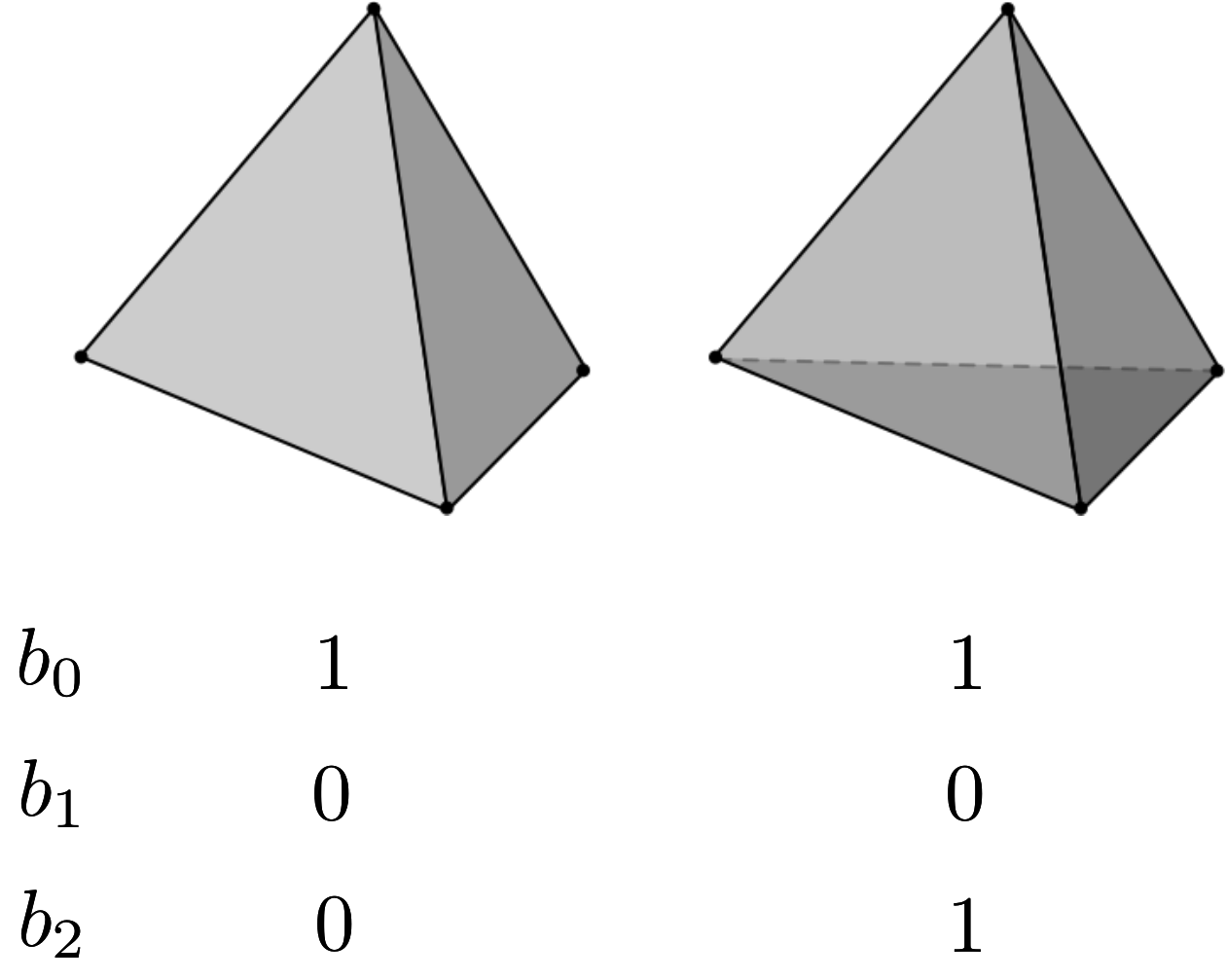}
    \caption{The Betti numbers of the simplicial ball and sphere.}
    \label{fig:Bettiex1}
\end{marginfigure}

\begin{marginfigure}
    \centering
    \includegraphics[width=.95\marginparwidth]{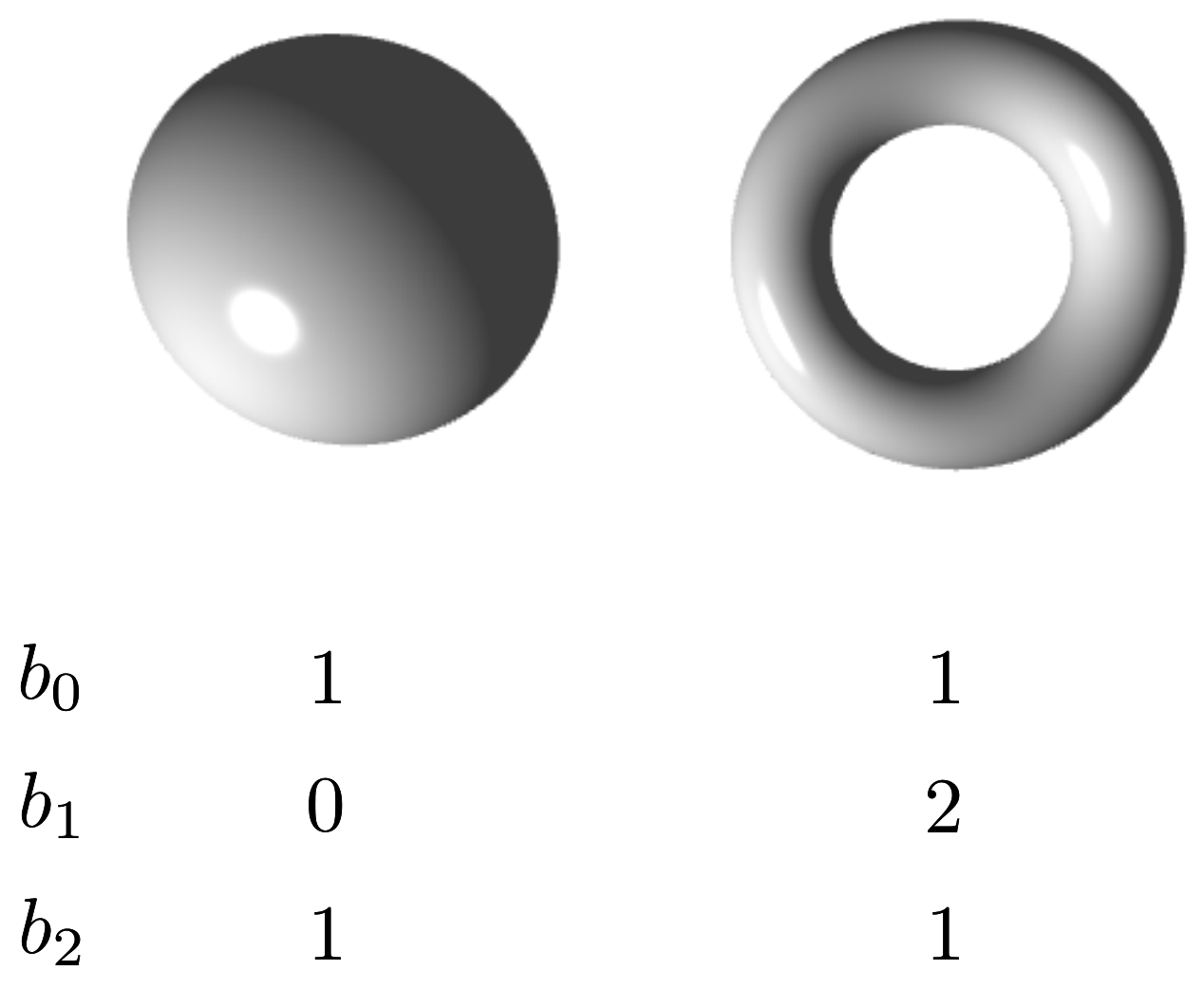}
    \caption{The Betti numbers of the ball and torus.}
    \label{fig:Bettiex2}
\end{marginfigure}

To get a sense what homology looks like in concrete cases, here are some simple examples: Figure \ref{fig:Bettiex1} shows a solid tetrahedron (that you can also think as an abstract $3$-simplex) and next to it the boundary of said shape with the interior empty. We see that both cases are a single connected component ($b_0=1$), while only the hollow shape has $b_2=0$. While computationally we will generally deal with simplicial complexes, remember that conceptually we are dealing with deformability, and if we inflate our hollow tetrahedron in the physical world we will arrive at something that looks like the surface of a sphere as seen in Figure \ref{fig:Bettiex2}. Next to it we depict the torus and its Betti numbers. The torus is characterized by two loops that cannot be collapsed as we have seen before and this is captured here by $b_1=2$. 
Sometimes homological information is not presented in terms of Betti numbers but rather in terms of the structure of the Homology group. In our case, the groups are finitely generated. The cycles (without bindaries) are the generators in this group. Hence we can draw upon the structure theorem for finitely-generated abelian groups with respect to direct sums which is as follows:
\begin{thm}
A {\bf finitely-generated abelian group} $A$ can be uniquely expressed in the form of direct sums of finite numbers of free cyclic groups $\mathbb{Z}$, and cyclic groups $\mathbb{Z}_{t_i}$ of finite period $t_I$ called {\bf Torsion}. The rank of the free group we will call {\bf Betti numbers}, The indices $t_i$ are not necessarily distinct prime and are called {\bf torsion coefficients}:
\begin{equation*}
A = \bigoplus\limits_{b}\mathbb{Z}\oplus \bigoplus\limits_{i}\mathbb{Z}_{t_i}
\end{equation*}
\end{thm}
\begin{marginfigure}
    \centering
    \includegraphics[width=.95\marginparwidth]{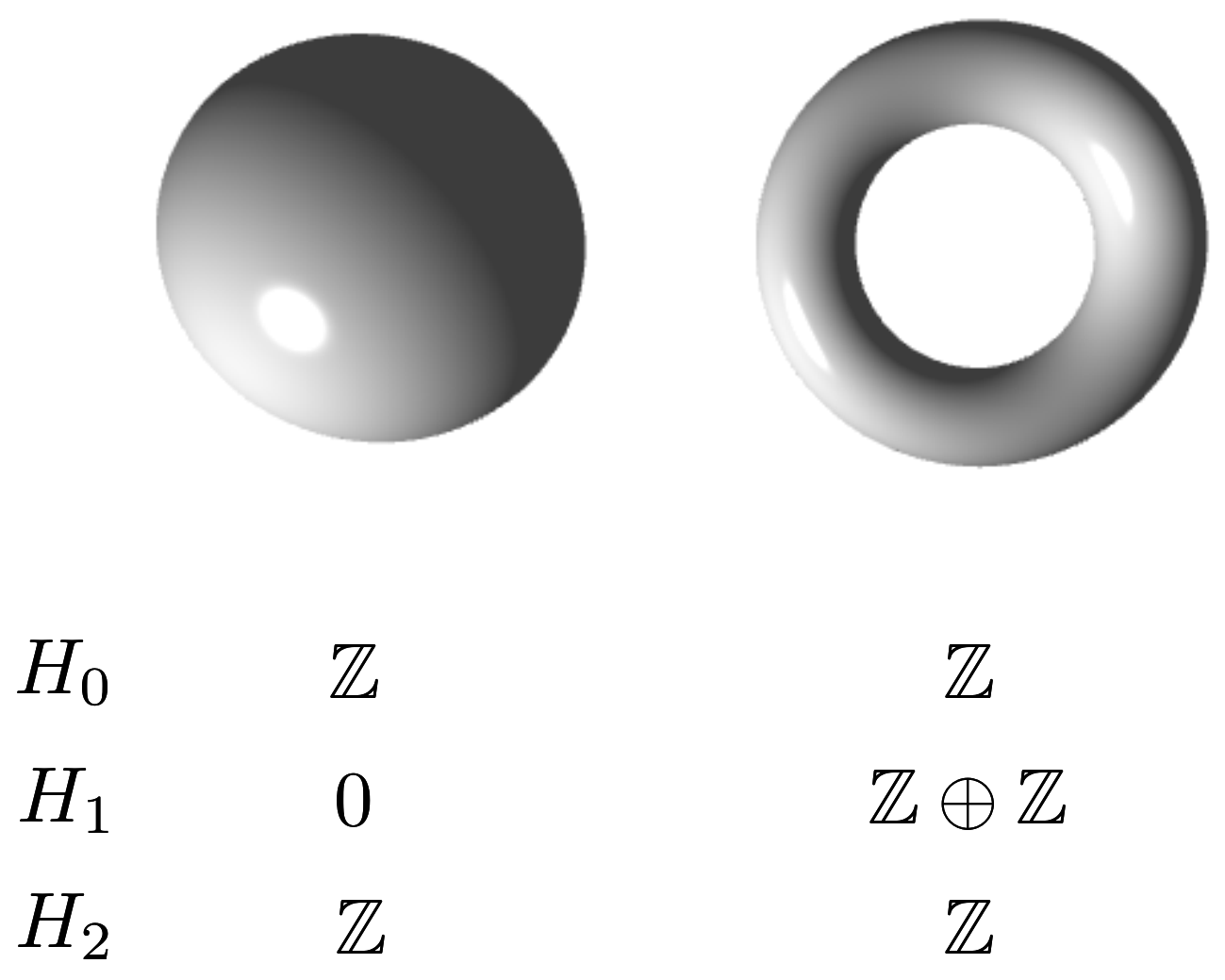}
    \caption{The homology groups of the ball and torus.}
    \label{fig:Bettiex3}
\end{marginfigure}
In this view, Betti numbers count the number of free groups $\mathbb{Z}$ and hence the homology groups for the sphere and torus written as shown in Figure \ref{fig:Bettiex3}. In our discussion we have ignored the {\em Torsion} coefficients that show up in this structure theorem. If we use our xor linear algebra, torsion never occurs, but it can occur if we consider oriented situations. However, so far torsion coefficients have not made a particular impression in applied topological examples, so we will continue to ignore them in this introduction. But it is good to know they exist!

\subsection{Persistent Homology}

The idea of persistence is fairly straightforward. We can compute homology for any simplicial complex, so of course we can compute homology for a sequence of additions to a simplicial complex. The Betti numbers will change as simplices are included, such as cycles are formed by closing the loop, or cycles become boundaries by filling simplices being added.

If we consider a complete simplicial complex $X$ but imagine that its piece were successively added (but nothing ever removed) we get a sequence of simplicial complexes $X_m$ starting with some starting simplex $X_0$. This sequence of inserting of simplices lead to a finite {\em filtration}:
\begin{marginfigure}
    \centering
    \includegraphics[width=.99\marginparwidth]{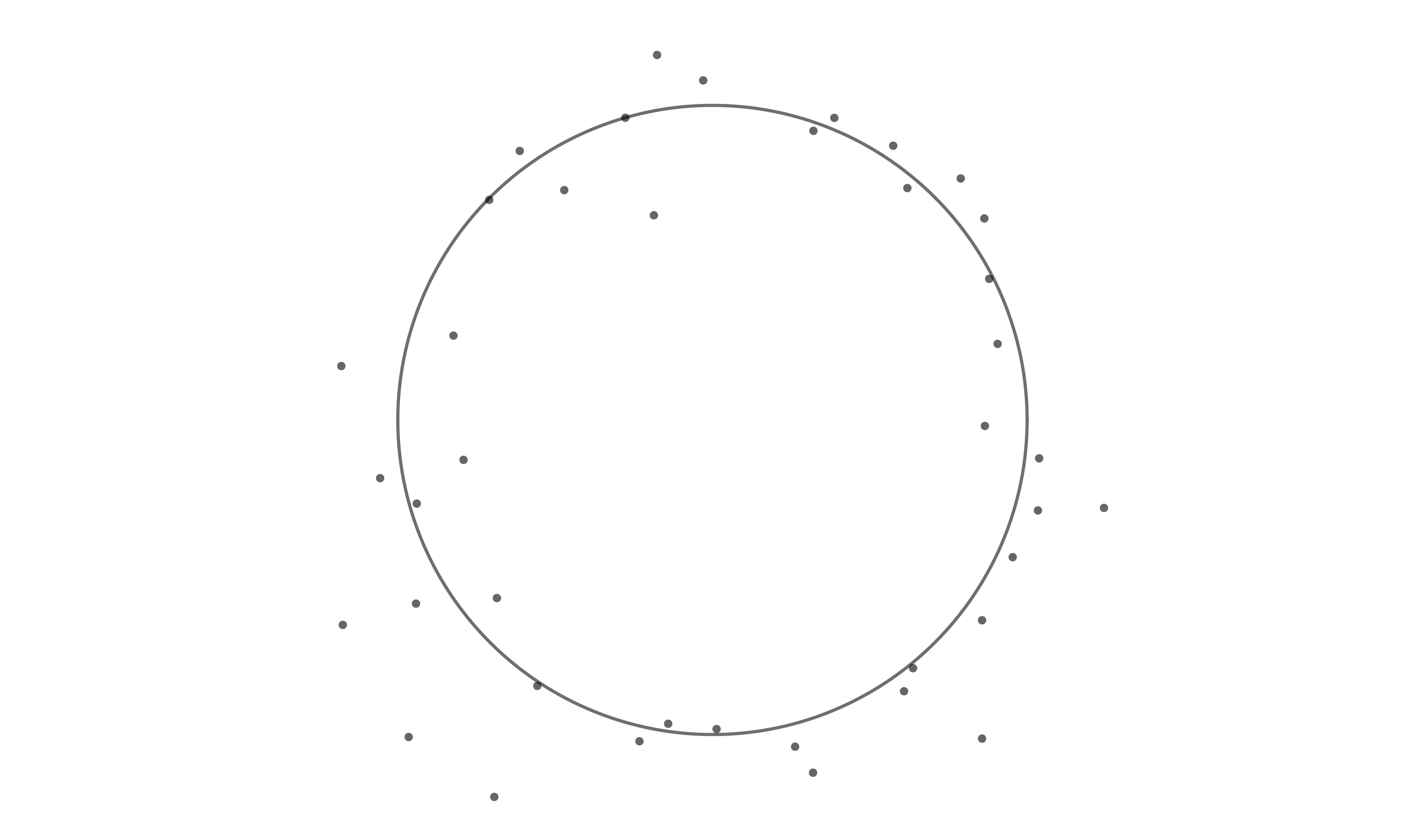}
    \includegraphics[width=.99\marginparwidth]{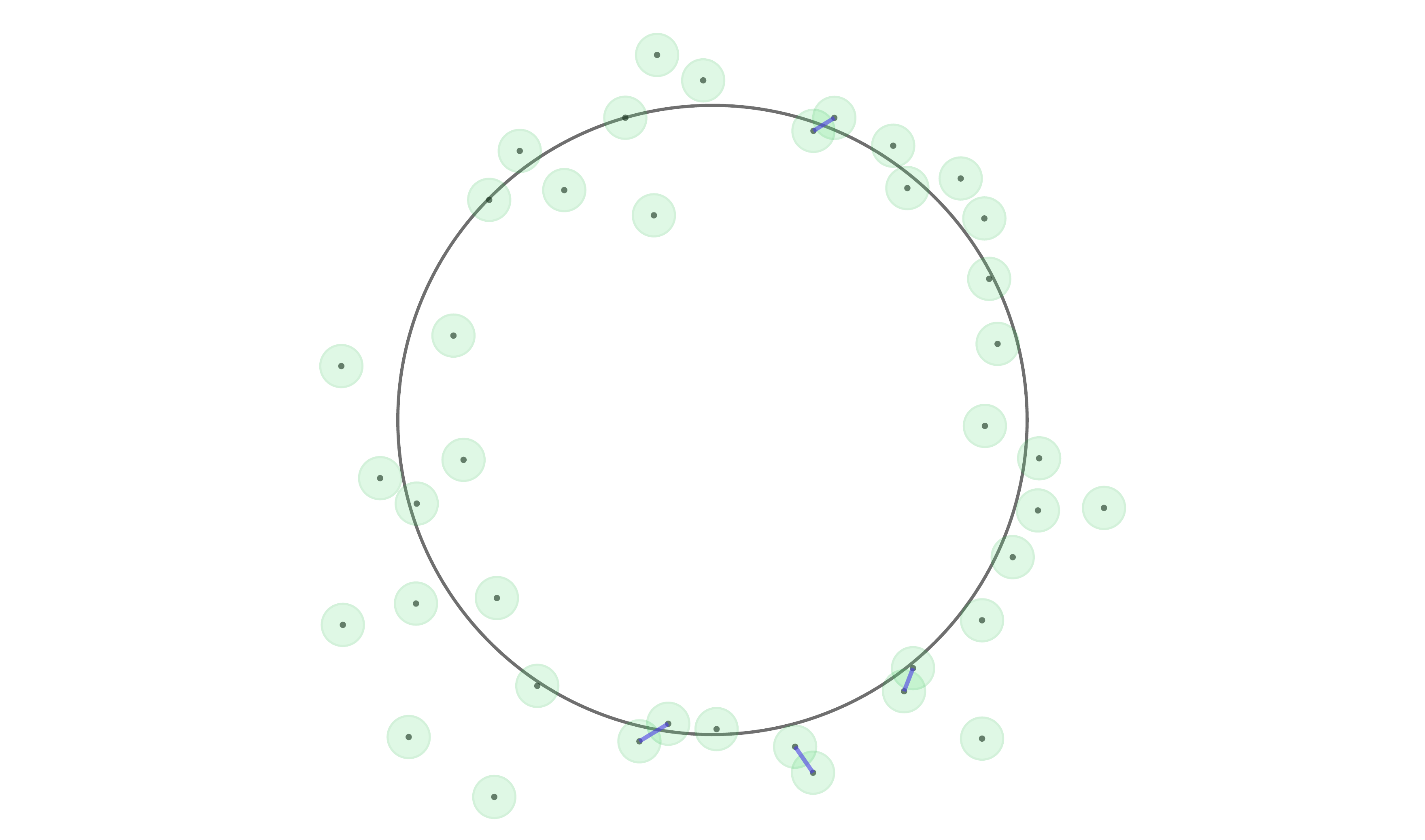}
    \includegraphics[width=.99\marginparwidth]{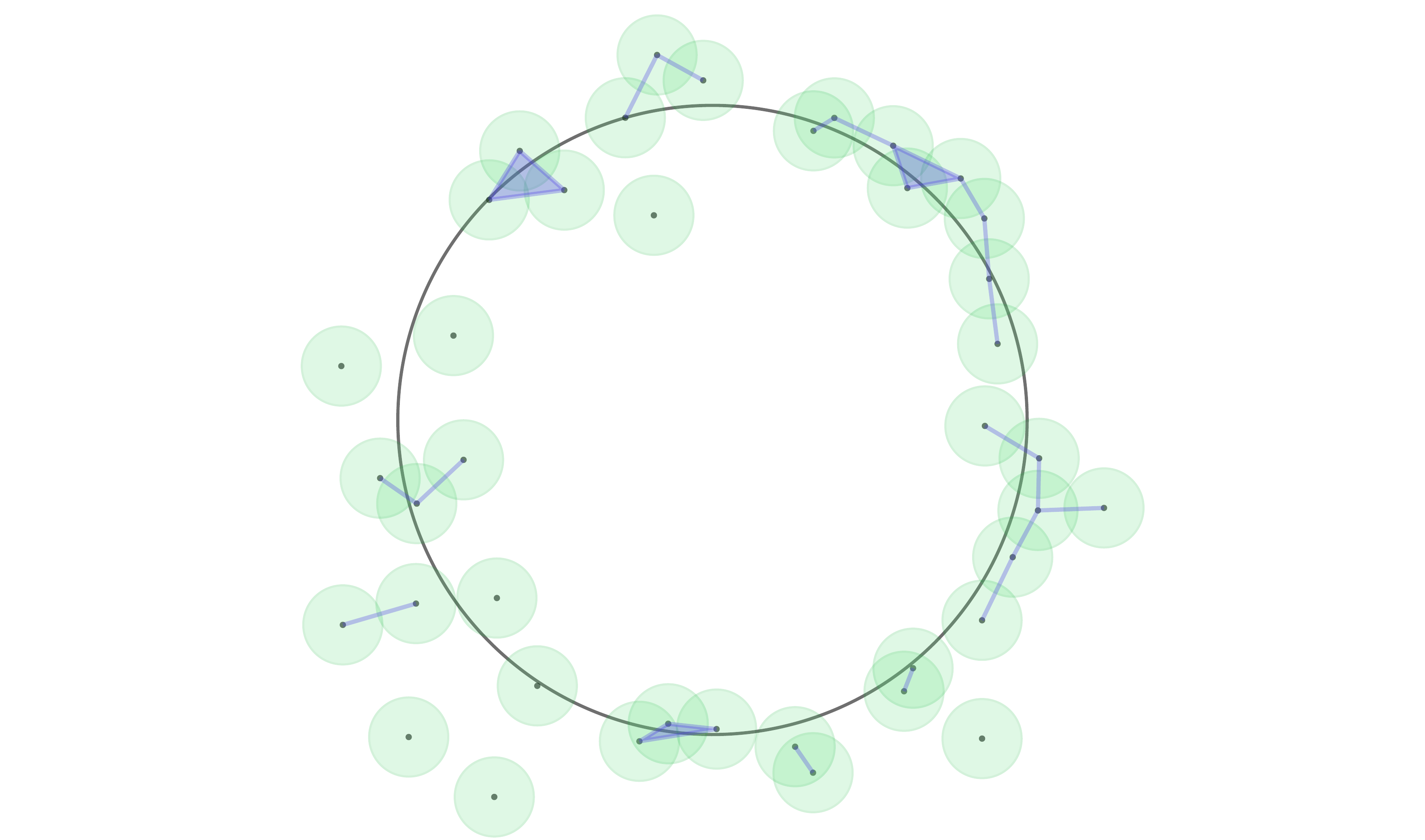}
    \includegraphics[width=.99\marginparwidth]{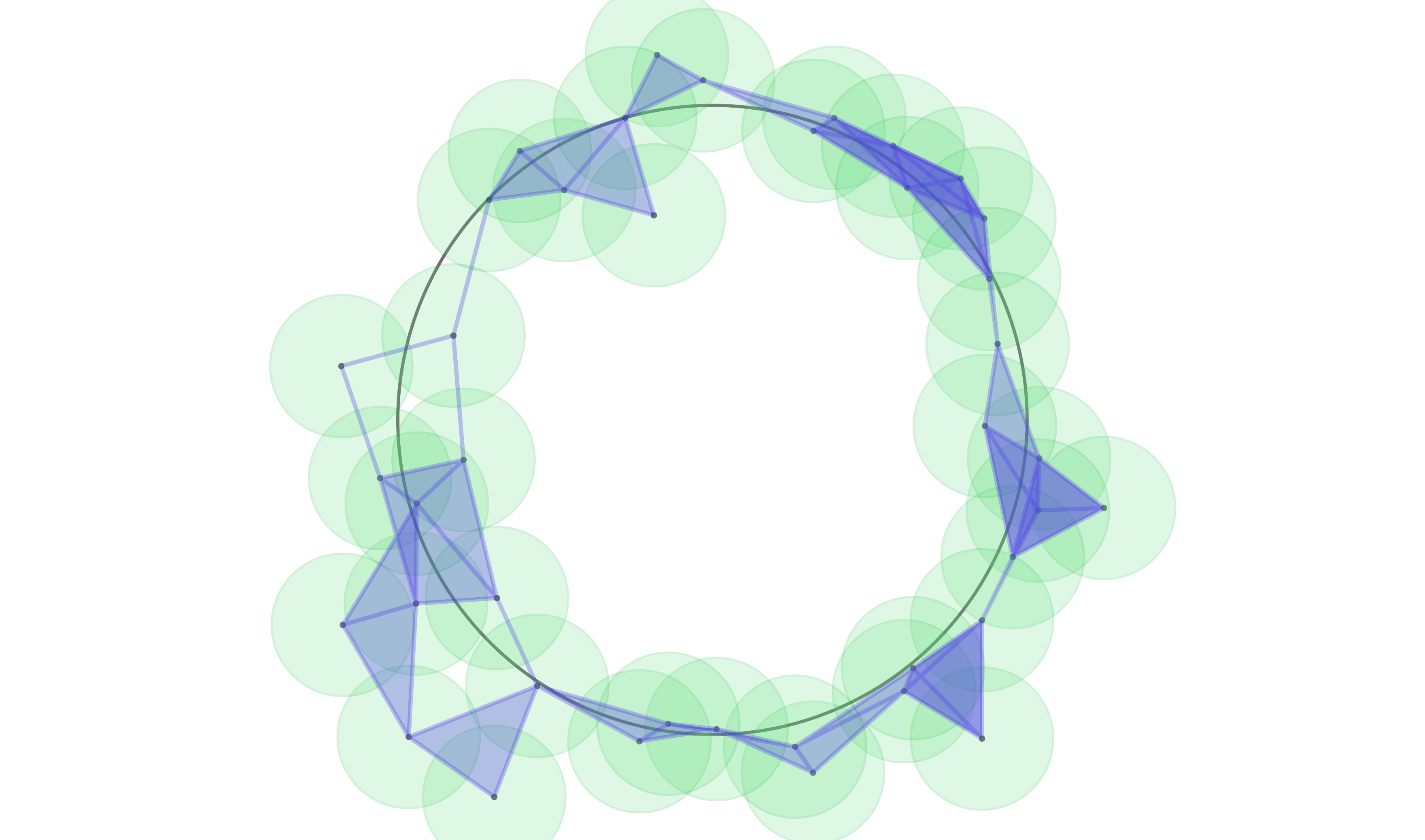}
    \includegraphics[width=.99\marginparwidth]{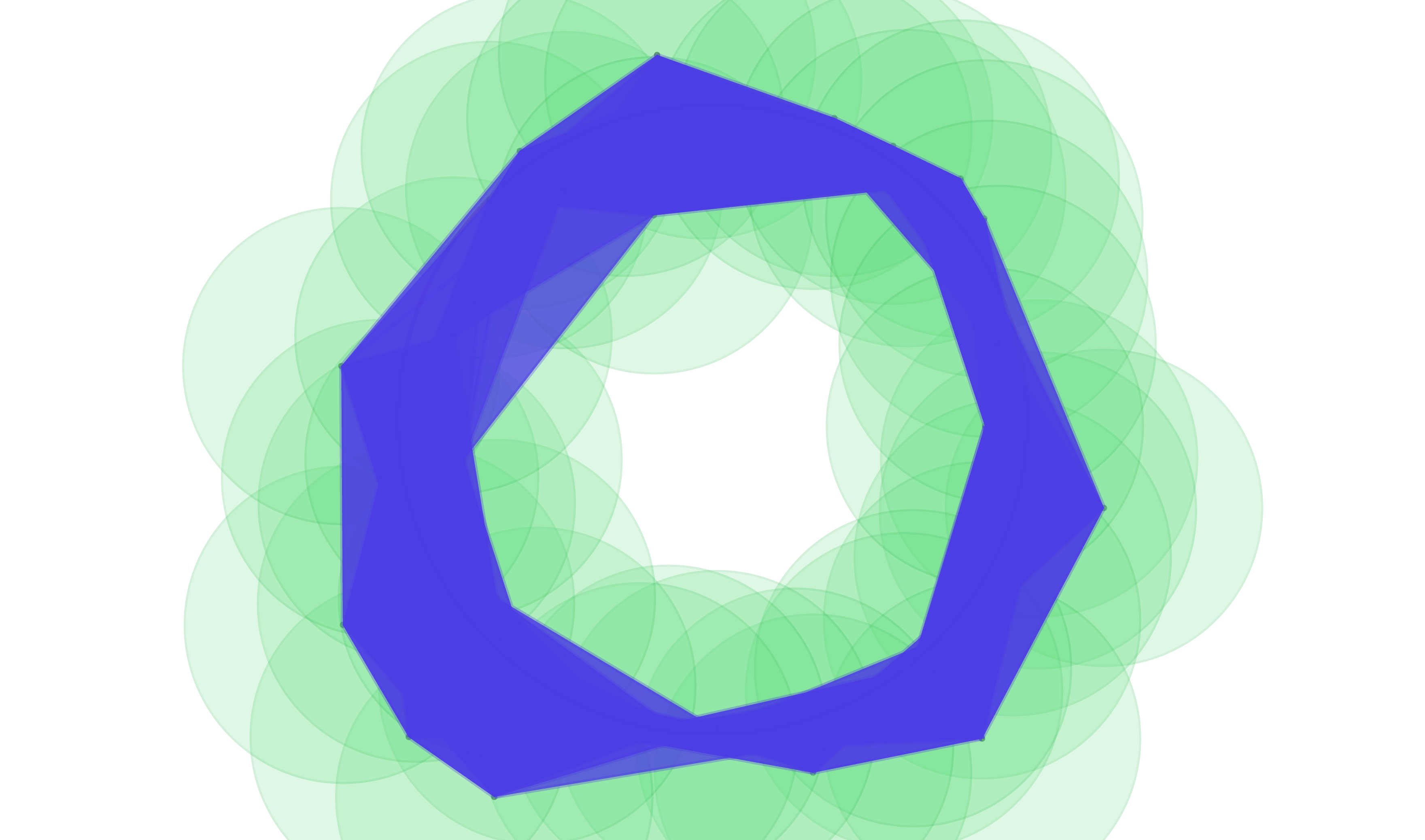}
    \caption{The Vietoris-Rips process. As growing radii touch, simplices are connected and become higher order. Notice that the {\em persistent} feature is the hole of the circle.}
    \label{fig:vietorisripscircle}
\end{marginfigure}
\begin{equation*}
0\xhookrightarrow{}X_0\xhookrightarrow{} X_1 \xhookrightarrow{} \cdots \xhookrightarrow{} X_{m-1} \xhookrightarrow{} X_m = X
\end{equation*}
Given that each $X_m$ is a simplicial complex on its own we can of course compute its homology with the algorithm we have seen. For example, we can compute the $n$-homology after each insertion. This will capture how the homology changes as we keep inserting simplices.
\begin{equation*}
\begin{tikzcd}[ampersand replacement=\&]
X_0 \arrow[hookrightarrow,r] \arrow[d, "H_n"] \& X_1 \arrow[d, "H_n"] \arrow[hookrightarrow,r]  \& \cdots \arrow[hookrightarrow,r] \& X_m \arrow[d, "H_n"] \\
H_n(X_0)  \& H_n(X_1) \& \cdots \& H_n(X_m) 
\end{tikzcd}
\end{equation*}
This process of course works for any sequence of inserting simplices, but persistence is usually introduced through some metric process that dictate how simplices are inserted. To get an intuition of this, consider starting out just with $0$-simplicies, that is a cloud of points. The process we are considering here is named after {\em Vietoris-Rips}. The rule is that if you grow a spherical neiborhood, and two neighborhoods touch, you add an edge. If three neighboorhoods touch you add a 2-simplex, for 4 a 3-simplex, and so fourth.
This process is depicted for an example of a sampled circle with noise in Figure \ref{fig:vietorisripscircle}. Imagine you do not know where the sampled point is from and you want to recover a possible topology that explains it. If it is from a circle, we should hopefully recover the circle topology.

The output of process can be represented visually as {\em barcodes}. They capture when a homological feature emerges and is destroyed. Features emerge as cycles form, and they are destroyed as cycles are filled in hence become cycles from boundaries. So, a barcode captures the birth and death of homology and persistence has its name by the intuition that salient topological features will {\em persist} longer than features that emerge from noisy process or are more fine grain or local. We see that when the big cycle first forms, there are a few more smaller cycles that also form due to the distribution of the random samples, but those smaller cycles fill in faster and hence disappear faster than the big cycle.

\begin{figure}[h]
\centering
    \includegraphics[width=.85\textwidth]{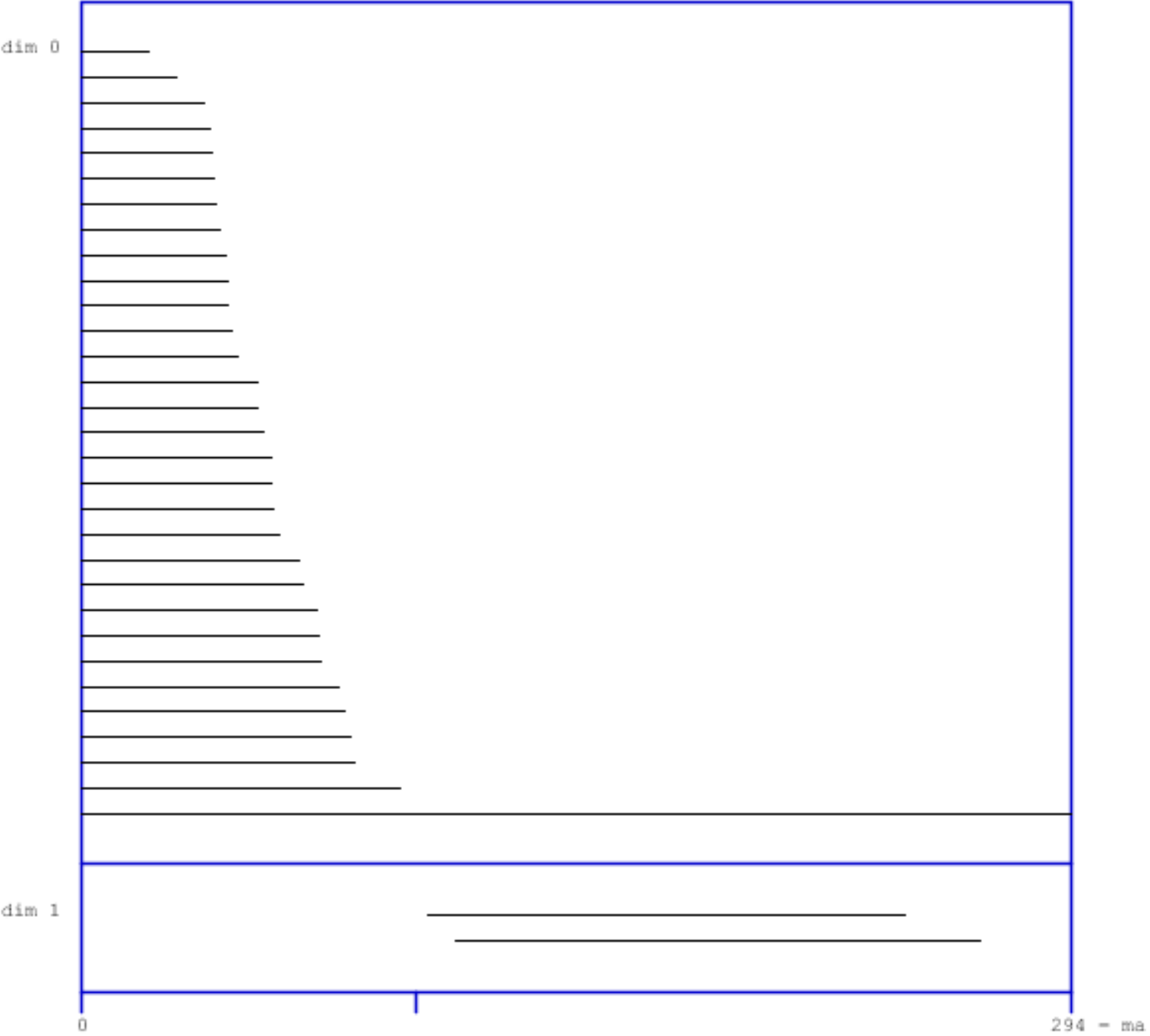}
\end{figure}
\begin{marginfigure}[-24\baselineskip]
    \centering
    \includegraphics[width=.95\marginparwidth]{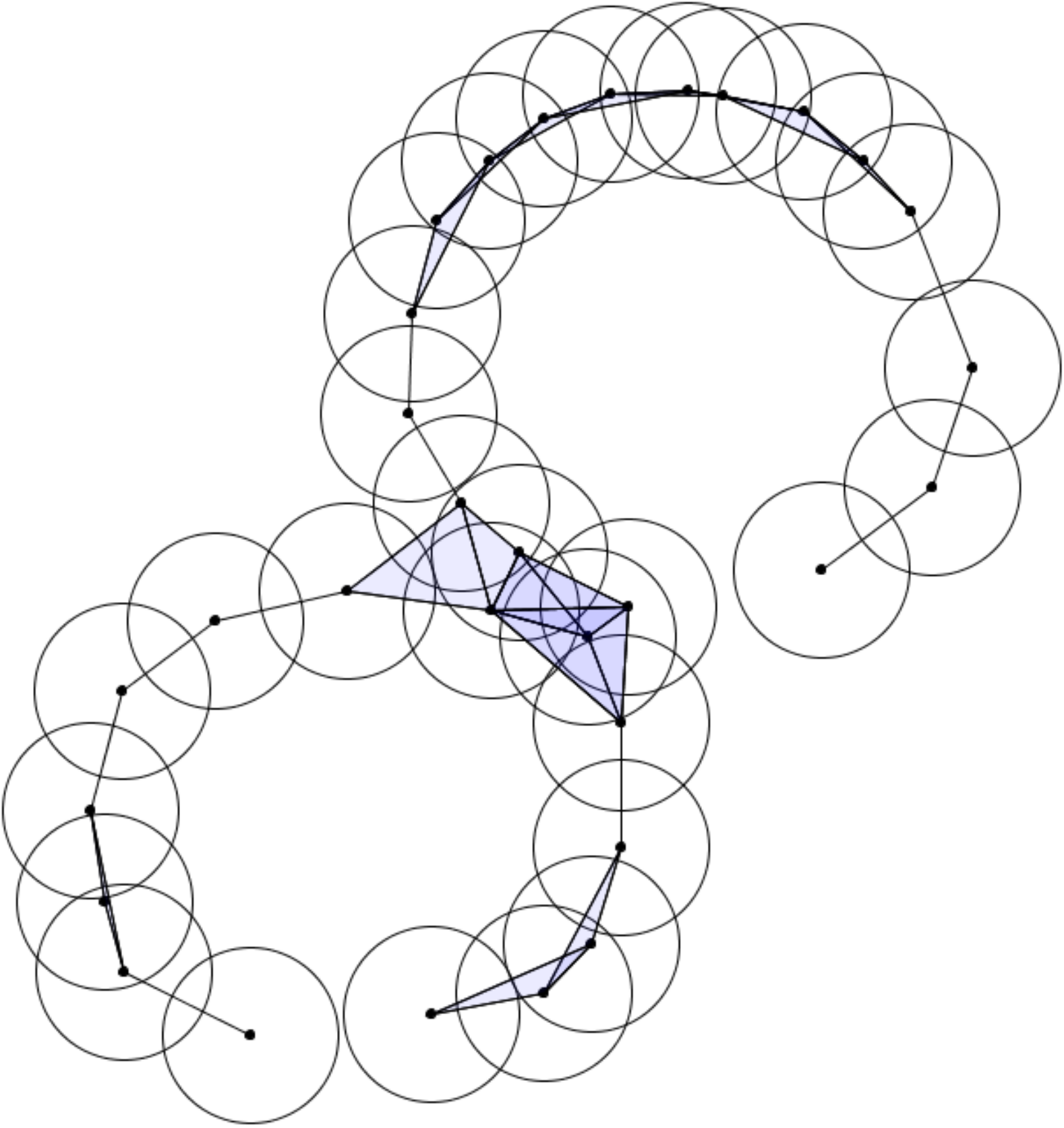}
    \caption{Barcodes for the $0$- and $1$-homology of a figure-8 style point cloud.}
    \label{fig:barcodes}
\end{marginfigure}

Figure \ref{fig:barcodes} shows a practical example of a bar code output from one of the many implementations that exist. This picture is captured from JavaPlex with the visual interface realised in Processing. The dataset here is a handdrawn pattern of a point cloud roughly capturing a figure-8 style arrangement. The $0$-homology captures the change in connected components, which shrinks as more and more points become connected. The $1$-homology indeed picks up two loops as one would hope. The blue marker at the bottom of the barcode shows the Vietoris-Rips radius used for the snapshot simplicial complex on the right. We see that at that instance the cycles have not quite closed yet.

\subsection{Embedding and Time-Series}

So far we have talked about point clouds, but audio is typically a sequence of samples without any particularly interesting geometric configuration. One way to make time series more geometrical and extract features from them is embedding the samples in a higher-dimensional space. 
\begin{figure}
    \centering
    \includegraphics[width=.95\textwidth]{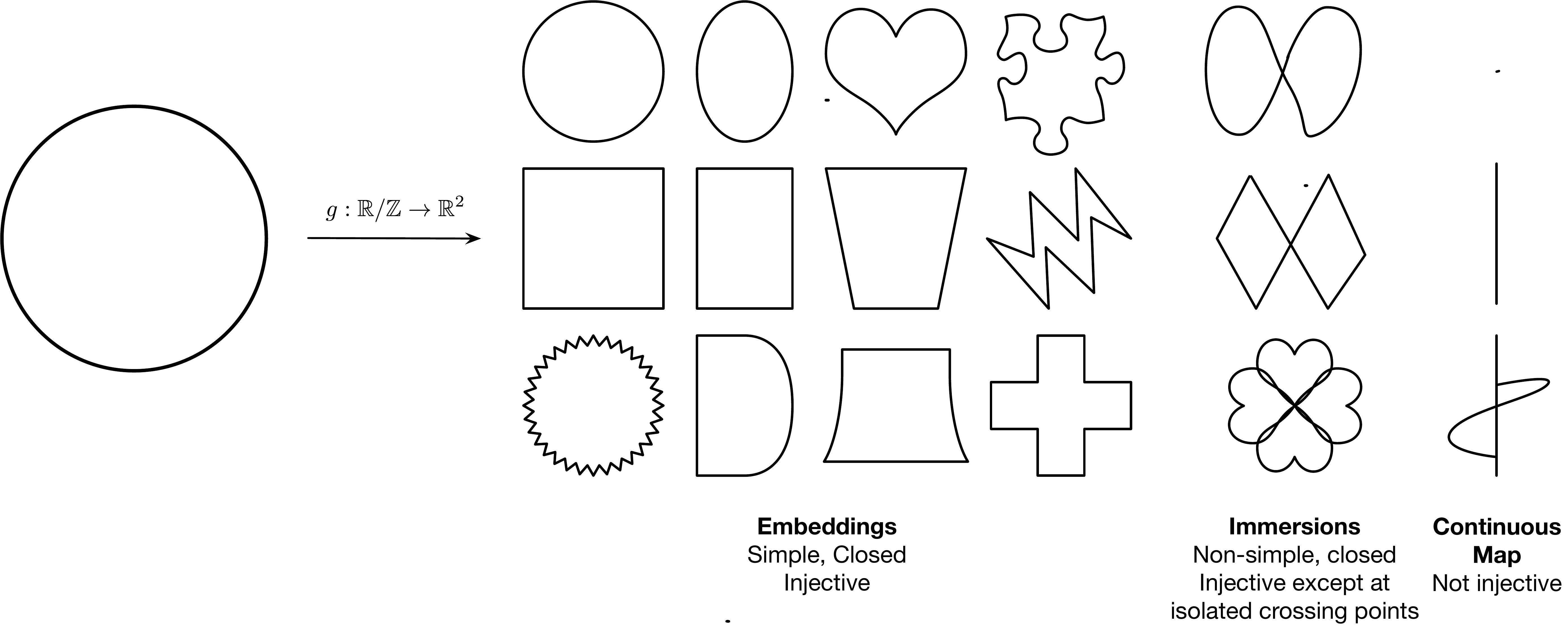}
    \caption{A topological circle can sit in the plane (left) without intersections (embedding), (middle) local intersections (immersion) and (right) lots of overlap (from a continuous map).}
    \label{fig:embedcircle}
\end{figure}
Another view that can be relevant relates to simplicial complexes that we defined combinatorially, and how to make them into geometric objects.
So there is a question of how to place a topological information into a space. Figure \ref{fig:embedcircle} shows a range of examples of how a topological circle can be placed into the plane. The leftmost group consists of curves that do not self-intersect. These are called {\em embeddings}. If there are isolated self-intersections it is called an {\em immersion}. However, a general continuous map does not need to be an immersion or embedding. Some examples of this are seen on the right in Figure \ref{fig:embedcircle}. Given that they are continuous maps from the circle, they have an underlying continuous parametrization in the topology of a circle.
\begin{marginfigure}
    \centering
    \includegraphics[width=.95\marginparwidth]{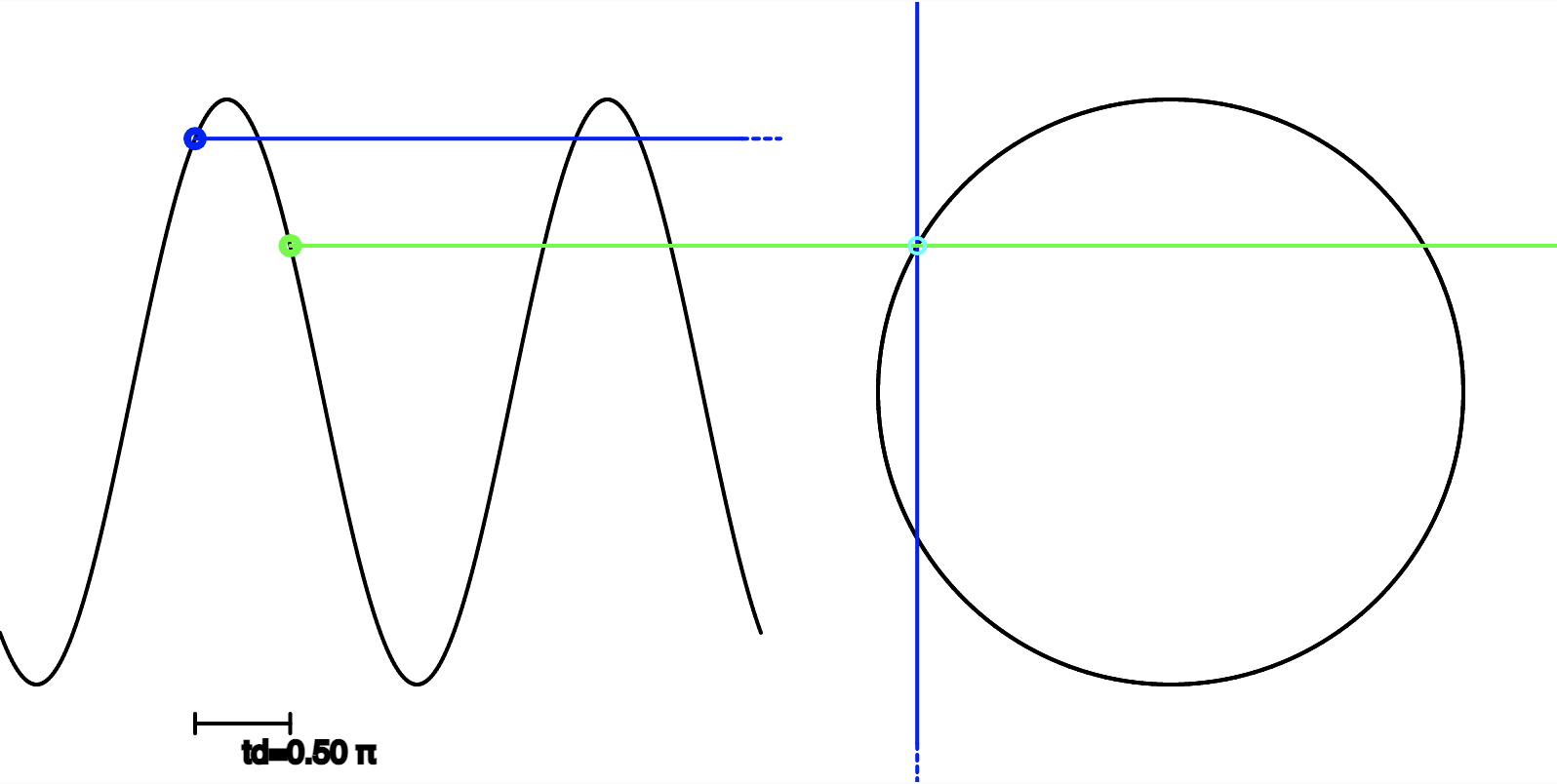}
    \includegraphics[width=.95\marginparwidth]{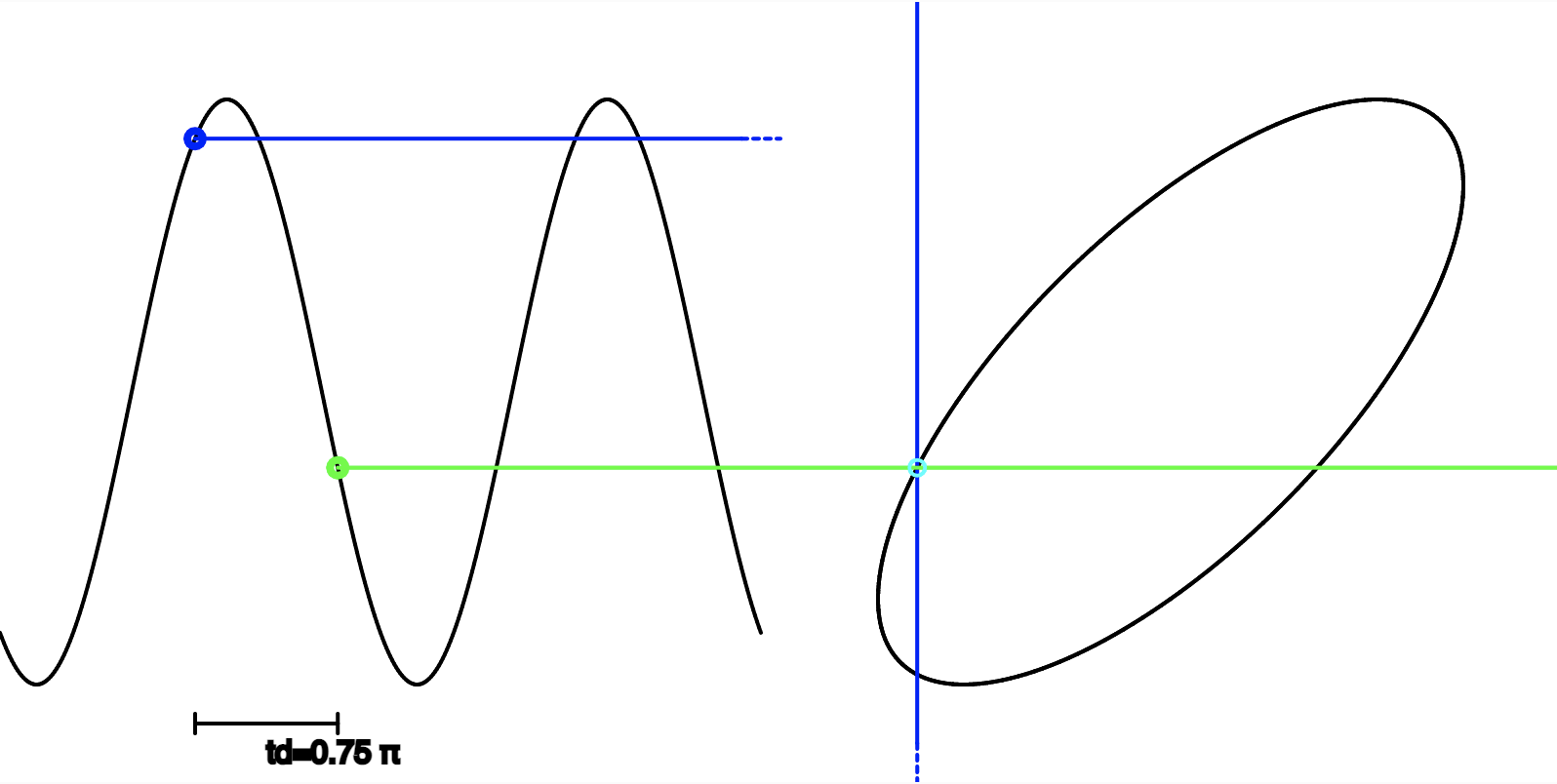}
    \includegraphics[width=.95\marginparwidth]{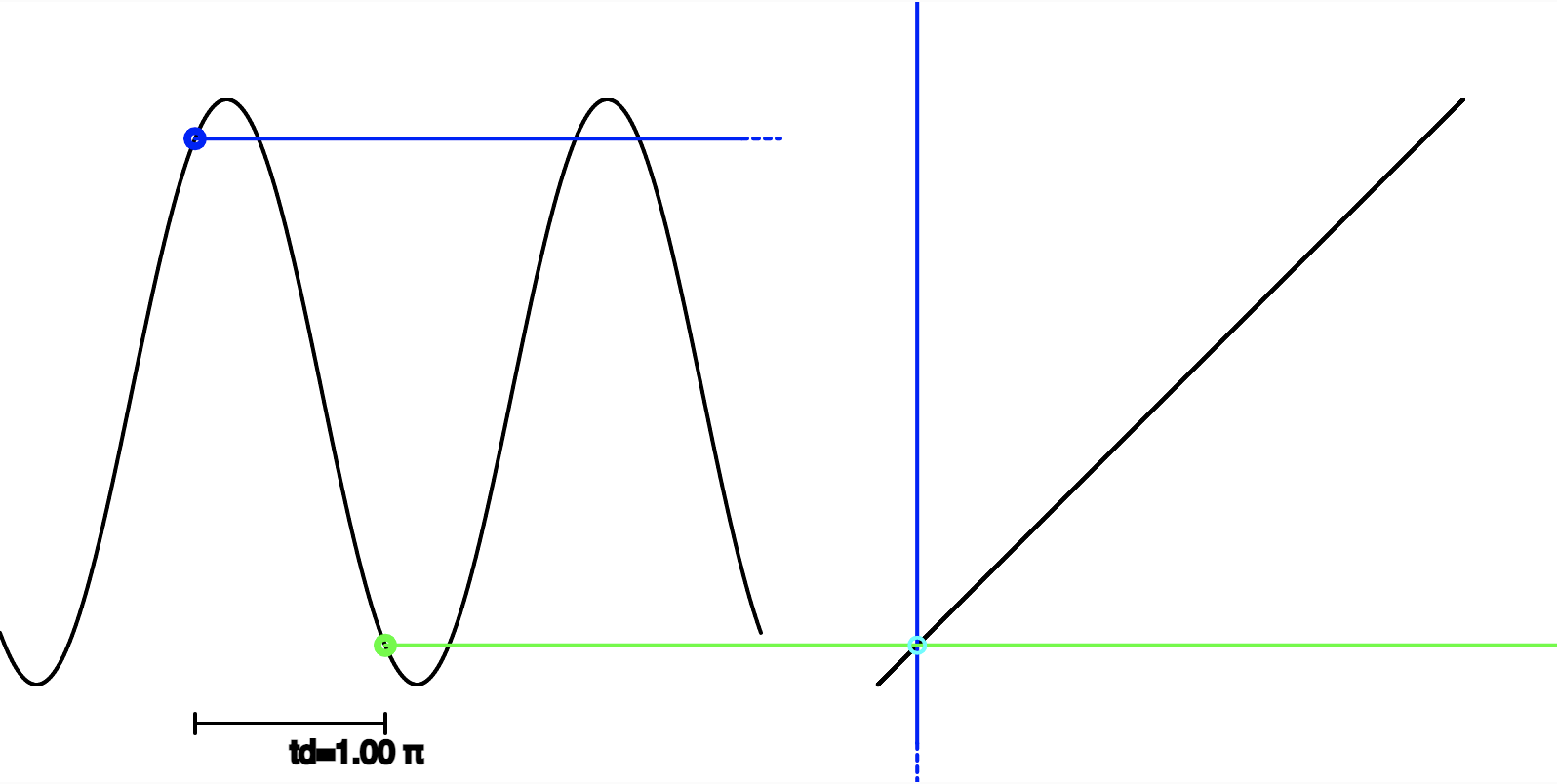}
    \caption{Embeddings of a sinusoid into the plane with one time delay with different delays. For multiples of $pi$ this does not actually give an embedding, both otherwise it does.}
    \label{fig:sinusembedding}
\end{marginfigure}
A particularly popular approach to embed a time-series is known as either time-delay or {\em Takens} embedding. The latter name comes from the work by Florens Takens who has proved that embeddings using these time delay constructions do exist as long as one embeds into a high enough dimensional space, and with some {\em genericity} assumptions. We will not cover the details here but will work on some intuitions.

Given a time function, one can sample the time functions at different delays. A {\em time-delay embedding} creates coordinates at fixed delayed intervals $t_d$ from a given point in time $t$. Hence we get coordinates $(x_0(t),x_1(t+t_d),\ldots,x_n(t+n t_d))$ for an $n$-dimensional embedding. We can view this as a short-term discrete window with a set unit delay of $t_d$.
Figure \ref{fig:sinusembedding} shows the time-delay embedding of a sinusoid into the plane. We should expect a time delay of $\frac{\pi}{2}$ to give us the circle. Other time delays give us circle topologies but they embed as skewed ovals of some sort. There is, however, a {\em degenerate} case, when the delay is an integer multiple of $\pi$ as depicted in the bottom case. Then we do not get an embedding but a line with multiplicity $2$ everywhere except at the extremes. Even adding higher dimensions, that is more delay coordinates at the same delay will never not keep this a line. This is perhaps the simplest example of why the embedding result of Takens needs genericity assumptions. Things are still good though, because even a tiny deviation will give us an embedding, and in general we expect that almost all time delays will give us an embedding.
\begin{marginfigure}
    \centering
    \includegraphics[width=.95\marginparwidth]{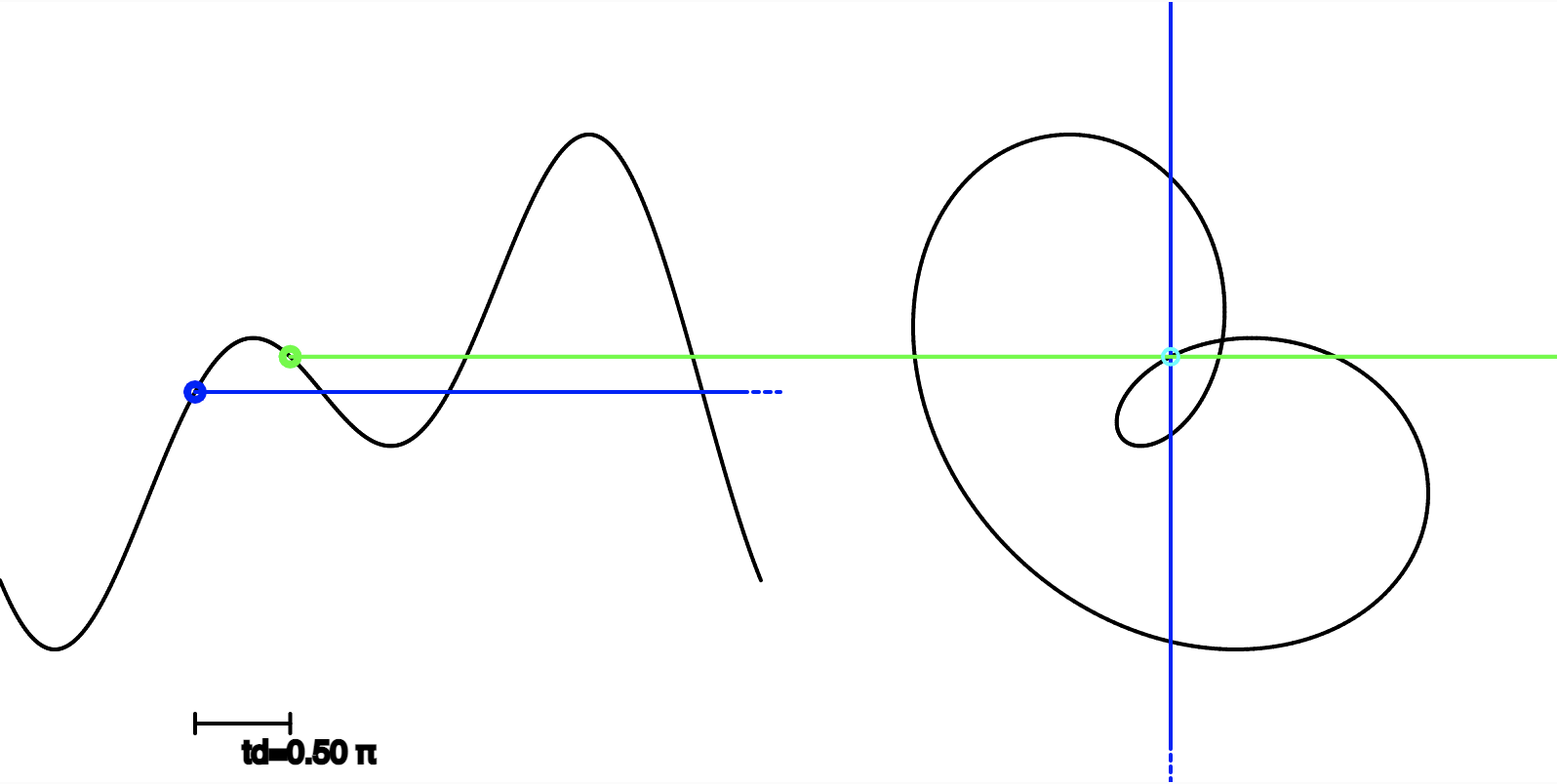}
    \includegraphics[width=.95\marginparwidth]{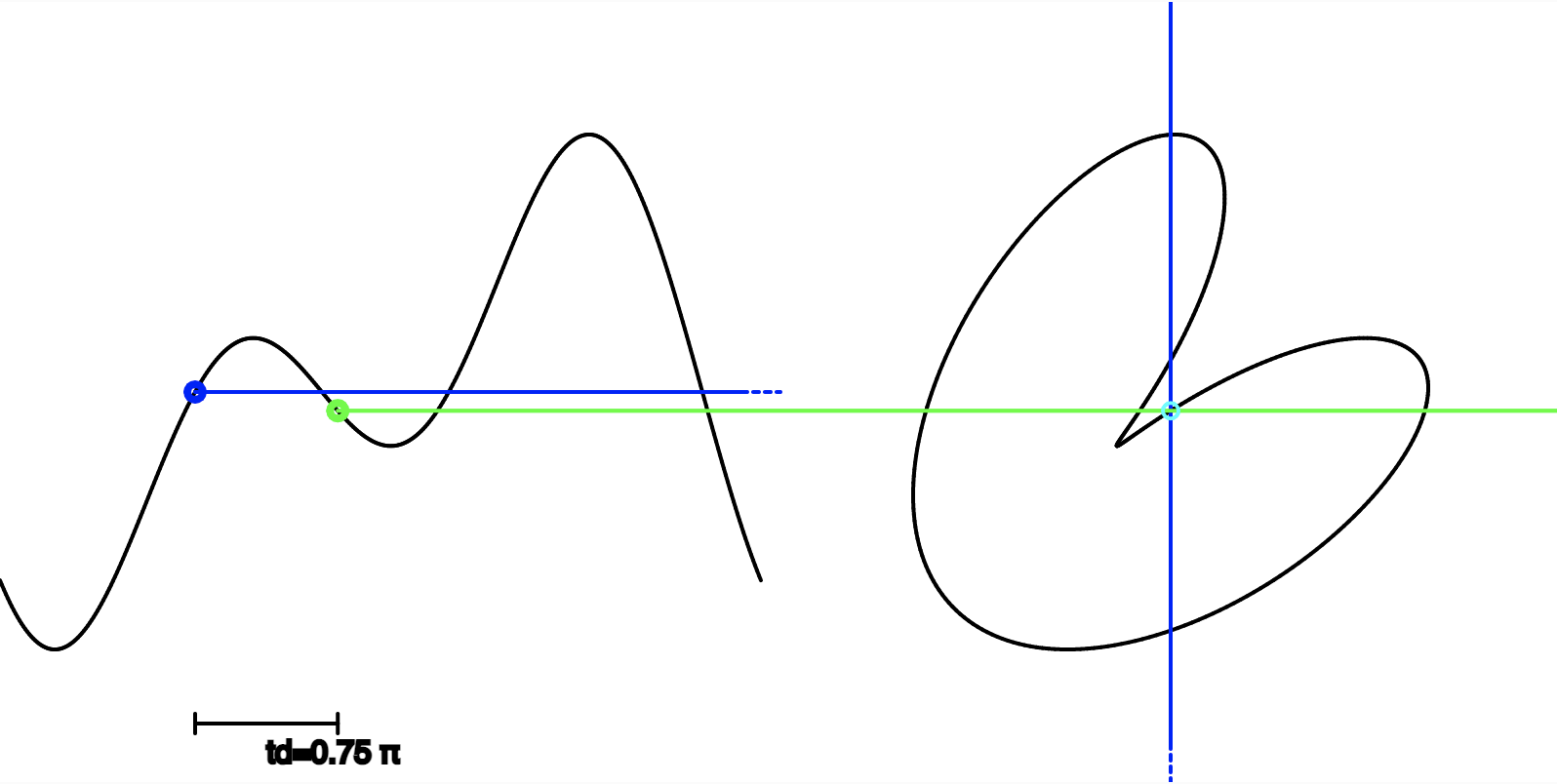}
    \includegraphics[width=.95\marginparwidth]{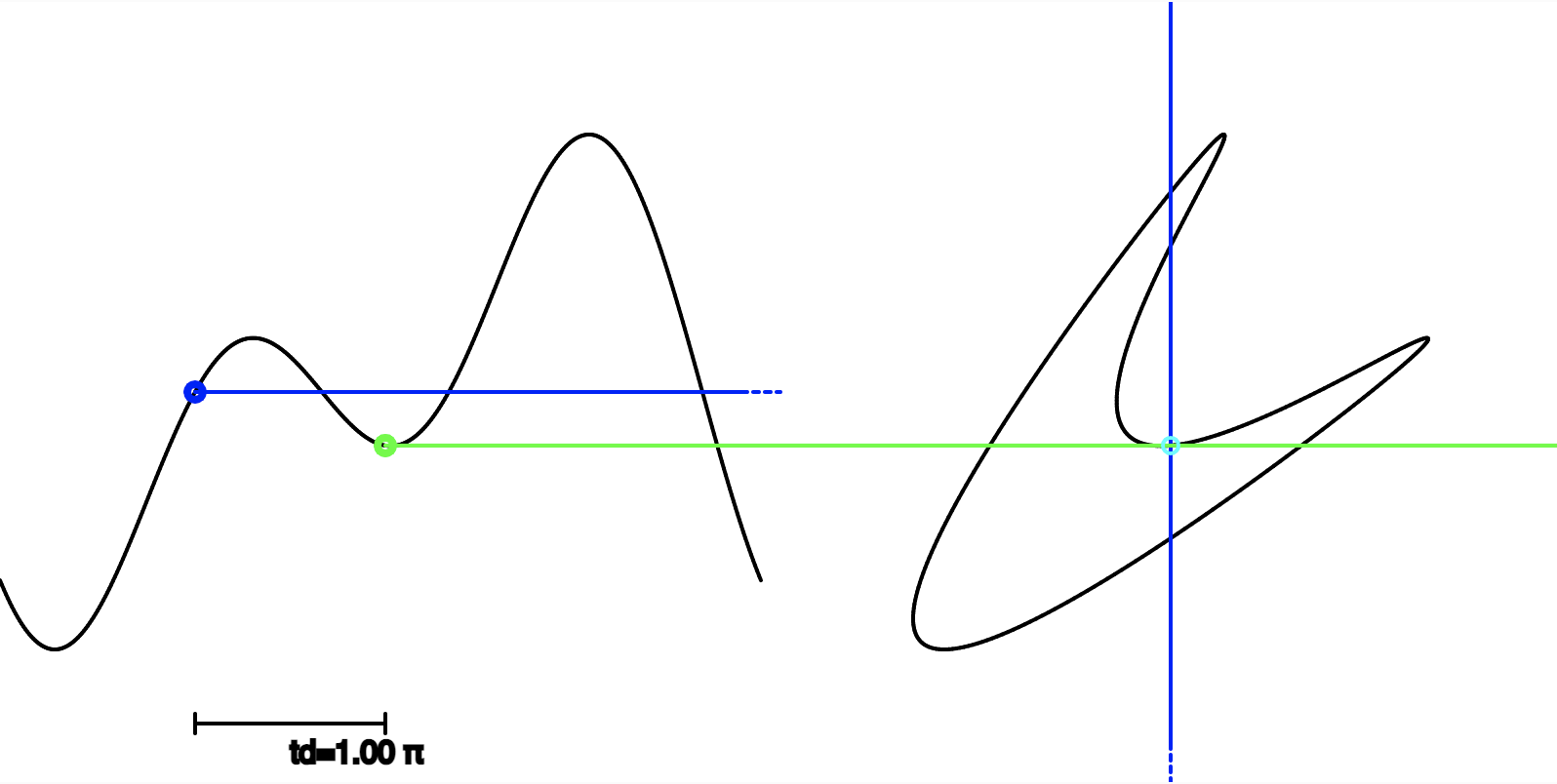}
    \caption{Embeddings of a non-sinusoidal function into the plane with one time delay with different delays.}
    \label{fig:twosinembedding}
\end{marginfigure}
Consider the next example, which is a somewhat more complicated waveform (Figure \ref{fig:twosinembedding}). The top embedding self-intersects, and even wiggling a little bit will not resolve this self-intersection, but will merely move it around somewhat. If we allow ourselves larger changes in time-delay we can however find delays at which it is an embedding, but it is not generically true for this function. However if you look at the lower two figures as side-projections of a 3-dimensional embedding it suggests that indeed lifting this curve into another dimension will resolve the crossing hence, generically, lead to embeddings. This is precisely what happens and the intuitions how crossings of complicated functions can be resolved by giving more dimensional space.

\section{Topological Filters via Sheaves}

Sheaves provide a way to attach data to a topological space. The mechanism is very general because what we mean here by data is very general. Sheaf theory is much more general than we will discuss here. Our sheaves are attached over simplicial complexes. Consider a simple example of a {\em line complex} which is the repeated alternation of a $0$-simplex and a $1$-simplex mirroring the pattern of a connected sampled line. 
%
\begin{equation*}
\begin{tikzcd}[scale=1.2,ampersand replacement=\&, column sep=large,/tikz/bullet/.style={circle,fill,inner
sep=2pt,label={[font=\bfseries]#1}},>=latex,execute at end picture={
    \draw [black,line width=1pt,->-/.list={1/2},shorten >=-2.5ex,shorten <=-2.5ex]
    (\tikzcdmatrixname-1-1) node[black,bullet] (a) {}
    -- (\tikzcdmatrixname-1-3) node[black,bullet] (b) {};
    \draw [black,line width=1pt,->-/.list={1/2},shorten >=-2.5ex,shorten <=-2.5ex]     (\tikzcdmatrixname-1-3) node[black,bullet] (b) {}
    -- (\tikzcdmatrixname-1-5) node[black,bullet] (c) {};
  }]
\textcolor{white}{A} \& \textcolor{white}{A} \&\textcolor{white}{A} \&\textcolor{white}{A} \& \textcolor{white}{A}\\[-15pt]
{\mathcal{X}_0} \arrow[r,phantom]{f} \& {\mathcal{X}_1} \&\arrow[l,phantom,swap]{f} \mathcal{X}_0 \arrow[r,phantom]{f} \& {\mathcal{X}_1} \& \arrow[l,phantom,swap]{f} \mathcal{X}_0
\end{tikzcd}
\end{equation*}
%
We can relate simplices via (co)face maps. In this example we pick the coface map $\delta$.\marginnote{Navigations between simplices in a line complex via coface maps.} 
%
\begin{equation*}
\begin{tikzcd}[scale=1.2,ampersand replacement=\&, column sep=large,/tikz/bullet/.style={circle,fill,inner
sep=2pt,label={[font=\bfseries]#1}},>=latex,execute at end picture={
    \draw [black,line width=1pt,->-/.list={1/2},shorten >=-2.5ex,shorten <=-2.5ex]
    (\tikzcdmatrixname-1-1) node[black,bullet] (a) {}
    -- (\tikzcdmatrixname-1-3) node[black,bullet] (b) {};
    \draw [black,line width=1pt,->-/.list={1/2},shorten >=-2.5ex,shorten <=-2.5ex]     (\tikzcdmatrixname-1-3) node[black,bullet] (b) {}
    -- (\tikzcdmatrixname-1-5) node[black,bullet] (c) {};
  }]
\textcolor{white}{A} \& \textcolor{white}{A} \&\textcolor{white}{A} \&\textcolor{white}{A} \& \textcolor{white}{A}\\[-15pt]
{\mathcal{X}_0} \arrow{r}{\delta} \& {\mathcal{X}_1} \&\arrow{l}[swap]{\delta}  \mathcal{X}_0 \arrow{r}{\delta} \& {\mathcal{X}_1} \& \arrow{l}[swap]{\delta}  \mathcal{X}_0
\end{tikzcd}
\end{equation*}
We will denote some data attached to a simplex by $\mathcal{S}$.
\begin{equation*}
\begin{tikzcd}[scale=1.2,ampersand replacement=\&, column sep=large,/tikz/bullet/.style={circle,fill,inner
sep=2pt,label={[font=\bfseries]#1}},>=latex,execute at end picture={
    \draw [black,line width=1pt,->-/.list={1/2},shorten >=-2.5ex,shorten <=-2.5ex]
    (\tikzcdmatrixname-2-1) node[black,bullet] (a) {}
    -- (\tikzcdmatrixname-2-3) node[black,bullet] (b) {};
    \draw [black,line width=1pt,->-/.list={1/2},shorten >=-2.5ex,shorten <=-2.5ex]     (\tikzcdmatrixname-2-3) node[black,bullet] (b) {}
    -- (\tikzcdmatrixname-2-5) node[black,bullet] (c) {};
  }]
\phantom{\cdots} \arrow[r,phantom] \& \phantom{0} \&\arrow[l,phantom] \mathcal{S}_{\phantom{o}} \arrow[r,phantom] \& \phantom{0} \& \arrow[l,phantom] \phantom{\cdots}\\[-15pt]
\textcolor{white}{A} \& \textcolor{white}{A} \&\textcolor{white}{A} \&\textcolor{white}{A} \& \textcolor{white}{A}\\[-15pt]
{\mathcal{X}_0} \arrow{r}{\delta} \& {\mathcal{X}_1} \&\arrow{l}[swap]{\delta}  \mathcal{X}_0 \arrow{r}{\delta} \& {\mathcal{X}_1} \& \arrow{l}[swap]{\delta}  \mathcal{X}_0
\end{tikzcd}
\end{equation*}
The definition of a sheaf requires that we attach data $\mathcal{S}$ to each simplex $\mathcal{X}_\bullet$.\marginnote{Sheaves consist of data $\mathcal{S}$ attached to each simplex.}  
%
\begin{equation*}
\begin{tikzcd}[scale=1.2,ampersand replacement=\&, column sep=large,/tikz/bullet/.style={circle,fill,inner
sep=2pt,label={[font=\bfseries]#1}},>=latex,execute at end picture={
    \draw [black,line width=1pt,->-/.list={1/2},shorten >=-2.5ex,shorten <=-2.5ex]
    (\tikzcdmatrixname-2-1) node[black,bullet] (a) {}
    -- (\tikzcdmatrixname-2-3) node[black,bullet] (b) {};
    \draw [black,line width=1pt,->-/.list={1/2},shorten >=-2.5ex,shorten <=-2.5ex]     (\tikzcdmatrixname-2-3) node[black,bullet] (b) {}
    -- (\tikzcdmatrixname-2-5) node[black,bullet] (c) {};
  }]
\cdots \arrow[r,phantom] \& \mathcal{S} \&\arrow[l,phantom] \mathcal{S}_{\phantom{o}} \arrow[r,phantom] \& \mathcal{S} \& \arrow[l,phantom] \cdots\\[-15pt]
\textcolor{white}{A} \& \textcolor{white}{A} \&\textcolor{white}{A} \&\textcolor{white}{A} \& \textcolor{white}{A}\\[-15pt]
{\mathcal{X}_0} \arrow{r}{\delta} \& {\mathcal{X}_1} \&\arrow{l}[swap]{\delta}  \mathcal{X}_0 \arrow{r}{\delta} \& {\mathcal{X}_1} \& \arrow{l}[swap]{\delta}  \mathcal{X}_0
\end{tikzcd}
\end{equation*}
A sheaf construction further requires that whenever there is a map between simplices, we have to provide a map between the attached data.\marginnote{Sheaves also require that for each map between simplices we provide a map between sheaf data.}  
%
\begin{equation*}
\begin{tikzcd}[scale=1.2,ampersand replacement=\&, column sep=large,/tikz/bullet/.style={circle,fill,inner
sep=2pt,label={[font=\bfseries]#1}},>=latex,execute at end picture={
    \draw [black,line width=1pt,->-/.list={1/2},shorten >=-2.5ex,shorten <=-2.5ex]
    (\tikzcdmatrixname-2-1) node[black,bullet] (a) {}
    -- (\tikzcdmatrixname-2-3) node[black,bullet] (b) {};
    \draw [black,line width=1pt,->-/.list={1/2},shorten >=-2.5ex,shorten <=-2.5ex]     (\tikzcdmatrixname-2-3) node[black,bullet] (b) {}
    -- (\tikzcdmatrixname-2-5) node[black,bullet] (c) {};
  }]
\cdots \arrow[r] \& \mathcal{S} \&\arrow[l] \mathcal{S}_{\phantom{o}} \arrow[r] \& \mathcal{S} \& \arrow[l] \cdots\\[-15pt]
\textcolor{white}{A} \& \textcolor{white}{A} \&\textcolor{white}{A} \&\textcolor{white}{A} \& \textcolor{white}{A}\\[-15pt]
{\mathcal{X}_0} \arrow{r}{\delta} \& {\mathcal{X}_1} \&\arrow{l}[swap]{\delta}  \mathcal{X}_0 \arrow{r}{\delta} \& {\mathcal{X}_1} \& \arrow{l}[swap]{\delta}  \mathcal{X}_0
\end{tikzcd}
\end{equation*}
Notice that typically we have more than one map pointing to the same data $\mathcal{S}$. This leads to a final requirement for sheaves. Informally we have to avoid that there is a conflict between these two maps. They have to in some suitable sense agree what $\mathcal{S}$ is. This can be thought of in two ways. The first is via composition. One can require that the two maps can be {\em composed}. Another way to think about this is to require that local data has to be {\em consistent}. With these three rules we have a full definition of a sheaf over a simplicial complex. In examples we will see how that is realized in practice soon.\marginnote{Sheaves maps are required to allow composition. Alternatively we can think of local sheaf data being required to be consistent.}  
%
\begin{equation*}
\begin{tikzcd}[scale=1.2,ampersand replacement=\&, column sep=large,/tikz/bullet/.style={circle,fill,inner
sep=2pt,label={[font=\bfseries]#1}},>=latex,execute at end picture={
    \draw [black,line width=1pt,->-/.list={1/2},shorten >=-2.5ex,shorten <=-2.5ex]
    (\tikzcdmatrixname-2-1) node[black,bullet] (a) {}
    -- (\tikzcdmatrixname-2-3) node[black,bullet] (b) {};
    \draw [black,line width=1pt,->-/.list={1/2},shorten >=-2.5ex,shorten <=-2.5ex]     (\tikzcdmatrixname-2-3) node[black,bullet] (b) {}
    -- (\tikzcdmatrixname-2-5) node[black,bullet] (c) {};
    \node [rounded corners,draw,dashed,
    inner xsep=-1pt, blue,line width=1.5pt,fit={(\tikzcdmatrixname-1-1) (\tikzcdmatrixname-1-2)}]{};
    \node [rounded corners,draw,dashed,
    inner xsep=-1pt, green,line width=1.5pt,fit={(\tikzcdmatrixname-1-2) (\tikzcdmatrixname-1-3)}]{};
  }]
\cdots \arrow[r] \& \mathcal{S} \&\arrow[l] \mathcal{S}_{\phantom{o}} \arrow[r] \& \mathcal{S} \& \arrow[l] \cdots\\[-15pt]
\textcolor{white}{A} \& \textcolor{white}{A} \&\textcolor{white}{A} \&\textcolor{white}{A} \& \textcolor{white}{A}\\[-15pt]
{\mathcal{X}_0} \arrow{r}{\delta} \& {\mathcal{X}_1} \&\arrow{l}[swap]{\delta}  \mathcal{X}_0 \arrow{r}{\delta} \& {\mathcal{X}_1} \& \arrow{l}[swap]{\delta}  \mathcal{X}_0
\end{tikzcd}
\end{equation*}
We are interested in one particular sheaf structure called {\em topological filters} as invented by Michael Robinson \cite{robinson2014topological}. A topological filter assume the typical structure that we assume a filter to have. It separates input, state and output. Later we will see that this structure can be used both for classical linear time-invariant filters as well as non-linear oscillatory structures.\marginnote{A topological filter has input, state, and output data connected by computation.}  
\begin{equation*}
\begin{tikzcd}[scale=1.2,ampersand replacement=\&, column sep=large,/tikz/bullet/.style={circle,fill,inner
sep=2pt,label={[font=\bfseries]#1}},>=latex,execute at end picture={
    \draw [black,line width=1pt,->-/.list={1/2},shorten >=-2.5ex,shorten <=-2.5ex]
    (\tikzcdmatrixname-4-1) node[black,bullet] (a) {}
    -- (\tikzcdmatrixname-4-3) node[black,bullet] (b) {};
    \draw [black,line width=1pt,->-/.list={1/2},shorten >=-2.5ex,shorten <=-2.5ex]     (\tikzcdmatrixname-4-3) node[black,bullet] (b) {}
    -- (\tikzcdmatrixname-4-5) node[black,bullet] (c) {};
    \node [label={[xshift=72pt,yshift=+8pt]Topological Filter}] (\tikzcdmatrixname-1-3) {};
  }]
\phantom{\cdots} \arrow[r,phantom] \& \phantom{0} \&\arrow[l,phantom] \mathcal{S}_i \arrow[r,phantom] \& \phantom{0} \& \arrow[l,phantom] \phantom{\cdots}\\
\phantom{\cdots} \arrow[r,phantom] \& \phantom{\mathcal{S}_{c}} \arrow[u,phantom] \arrow[d,phantom]  \&\arrow[l,phantom]{r} \mathcal{S}_{s} \arrow{u}{i} \arrow{d}{o} \arrow[r,phantom]{s} \& \phantom{\mathcal{S}_{c}} \arrow[u,phantom] \arrow[d,phantom] \& \arrow[l,phantom]  \phantom{\cdots}\\
\phantom{\cdots} \arrow[r,phantom] \& \phantom{0} \&\arrow[l,phantom] \mathcal{S}_o \arrow[r,phantom] \& \phantom{0} \& \arrow[l,phantom] \phantom{\cdots}\\[-15pt]
\textcolor{white}{A} \& \textcolor{white}{A} \&\textcolor{white}{A} \&\textcolor{white}{A} \& \textcolor{white}{A}\\[-15pt]
{\mathcal{X}_0} \arrow{r}{\delta} \& {\mathcal{X}_1} \&\arrow{l}[swap]{\delta}  \mathcal{X}_0 \arrow{r}{\delta} \& {\mathcal{X}_1} \& \arrow{l}[swap]{\delta}  \mathcal{X}_0
\end{tikzcd}
\end{equation*}
By the rules of sheaves we replicate this structure over all simplices.
\begin{equation*}
\begin{tikzcd}[scale=1.2,ampersand replacement=\&, column sep=large,/tikz/bullet/.style={circle,fill,inner
sep=2pt,label={[font=\bfseries]#1}},>=latex,execute at end picture={
    \draw [black,line width=1pt,->-/.list={1/2},shorten >=-2.5ex,shorten <=-2.5ex]
    (\tikzcdmatrixname-4-1) node[black,bullet] (a) {}
    -- (\tikzcdmatrixname-4-3) node[black,bullet] (b) {};
    \draw [black,line width=1pt,->-/.list={1/2},shorten >=-2.5ex,shorten <=-2.5ex]     (\tikzcdmatrixname-4-3) node[black,bullet] (b) {}
    -- (\tikzcdmatrixname-4-5) node[black,bullet] (c) {};
  }]
\cdots \arrow[r,phantom] \& \mathcal{S}_i \&\arrow[l,phantom] \mathcal{S}_i \arrow[r,phantom] \& \mathcal{S}_i \& \arrow[l,phantom] \cdots\\
\cdots \arrow[r,phantom] \& \mathcal{S}_{s} \arrow[u] \arrow[d]  \&\arrow[l,phantom]{r} \mathcal{S}_{s} \arrow{u}{i} \arrow{d}{o} \arrow[r,phantom]{s} \& \mathcal{S}_{s} \arrow[u] \arrow[d] \& \arrow[l,phantom]  \cdots\\
\cdots \arrow[r,phantom] \& \mathcal{S}_o \&\arrow[l,phantom] \mathcal{S}_o \arrow[r,phantom] \& \mathcal{S}_o \& \arrow[l,phantom] \cdots\\[-15pt]
\textcolor{white}{A} \& \textcolor{white}{A} \&\textcolor{white}{A} \&\textcolor{white}{A} \& \textcolor{white}{A}\\[-15pt]
{\mathcal{X}_0} \arrow{r}{\delta} \& {\mathcal{X}_1} \&\arrow{l}[swap]{\delta}  \mathcal{X}_0 \arrow{r}{\delta} \& {\mathcal{X}_1} \& \arrow{l}[swap]{\delta}  \mathcal{X}_0
\end{tikzcd}
\end{equation*}
And we have to provide maps between all the data.

\begin{equation*}
\begin{tikzcd}[scale=1.2,ampersand replacement=\&, column sep=large,/tikz/bullet/.style={circle,fill,inner
sep=2pt,label={[font=\bfseries]#1}},>=latex,execute at end picture={
    \draw [black,line width=1pt,->-/.list={1/2},shorten >=-2.5ex,shorten <=-2.5ex]
    (\tikzcdmatrixname-4-1) node[black,bullet] (a) {}
    -- (\tikzcdmatrixname-4-3) node[black,bullet] (b) {};
    \draw [black,line width=1pt,->-/.list={1/2},shorten >=-2.5ex,shorten <=-2.5ex]     (\tikzcdmatrixname-4-3) node[black,bullet] (b) {}
    -- (\tikzcdmatrixname-4-5) node[black,bullet] (c) {};
  }]
\cdots \arrow{r} \& \mathcal{S}_i \&\arrow{l} \mathcal{S}_i \arrow{r} \& \mathcal{S}_i \& \arrow{l} \cdots\\
\cdots \arrow{r} \& \mathcal{S}_{s} \arrow{u} \arrow{d}  \&\arrow{l}{r} \mathcal{S}_{s} \arrow{u}{i} \arrow{d}{o} \arrow{r}{s} \& \mathcal{S}_{s} \arrow{u} \arrow{d} \& \arrow{l}  \cdots\\
\cdots \arrow{r} \& \mathcal{S}_o \&\arrow{l} \mathcal{S}_o \arrow{r} \& \mathcal{S}_o \& \arrow{l} \cdots\\[-15pt]
\textcolor{white}{A} \& \textcolor{white}{A} \&\textcolor{white}{A} \&\textcolor{white}{A} \& \textcolor{white}{A}\\[-15pt]
\mathcal{X}_0 \arrow{r}{\delta} \& \mathcal{X}_1 \&\arrow{l}[swap]{\delta} \mathcal{X}_0 \arrow{r}{\delta} \& \mathcal{X}_1 \& \arrow{l}[swap]{\delta} \mathcal{X}_0
\end{tikzcd}
\end{equation*}

\begin{figure}
    \centering
    \includegraphics[width=.75\textwidth]{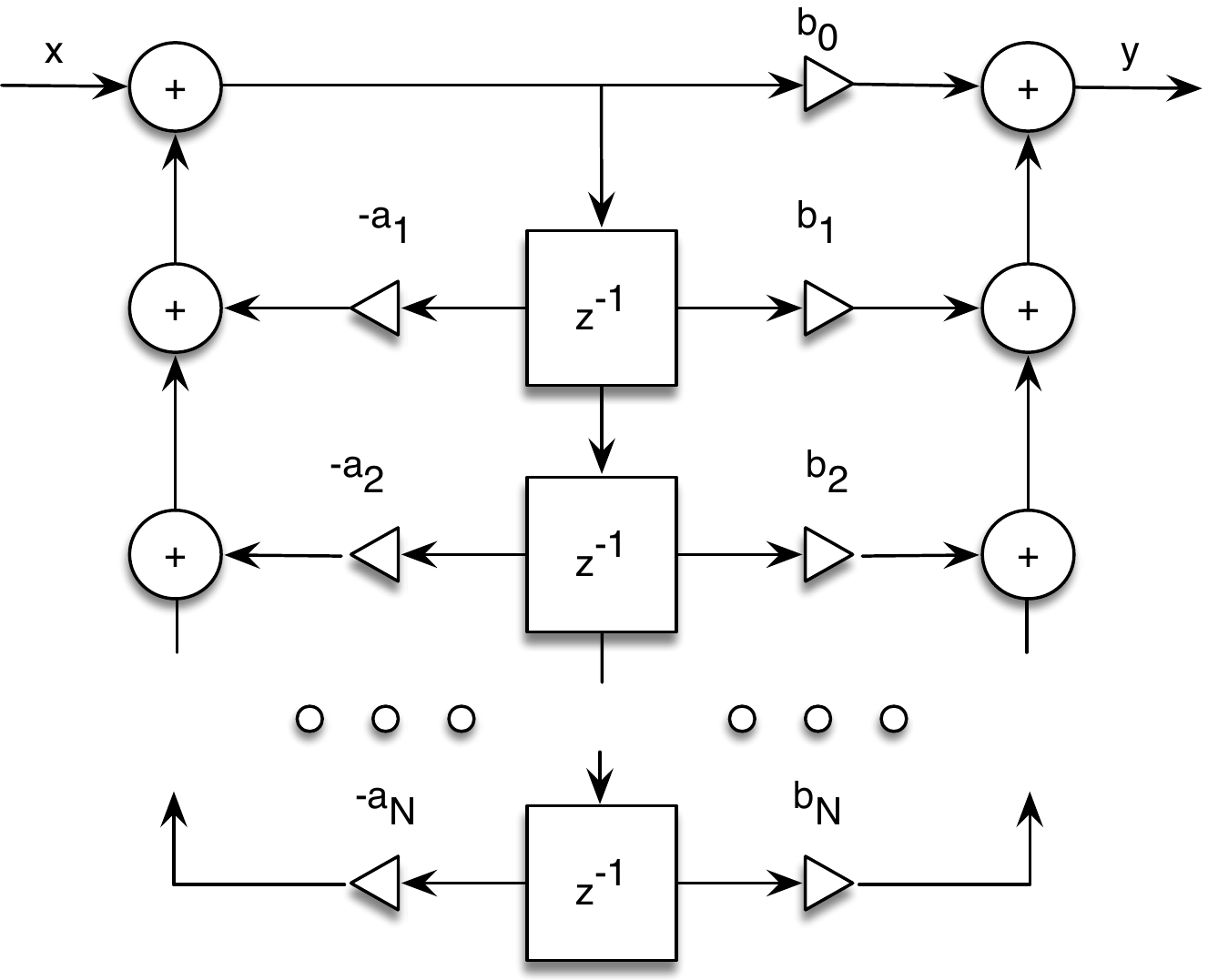}
    \caption{General LTI filter in Direct-II form.}
    \label{fig:iirfilter}
\end{figure}

To understand how to place general linear time-invariant filters into this sheaf formalism consider, the structure of the Direct-II form is particularly convenient (Figure \ref{fig:iirfilter}) for the filter equation:
\begin{align*}
 y&=\sum_{i=1}^N{b_i\cdot x_{N-i}} + b_0\cdot x +\sum_{j=1}^{N}{-a_j\cdot y_{N-j}}\label{eq:geniir}
\end{align*}
%
This filter structure suggests that our sheaf state should capture the shift register information. We can model this as a finite vector space $\mathbb{R}^N$. At a local position, we have a given input $i$ that is injected into the state and leading to an augmented state $\mathbb{R}^{N+1}$. From this we can derive the output map $o$ being the feedforward computation of the filter. The map $s$ captures both the shift of the register and the feedback computation. Maps $i$ and $r$ simple inject information into the next step. Given that there is no input over $1$-simplices the state over them is $\mathbb{R}^N$. Input and output in this construction are isolated audio samples in $\mathbb{R}$. Hence we get the following sheaf structure for general linear time invariant filter formulated as topological filters:\marginnote{LTI filters maps for topological filters.}  
\begin{equation*}
\begin{tikzcd}[scale=1.2,ampersand replacement=\&, column sep=large,/tikz/bullet/.style={circle,fill,inner
sep=2pt,label={[font=\bfseries]#1}},>=latex,execute at end picture={
    \draw [black,line width=1pt,->-/.list={1/2},shorten >=-2.5ex,shorten <=-2.5ex]
    (\tikzcdmatrixname-4-1) node[black,bullet] (a) {}
    -- (\tikzcdmatrixname-4-3) node[black,bullet] (b) {};
    \draw [black,line width=1pt,->-/.list={1/2},shorten >=-2.5ex,shorten <=-2.5ex]     (\tikzcdmatrixname-4-3) node[black,bullet] (b) {}
    -- (\tikzcdmatrixname-4-5) node[black,bullet] (c) {};
  }]
\cdots \arrow{r} \& 0 \&\arrow{l} \mathbb{R} \arrow{r} \& 0 \& \arrow{l} \cdots\\
\cdots \arrow{r} \& \mathbb{R}^N \arrow{u} \arrow{d}  \&\arrow{l}{r} \mathbb{R}^{N+1} \arrow{u}{i} \arrow{d}{o} \arrow{r}{s} \& \mathbb{R}^N \arrow{u} \arrow{d} \& \arrow{l}  \cdots\\
\cdots \arrow{r} \& 0 \&\arrow{l} \mathbb{R} \arrow{r} \& 0 \& \arrow{l} \cdots\\[-15pt]
\textcolor{white}{A} \& \textcolor{white}{A} \&\textcolor{white}{A} \&\textcolor{white}{A} \& \textcolor{white}{A}\\[-15pt]
\phantom{\mathcal{X}_0} \arrow[r,phantom]{f} \& \phantom{\mathcal{X}_1} \&\arrow[l,phantom,swap]{f} \phantom{\mathcal{X}_0} \arrow[r,phantom]{f} \& \phantom{\mathcal{X}_1} \& \arrow[l,phantom,swap]{f} \phantom{\mathcal{X}_0}
\end{tikzcd}
\end{equation*}
The following sheaf maps are equivalent to the filter equation:
\begin{align*}
\begin{split}
s: (x_0,x_1,...,x_{N-1},x) \rightarrow& (x_1,x_2,...,x_{N-1},\\&  x+\sum_{j=1}^{N}{-a_j\cdot x_{N-j}})
\end{split}\\
r: (x_0,x_1,...,x_{N-1},x) \rightarrow& (x_0,x_1,...,x_{N-1})\\
i: (x_0,x_1,...,x_{N-1},x) \rightarrow& (x)\\
o: (x_0,x_1,...,x_{N-1},x) \rightarrow& (b_0 \cdot x + \sum_{i=1}^N{b_i\cdot x_{N-i}})
\end{align*}
Maps $i$ and $r$ just copy information around, so we are left with maps $s$ and $o$ to do actual computation. The $s$-map captures the dynamical update of the filter, while the $o$ captures the static feed-forward computation locally. The sheaf construction alone allows us to read of properties of digital filters. For example the feed-forward computation is unconditionally stable because it does no longer interact with any other data on the sheaf, as it is surrounded by $0$-maps on the output side. With this construction any general IIR filter can be associated with a topological path construction.

But the construction is not confined to linear filters.\marginnote{Sheaf maps can be nonlinear.}  Highly nonlinear computations can also be treated in this framework. If instead of vectors in a vector space we consider ensembles of circle and maps between them, we can capture many oscillatory sound synthesis methods and topologize them in essentially the same way. Consider this sheaf structure:\marginnote{Topological filters using maps between ensemble of circles.} 

\begin{equation*}
\begin{tikzcd}[scale=1.2,ampersand replacement=\&, column sep=large,/tikz/bullet/.style={circle,fill,inner
sep=2pt,label={[font=\bfseries]#1}},>=latex,execute at end picture={
    \draw [black,line width=1pt,->-/.list={1/2},shorten >=-2.5ex,shorten <=-2.5ex]
    (\tikzcdmatrixname-4-1) node[black,bullet] (a) {}
    -- (\tikzcdmatrixname-4-3) node[black,bullet] (b) {};
    \draw [black,line width=1pt,->-/.list={1/2},shorten >=-2.5ex,shorten <=-2.5ex]     (\tikzcdmatrixname-4-3) node[black,bullet] (b) {}
    -- (\tikzcdmatrixname-4-5) node[black,bullet] (c) {}; 
  }]
\cdots \arrow{r} \& 0 \&\arrow{l} \mathbb{S}^1 \arrow{r} \& 0 \& \arrow{l} \cdots\\
\cdots \arrow{r} \& (\mathbb{S}^1)^N \arrow{u} \arrow{d}  \&\arrow{l}{r} (\mathbb{S}^1)^{N+1} \arrow{u}{i} \arrow{d}{o} \arrow{r}{s} \& (\mathbb{S}^1)^N \arrow{u} \arrow{d} \& \arrow{l}  \cdots\\
\cdots \arrow{r} \& 0 \&\arrow{l} \mathbb{R} \arrow{r} \& 0 \& \arrow{l} \cdots\\[-15pt]
\textcolor{white}{A} \& \textcolor{white}{A} \&\textcolor{white}{A} \&\textcolor{white}{A} \& \textcolor{white}{A}\\[-15pt]
\phantom{\mathcal{X}_0} \arrow[r,phantom]{f} \& \phantom{\mathcal{X}_1} \&\arrow[l,phantom,swap]{f} \phantom{\mathcal{X}_0} \arrow[r,phantom]{f} \& \phantom{\mathcal{X}_1} \& \arrow[l,phantom,swap]{f} \phantom{\mathcal{X}_0}
\end{tikzcd}
\end{equation*}

Perhaps one of the more prominent examples of oscillatory phenomena is {\em Frequency Modulation} (or just {\em FM}).\marginnote{Sheaf maps for frequency modulation.} 

\begin{align*}
\begin{split}
s: (x_0,z_0,x) \rightarrow& (x_0 + \Omega+H\sin(2\pi z_0) + x \mod 1 \\
&,z_0+\omega_m \mod 1)
\end{split}\\
r: (x_0,z_0,x) \rightarrow& (x_0,z_0)\\
i: (x_0,z_0,x) \rightarrow& (x)\\
o: (x_0,z_0,x) \rightarrow& (sin(2\pi x_0+\phi))
\end{align*}

But whatever the computations are that are formalized in this way, we can now associate them with topological construction and perform the computations.
\begin{marginfigure}[8\baselineskip]
    \centering
    \includegraphics[width=.95\marginparwidth]{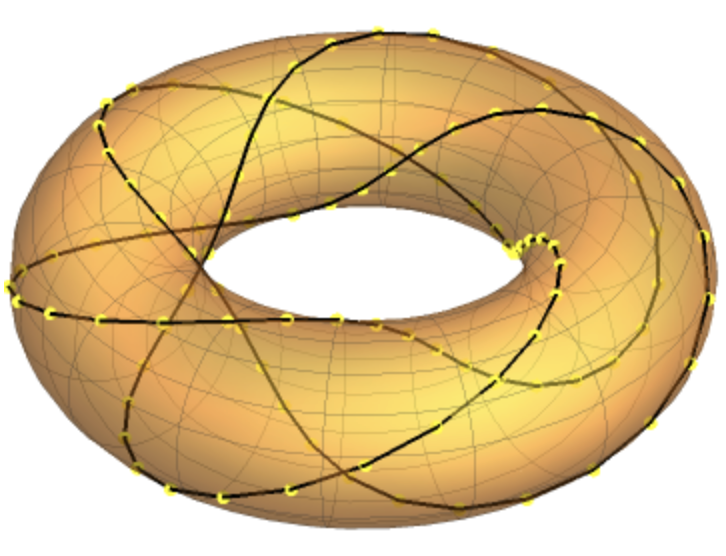}
    \includegraphics[width=.95\marginparwidth]{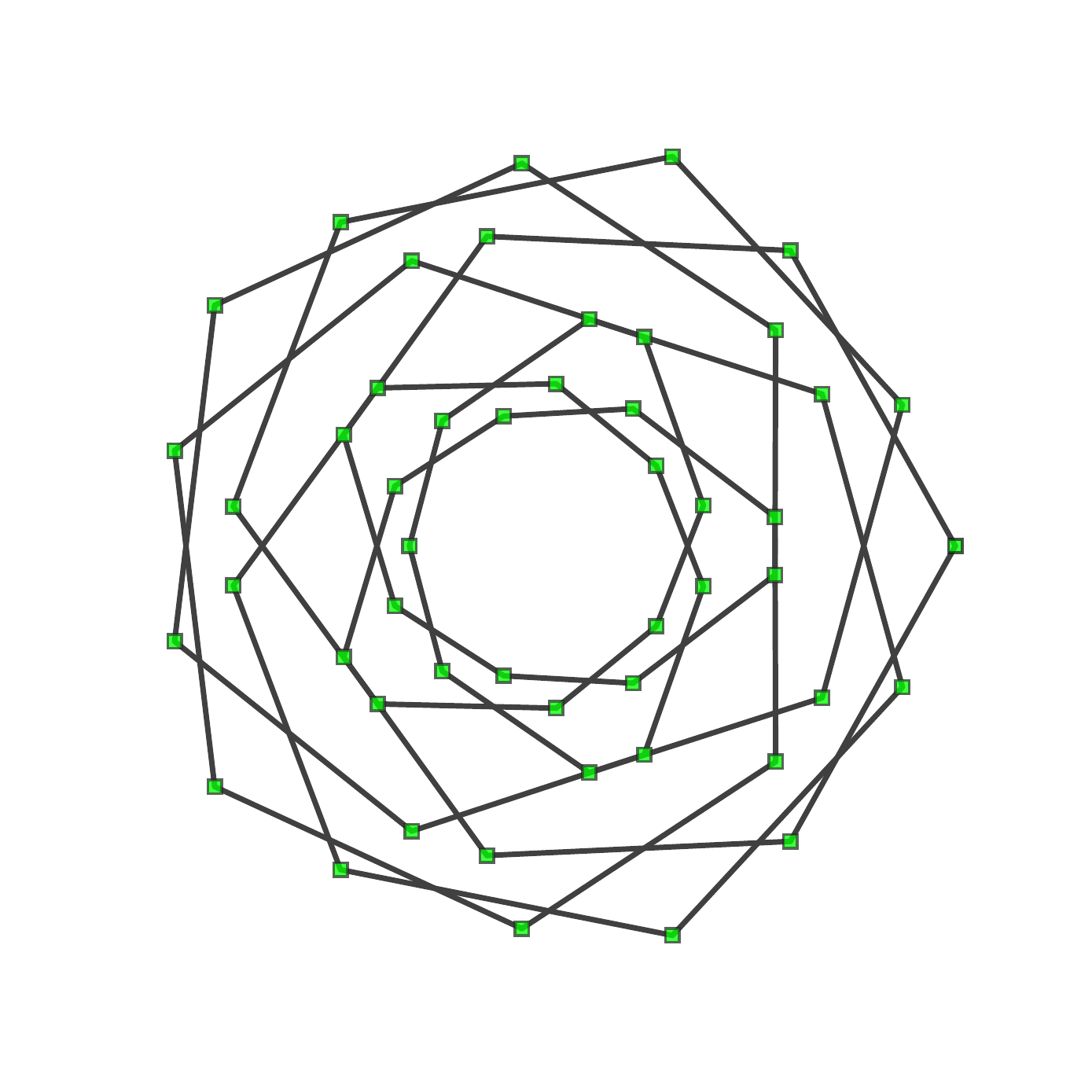}
    \caption{A close line complex on a torus and its top projection. It gives variation in distance that is used to render sound.}
    \label{fig:toruspath}
\end{marginfigure}

Figure \ref{fig:toruspath} shows a line path that winds around a torus and closes onto itself. We can use this to create metric information that specifies the distance between sample points. Standard digital filter algorithms, such as simple resonators using BiQuad filters, or plugged string simulations using Karplus-Strong can be realized using the LTI-based sheaf filter, and FM and many other oscillatory methods can be realized using the circle map sheaf.

\section{Epilogue --- More to know that we did not cover}

The fields of topological signal processing and topological data analysis are much richer than could be covered here. For example topics such as {\em Graph Signal Processing}, {\em Combinatorial Hodge Theory}, {\em Sheaf Cohomology}, {\em Topological Features in Machine Learning} have all been left out completely. All of them are however interesting and worthwhile to learn about.

\subsection{Vistas for future research}

Despite its substantial history topological data analysis and topological signal processing have left many open research avenues ripe for exploration. In particular in the real of audio signal processing there remains much to be done. Here are but two of many possible topics of interest: {\em Topological Harmonic Analysis - Build bridge between Persistence and Fourier}, {\em Sheaves over higher order topologies}.

\subsection{Further Reading}
To learn more details and cover gaps in the material discussed here, the following are good yet fairly accessible long form treatments.
\begin{itemize}
\item Edelsbrunner \& Harrer, Computational Topology, 2010.
\item Robinson, Topological Signal Processing, 2014.
\item Ghrist, Elementary Applied Topology, 2014.
\item Perea, Jose A. "Topological time series analysis." Notices of the AMS 66:5 (2019).
\end{itemize}

\subsection{Software Recommendations}
The landscape of software implementation of topological algorithms is already fairly vast and rapidly expanding. This is a very small selection curated for either ease of use, relevance to signal processing, or performance.
\begin{itemize}
    \item (Java)Plex (C++,Java) - Persistent Homology: Accessible with visual frontent (Processing/Java)
    \item Ripser (C++) - Persistent Homology: Fast
    \item Teaspoon (Python) - Assorted methods including Topological Signal Processing
\end{itemize}

\section{Acknowledgements}
I owe Gianpaolo Evangelista, chair of DAFx-2022 my gratitude for offering the time to present this material during two tutorial sessions at the conference.  These notes have benefitted from feedback of numerous tutorial attendees, in particular Lauri Savioja for extensive discussion and helpful suggestions for improvements. Figure \ref{fig:embedcircle} is from a 2022 DAFx paper by the author. Figures \ref{fig:pacmantorus} and \ref{fig:toruscover} are snapshots from very nice movie visualizations made by Ajeet Gary, used here and in the DAFx tutorial presentation with his kind permission. Last but not least the material presented is part of a larger book project with the generous support of a Simon Guggenheim Foundation fellowship. 
Figure \ref{fig:simplicialcomplex} is a public domain figure from wikipedia.

\bibliography{main.bib} 
\bibliographystyle{plainnat}

\end{document}